\documentclass[10pt,preprint]{aastex}

\slugcomment{accepted for publication in the Astrophysical Journal}

\shorttitle{Silicate Features in the PKS 1830-211 Absorber}
\shortauthors{Aller et al.}

\begin{document}

\title{Interstellar Silicate Dust in the z=0.89 Absorber Towards PKS 1830-211: Crystalline Silicates at High Redshift?}

\author{Monique C. Aller}
\affil{Department of Physics \& Astronomy, University of South Carolina, 712 Main Street, Columbia, SC 29208, USA}
\email{ALLERM@mailbox.sc.edu}

\author{Varsha P. Kulkarni}
\affil{Department of Physics \& Astronomy, University of South Carolina, 712 Main Street, Columbia, SC 29208, USA}

\author{Donald G. York}
\affil{Department of Astronomy \& Astrophysics, University of Chicago, 5640 S. Elis Ave., Chicago, IL 60637, USA}

\author{Giovanni Vladilo}
\affil{Osservatorio Astonomico di Trieste, Via Tiepolo 11 34143 Trieste, Italy}

\author{Daniel E. Welty}
\affil{Department of Astronomy \& Astrophysics, University of Chicago, 5640 S. Elis Ave., Chicago, IL 60637, USA}

\and

\author{Debopam Som}
\affil{Department of Physics \& Astronomy, University of South Carolina, 712 Main Street, Columbia, SC 29208, USA}

\begin{abstract}
We present evidence of a $>$10-$\sigma$ detection  of the 10~$\micron$ silicate dust absorption feature in the spectrum of the gravitationally lensed quasar PKS 1830-211, produced by a foreground absorption
system at redshift 0.886. We have examined more than 100 optical depth templates, derived from both observations of Galactic and extragalactic sources and laboratory measurements,
in order to constrain the chemical structure of the silicate dust. We find that the best fit to the observed absorption profile is produced by laboratory crystalline olivine, with a corresponding peak optical depth of 
$\tau_{10}=0.27\pm0.05$. The fit is slightly improved upon by including small contributions from additional materials such as silica, enstatite, or serpentine, which suggests that
the dust composition may consist of a blend of crystalline silicates. 
Combining templates for amorphous and crystalline silicates, we find that the fraction of crystalline silicates needs to be at least 95\%. 
Given the rarity of extragalactic sources with such a high degree of silicate crystallinity, we also explore the possibility that the observed spectral 
features are produced by amorphous silicates in combination with other molecular or atomic transitions, or by foreground source contamination. 
While we cannot rule out these latter possibilities, they lead to much poorer profile fits than for the crystalline olivine templates. 
If the presence of crystalline interstellar silicates in this distant galaxy is real, it would be highly unusual, given that the Milky Way interstellar matter contains essentially only amorphous silicates. It is possible that 
the z=0.886 absorber towards PKS 1830-211, well known for its high molecular content, has a unique star-forming environment that enables crystalline silicates to form and prevail.
\end{abstract}


\section{INTRODUCTION}

Dust is a significant constituent of the observable universe, both impacting the appearance of galaxies and influencing their evolution.
It affects the derived properties of some galaxies by attenuating the shorter-wavelength radiation emitted by stars and ionized
gases, and re-emitting this radiation at longer wavelengths; more than 30\% of the energy emitted as starlight is re-radiated in the IR by dust \citep{Bernstein}. 
Furthermore, dust influences physical processes in the interstellar medium (ISM) ranging from heating, cooling, and ionization, to those dictating the production of molecules and the formation of stars. 
Thus, understanding the nature of extragalactic interstellar dust grains is essential for characterizing the chemical evolution of galaxies, and for correcting the observations of local and high redshift objects used
in studies spanning topics from galaxy morphology to cosmology. 

Despite the fundamental importance of dust in analyses characterizing galaxy evolution, however, there remain many open questions about the distribution and composition of dust in non-local galaxies, and 
about the possible cosmological evolution of dust properties. While significant evidence of dust-enshrouded star-formation has been found in mid/far-IR and sub-mm studies of galaxies at moderate redshifts 
\citep{Barger,Elbaz,Chary}, detailed extinction curves have only been measured for a small number of local group galaxies, such as the Milky Way (MW), SMC, and LMC, and for a subset of 
local starburst galaxies \citep{Pei,Calzetti}. The local ISM can be studied both directly, through extinction and IR emission/absorption, and indirectly, via depletions inferred from gas-phase abundances.
These studies reveal environmental variations in even local dust grain properties, such as differences
in the dust-to-gas ratios, UV extinctions, and depletion patterns observed in the lower metallicity Magellanic Clouds \citep{Welty01,Gordon03,Sofia06}, that are not fully understood.
At higher redshift our understanding is even more incomplete, with limited direct  knowledge about the detailed properties of dust in regular galaxies.
The simplistic assumption that high-redshift dust grains are similar in size and composition to local dust
may impact both derived galaxy properties, particularly in on-going large-scale galaxy surveys (e.g. SDSS, COSMOS), and techniques which are sensitive to the effects of dust attenuation, such as the 
Type Ia supernova studies used to infer the acceleration of the cosmic expansion [e.g., \citet{Aguirre99a,Aguirre99b,Riess}]. 

One technique which may be exploited to provide insight into the chemical composition of high-redshift ISM dust is the examination of absorption line systems in quasar spectra; particularly those in 
damped Lyman-alpha (DLA) absorption systems. These neutral-hydrogen rich systems (log N$_{HI}>$20.3) contain a substantial fraction of the neutral gas in galaxies \citep{Wolfe95,Storrie00}, and
are considered to be the best indicators of the chemical content of high redshift galaxies \citep{Pettini94,Prochaska03,Kulkarni05}. Furthermore, since DLAs are selected solely by gas cross-section, rather
than by galaxy brightness, they provide a relatively unbiased direct probe of dust in high redshift systems \citep{Ostriker90,Fall93}. Evidence for the presence of dust in DLA systems comes from both
measured depletions of refractory elements, and from the detected reddening of background quasars \citep{Pei91,Pettini97}; a study of $>$800 SDSS quasar spectra has found clear evidence of an association of quasar
reddening with absorption line strengths at 1$<$z$<$2 \citep{York06}. Dust in DLAs is also indirectly suggested by recent analyses 
\citep{Kulkarni05,Kulkarni07,Kulkarni2,Peroux06,Peroux10,Prochaska06} which find that the majority of DLAs exhibit low metallicities and low star-formation rates (SFRs). 
This finding is in contradiction to chemical evolution model predictions \citep{Pei99,Somerville} that are based on the cosmic star-formation history as inferred
from galaxy imaging surveys (e.g. the Hubble Deep Field, \citet{Madau96}). A scenario \citep{Fall93,Boisse98,Vladilo05} in which the more dusty and more metal-rich DLAs provide greater obscuration of their background 
quasars, hence impeding their detection, could explain these findings. 

One of the primary open questions about high-redshift dust, which may be probed using DLA systems, pertains to the chemical composition and size distribution of the dust grains. 
Models for local dust generally assume a mixture of carbonaceous, silicate, and metallic/oxide grains with a given distribution of grain sizes. 
In the Milky Way, about 70\% of the core mass of interstellar dust is found in silicate grains \citep{greenli,sildust}. These silicate grains are  produced only when oxygen is more abundant than carbon; the excess
of silicate dust measured in the MW galactic center has generally been attributed to a shortage of carbon-rich stars \citep{RocheA,Whittet}.
At high redshift, however, most studies have focused only on characterizing carbonaceous dust, by probing either the 2175~\AA~feature or the shape of the rest-frame UV extinction curve. 
Recent studies, by our group, examining quasar absorbers using the Spitzer InfraRed Spectrograph (IRS) though, have found clear detections of the 10~$\micron$ silicate dust feature in these objects, with a $\gtrsim$10-$\sigma$
detection in the AO 0235+164 DLA system \citep{Kulkarni1,Kulkarni2}. These analyses suggest that the overall shape of the silicate absorption feature is best fit by an optical depth profile template derived from either
laboratory amorphous olivine or from diffuse Galactic interstellar clouds, rather than from dense molecular clouds. 
These studies also find that the silicate feature is about 2-3 times as deep as expected from an extrapolation
of the $\tau_{10}$-A$_V$ relationship for MW diffuse ISM clouds, suggesting that these systems may be probing dust in the inner regions of their respective absorbing host galaxies, rather than dust on the outskirts.

In light of this recent success in detecting silicate dust in DLAs towards distant, moderately reddened quasars, we have now selected the 10~$\micron$ silicate feature in the PKS 1830-211 quasar absorption system for study. 
This particular quasar absorption system has been targeted for several reasons.  
First, the illuminating z=2.507 quasar is strongly lensed by a foreground (z=0.886), late-type (Sbc), low-inclination spiral, which is the presumed absorber host galaxy \citep{winn,Courbin,Wiklind}.
This lensing produces two distinct images of the quasar; a brighter (NE) component with a magnification of 5.9, and a second (SW) component which provides 1.5 times less magnification and exhibits signatures of 
dustier, more molecule-rich material \citep{winn}. Towards the NE component and in the absorber disk N$_{HI}\sim$2-3$\times$10$^{21}$ cm$^{-2}$,  assuming T$_{spin}$=100K \citep{Chengalur,Koopman}, with
a density of n$(H_2)=1700-2600$~cm$^{-3}$ towards the more obscured SW component \citep{Henkel09}.
While the spatial resolution of the Spitzer IRS is unable to separate these two lines of sight through the host galaxy, their integrated combination
provides a significantly stronger quasar continuum detection, and resultant signal-to-noise ratio (SNR) for the silicate feature, than has been obtained for the previous analyses of silicate dust in DLAs 
\citep{Kulkarni1,Kulkarni2}. 
Second, while the majority of DLAs have relatively low amounts of dust, this source is part of a small subset of quasar absorbers found through radio/mm surveys to be a molecule-rich, 21-cm (HI) absorber \citep{Chengalur}, 
indicative of an especially dusty and chemically more evolved quasar absorption\ system. 
Further evidence for the high dust content is provided by the relatively high measured absorber-frame extinction (A$_V$$<$4.1) and by the
ratio of the total to selective extinction (R$_V$=6.34$\pm$0.16) for this system \citep{Falco99}. The high absorber rest-frame reddening of E(B-V)=0.57$\pm$0.13, with $\Delta$E(B-V)=3.00$\pm$0.13 between the two quasar sight lines, 
is more typical of reddening values found for sub-mm sources [$\sim$0.55, \citep{Calzetti01}] than for 
either Lyman-break galaxies [0-0.45 with a median of 0.16 at z$\sim$3; \citet{Shapley01,Papovich01}] or MgII absorbers [0.002; \citet{York06}]. 
Thus, dusty absorbers such as this system may provide part of an evolutionary link in the sequence of SFRs, masses, metallicities, and dust content
connecting the generally low-SFR, metal- and dust-poor general DLA population (detected in absorption) to the actively star-forming metal- and dust-rich Lyman-break galaxies (detected in emission). 
Third, the PKS 1830-211 system is known to exhibit mm/sub-mm absorption features revealing the presence of a multitude of atoms, such as [C I] \citep{Bottinelli}, and molecules, in many cases in multiple transitions, including 
HCN, HCO$^+$, ortho-H$_2$O, HNC, ortho-NH$_3$, N$_2$H$^+$, H$^{13}$CO$^+$, H$^{13}$CN, CS, CO, $^{13}$CO, and C$^{18}$O [e.g. \citet{WiklindN, Wiklind,Bottinelli,Henkel,Menten,Muller,Muller11}]. 
 \citet{Muller06,Muller11} have found isotopic ratios ($^{16}$O/$^{18}$O, $^{17}$O/$^{18}$O, $^{14}$N/$^{15}$N, and $^{32}$S/$^{34}$S) for this absorber which differ from those in the Milky Way.
This abundance of molecules is suggestive of a chemically rich environment for the quasar absorption system, which we anticipate may be characteristic of systems which contribute significantly to the global metal content. 

We begin our paper with a description of our observations and data reductions in \S\ref{DATA}, and then in \S\ref{ANALYSIS} present our analysis utilizing the laboratory and 
observationally-derived silicate profile templates to estimate the optical depth in the 10~$\micron$ silicate absorption feature towards PKS 1830-211. We discuss
in \S\ref{CONCL} the implications of our finding that the prominently detected 10~$\micron$ silicate dust feature present in the PKS 1830-211 absorber has a 
crystalline silicate composition. Finally, we summarize our results in \S\ref{SUMMARY}. In Appendix~\ref{normAP} we discuss the impacts of variations in the adopted quasar-continuum normalization
on our results, in Appendix~\ref{caveat} we address two minor observational caveats which could potentially affect our analysis, and in Appendix~\ref{expandedAP} we consider a broader range of observational and laboratory
based optical depth profiles.

\section{DATA \& ANALYSIS}\label{DATA}

\subsection{Observations and Reductions}\label{clean}
Our observations, summarized in Table~\ref{obs}, were obtained using the InfraRed Spectrograph (IRS) on the Spitzer Space Telescope \citep{Spitzer}. These data 
consist of 69 individual spectral exposures, each observed in two nod positions, totaling 198 minutes (3.3 hours) of integration on-target, centered at $(\alpha,\delta)_{J2000}$=(18h33m39.89s, -21d03m40.4s).
The two nods, obtained using the IRS Staring mode, facilitate both background subtraction and the identification of spectral features induced by bad pixels. 
Given the spectral breadth of the silicate absorption features in previously studied quasar absorption systems \citep{Kulkarni1,Kulkarni2}, we chose to utilize the low-resolution mode (R$\sim$60-130),
rather than the high-resolution mode (R$\sim$600), in order to reach a higher sensitivity \citep{IRS}.
In addition to the LL1/2 and SL1/2 spectra, bonus (third) order SL and LL spectra were obtained, which facilitate joining the first and second order spectra.
Our complete data set spans a spectral region from 5.13 to 39.90~$\micron$; this broad wavelength coverage is crucial for the accurate determination of the underlying quasar continuum which we 
require to measure the strength of the broad silicate features.

The IRS slit cannot individually distinguish the two lensed PKS 1830-211 quasar images; thus, we cannot probe two distinct lines of sight through the absorber. 
The lower-wavelength data (SL) are obtained with a  3.\arcsec6-3.\arcsec7 x  57\arcsec slit and a 1.\arcsec8/pixel scale Si:As CCD, 
while the longer-wavelength data (LL) are obtained with a 10.\arcsec5-10.\arcsec7 x 168\arcsec slit and a 5.\arcsec1/pixel scale Si:Sb CCD \citep{IRS}.
Given this resolution, we are unable to spatially differentiate the two lensed images, which are separated by ($\Delta\alpha,\Delta\delta$)=($-0.\arcsec642,-0.\arcsec728$) and embedded in a fainter Einstein ring \citep{winn},
and instead we address, in this study, the composite spectrum which constrains  the \textit{average} dust properties within the quasar absorber. 
The NE lensed quasar image is significantly brighter at all wavelengths \citep{Lehar,Courbin,winn}, 
and we assume that our continuum is dominated by this
line of sight. The weaker (SW) component is more heavily dust-obscured, since it passes through a region of enhanced star formation in the western spiral arm of the host \citep{winn}, 
and will contribute more significantly to our measured silicate dust absorption, as it
does in detections of molecular absorption [see e.g. \citet{Lehar,Frye}]. 
Furthermore, we note that several neighboring objects are spatially inseparable from our quasar spectrum. These sources include [see illustrations in \citet{Courbin} and \citet{winn}] 
an M4 dwarf star, a foreground galaxy at z$\sim$0.19, and possibly a second star and a second galaxy close to the z=0.886 absorber host.
As explained in Appendix~\S\ref{foreground}, none of these other sources are expected to contribute significantly to our observed spectrum.

Our analysis utilizes the Basic Calibrated Data (BCD) reduced spectra, and associated uncertainties and bad pixel masks, produced by version 18.7.0 of the Spitzer automatic pipeline. 
These spectra have been bias and dark subtracted, and
a characteristic flat field has been applied. Additionally, the data have been corrected for systematic effects such as droop and dark-current drift, and cosmic ray hits have been identified and flagged
in the corresponding masks.

We have further processed the Spitzer BCD spectra by utilizing \textit{IRSCLEAN (v1.9-2.0)} to identify and remove rogue pixels and cosmic rays, and to reduce the background noise in the frames.
In addition to replacing bad/rogue pixels identified by the masks and iterative algorithms, we have implemented a slightly non-traditional application of the \textit{IRSCLEAN} 
software in which we also manually identified for replacement all pixels in the spectra backgrounds with visibly high/low signal in order to smooth the background.
We did not apply \textit{IRSCLEAN}-replacements, even for noted bad pixels, within the quasar spectrum or its environs, 
in order to prevent any spurious spectral signatures from being added to our data by the algorithm. 
We then manually removed bad/rogue pixels along or near the quasar spectrum using the \textit{epix} task in IRAF, and replacing points with the median of the surrounding pixels. For pixels along the quasar spectrum, the unweighted
average of bordering points along the spectral axis was used instead of the median. 
We were generally conservative with identifying and replacing these pixels, and only targeted strongly discrepant points, e.g. those pixels with saturated flux or isolated pixels with substantially higher/lower values than neighboring pixels. Pixels producing all other spectral features were retained as plausibly physical in origin, and 
those which were produced by transitory events, such as cosmic rays, or pixel imperfections were effectively removed when we combined the data from the two nod positions. 

We combined the cleaned individual BCD exposures in each nod position using the IDL script \textit{coad}, which automatically propagates the uncertainties and mask files, and then subtracted the backgrounds
using the IRAF task \textit{imarith}, propagating the uncertainties in quadrature. For the SL data we subtracted one nod position from the other, in order to remove the background contribution. For the LL data, however,
there exists an anomalous scattered light feature (see Appendix~\ref{scatlight}). This spatially-broad and intense feature prevented us from subtracting the two nods to remove the background. Therefore, we combined with \textit{coad} all of the off-target LL1 frames (obtained while the quasar was in the LL2 order) and vice-versa to create off-target background frames, which were then subtracted from the spectra in each nod position.
We applied the IRAF task \textit{fixpix} to smooth over two weak foreground stellar spectra present in the off-target LL1 spectra utilized for background subtraction. 

\subsection{Spectral Extraction and Continuum Normalization}\label{spice}
We extracted the one-dimensional spectra for each order using the Spitzer IRS Custom Extractor (\textit{SPICE}) algorithm, set for a point source extraction with optimal-weighting, using the default extraction aperture in each order,
with no subsequent corrections for fringing.
Our final 1-dimensional spectra were extracted with a varying spatial width over the dispersion axis, as per the standard \textit{SPICE} templates, sampling a broader spatial extent at the longest wavelengths within each order. 
A visual inspection of the two-dimensional spectra prior to extraction showed no apparent fringing in the SL2, SL1, or LL2 spectra, and only weak fringing in LL1.
The \textit{IRSFRINGE} manual stipulates that fringes are not spectrally resolved in the SL module, while in the LL module fringes resulting from one of the filters are present in the raw un-calibrated data,
 with an amplitude of 3-10\%, depending on wavelength.
These are largely corrected through the applied pipeline flat-fielding. As a test, we applied \textit{IRSFRINGE (v. 1.1)}
to our extracted LL data and found that the algorithm did not automatically
detect any significant fringes, and when forced to correct for any measurable fringes, produced only a negligible statistical and visual improvement in the data quality, with small spurious spectral features resulting in a few cases.
Therefore, we have applied no fringe corrections to our extracted spectra. 

In order to improve the signal-to-noise ratio (SNR) for our spectra, we have combined the one-dimensional extracted spectra from each of the two nod positions in an unweighted average.
The flux density uncertainties have been propagated in quadrature, and reflect both the measurement uncertainty estimated through the BCD pipeline reductions,
and the scatter of the two nods relative to their mean. These uncertainties are dominated by variations between the two nod positions. Although the spectra extracted from the two nod positions are generally consistent, there exist
regions with small discrepancies where, e.g., one of the nods exhibits a ``convex" feature, while the other exhibits a ``concave" feature. These small variations are at most 10\% relative to the mean spectrum 
in the SL2, SL1, and LL2 data. In the LL1 data, a discrepancy on the order of $\pm$15-20\% is observed over an extended spectral region from $\sim$22-26~$\micron$; this feature does not affect our measurements of
the 10 and 18~$\micron$ features in the z=0.886 absorber. After consultation with the Spitzer Science Center staff, we conclude that
the origin of this feature is likely scattered light, as discussed in Appendix~\ref{scatlight}. 

\begin{figure}
\epsscale{1.0}
\plottwo{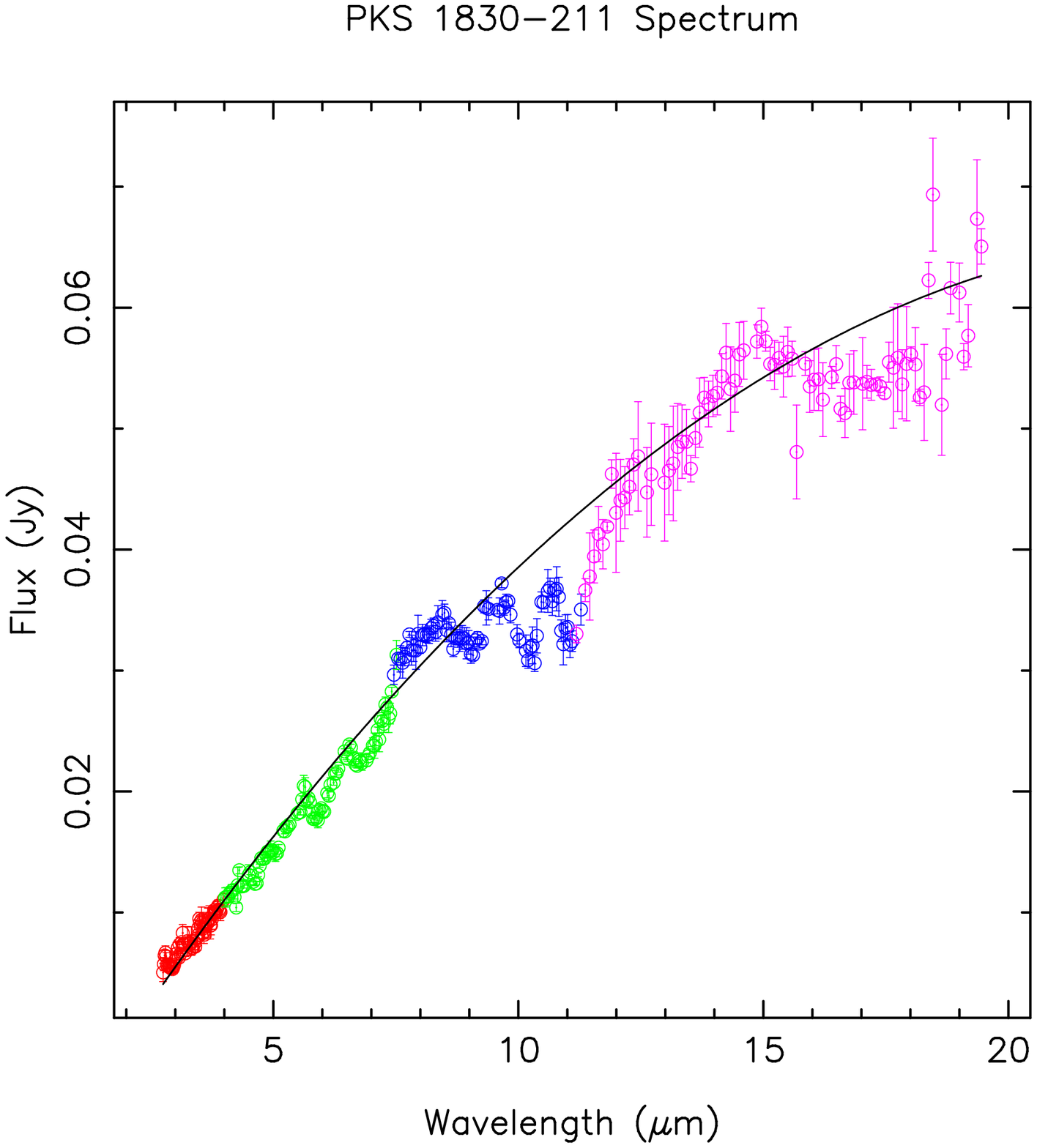}{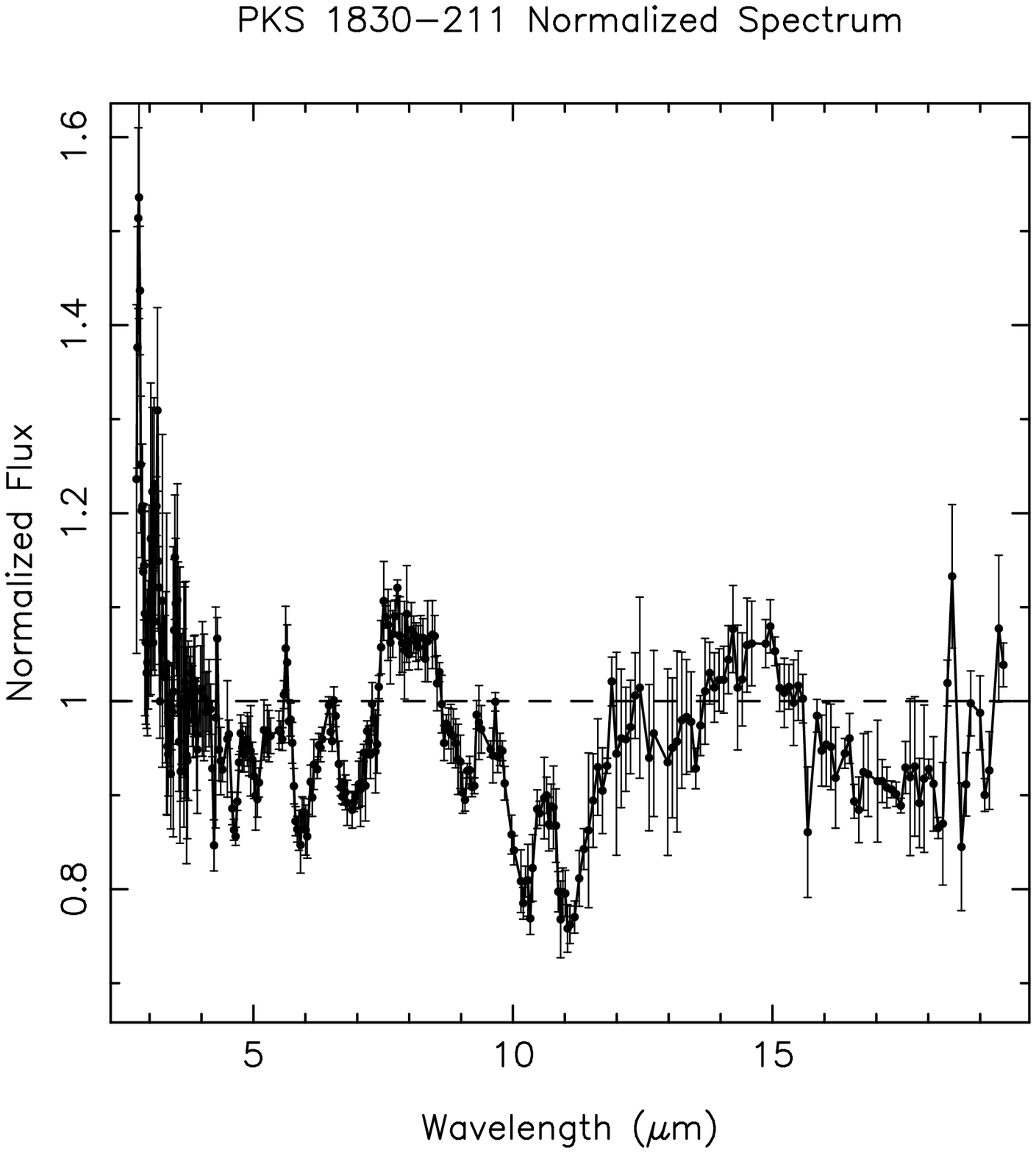}
\caption{
PKS 1830-211 spectrum. Left: Observed flux versus absorber rest-frame (z=0.886) wavelength with the colors denoting individual spectral orders as follows: red-SL2; green-SL1; blue-LL2; and magenta-LL1.
The black line depicts the 3rd order Chebyshev polynomial fit to the quasar continuum used to normalize the spectrum. 
Right: The resultant normalized spectrum, as a function of absorber-rest-frame wavelength, with the dashed line denoting unity.}\label{spectrum}
\end{figure}

We have manually trimmed the nod-combined spectra to eliminate lower quality data points, and subsequently we normalized and joined the four SL-LL spectra to construct a single, continuous
spectrum for the PKS 1830-211 system. These trimming and joining procedures are detailed in Appendix~\ref{joinspect}.  
Our combined spectrum has then been shifted to the rest frame of the quasar absorber host, assuming an absorber redshift of 0.886.
The final combined spectrum is illustrated in the left panel of Figure~\ref{spectrum}, with the individual SL and LL orders differentiated by color. 
There, and in all other figures, we depict the absorber-rest-frame wavelength, unless otherwise noted.

In order to search for and study the 10 and 18~$\micron$ silicate features, we have normalized our spectrum by a fit to the quasar continuum.
As detailed in Appendix~\ref{normspect}, we explored a range of approximately 20 different normalization fits, and identified a third order Chebyshev polynomial as producing the best fit to the
underlying continuum. This quality assessment was based on a combination of a visual inspection of the continuum fits and resultant continuum-normalized spectra, and an examination
of the relative contributions by the continua to the reduced-chi-squared values for the template fits used in characterizing the optical depth and shape of the silicate absorption profile (for details, see \S\ref{ANALYSIS}).
Our final normalization is illustrated with a black line in the left panel of Figure~\ref{spectrum}, with the resultant normalized spectrum of PKS 1830-211 depicted in the right panel.
As is evident in this figure, there is a broad absorption feature present at $\sim$10~$\micron$, as well as a shallower absorption feature near 18~$\micron$. These features are associated
with silicate dust in the z=0.886 foreground absorber, as discussed in the remainder of the paper. Additionally, there are several weaker absorption features in the 5-8~$\micron$ region, and some
emission features both longward and shortward of the 10~$\micron$ silicate absorption feature; the possible origins of these features are discussed in Appendix~\ref{residbump}, as they are not
the primary focus of this analysis.
\section{RESULTS}\label{ANALYSIS}
The z=0.886 absorber towards PKS 1830-211 produces prominent silicate absorption features near both 10 and 18~$\micron$, as is clearly illustrated in Figure~\ref{spectrum}.
We have measured the equivalent width of the 10~$\micron$ feature directly from this rest-frame, continuum-normalized spectrum, adopting a ``nudge factor" of 0.3 \citep{SS92} in estimating the continuum-fitting uncertainty.  
The total uncertainty associated with the calculated equivalent width
includes the uncertainty from photon noise combined in quadrature with the uncertainty resulting from fitting the continuum. 
We find that the equivalent width of the 10~$\micron$ feature is 
$W_{rest}=0.366\pm0.029$~$\micron$, resulting in a detection significance of 12.7-$\sigma$.
If we exclude the contribution to the uncertainty from the continuum fitting, this significance would rise to a 28.4-$\sigma$ detection. 
We have not estimated the equivalent width associated with the 18~$\micron$ absorption feature, since it lies at the edge of our spectral range and may not be fully covered by the current data set.

\subsection{Template Profile Fitting Procedures}\label{fitting}
In order to understand the physical origin of the 10~$\micron$ silicate feature, and the chemical composition of the dusty material, 
we have fitted a series of template optical depth profiles (see Tables~\ref{templatesOBS} and \ref{templatesLAB}) to
our observed spectrum. Our fitting assumes simple radiative transfer through the cloud, 
such that $I/I_0 = \exp[-\tau]$ where $\tau\equiv{a_{n}\tau_{norm}}$; $\tau_{norm}$ is the optical depth profile for the template, 
normalized to have a maximum peak depth of 1.0 over the full spectral extent of profile specified in Tables~\ref{templatesOBS}-\ref{templatesLAB}. 
The peak optical depth normalization factor ($a_n$) for each template profile is selected to produce the minimum reduced chi-squared ($\chi_r^2$). 
In order to obtain the associated 1-$\sigma$ error bars, we have (cubic-spline) interpolated the two values in the $\chi_r^2$ vs. $\tau$ curve for which $\chi_r^2-\chi_{r,min}^2=1$. 
The primary fitting has been performed over the 10~$\micron$ spectral region, i.e. 8.6-12.5~$\micron$. This fitting region has been selected to focus on contributions to the $\chi_r^2$ from portions of the spectrum near the absorption feature.
For those objects in which the template profile additionally covers
the 18~$\micron$ silicate feature, we have performed a second fit, termed the ``full" fit, extending from 8.0-19.45~$\micron$. 
This extended fitting region includes contributions to the $\chi_r^2$ from not only the 10 and 18~$\micron$ features, but also from
the region of the continuum between these two features, which as discussed in Appendix~\ref{normAP} may also contain some emission from non-silicate sources. 
In those template profiles for which only a more limited wavelength range is covered, we have reduced this expanded fitting range appropriately, as detailed in Table~\ref{fits}.

\begin{figure}
\epsscale{1.}
\plotone{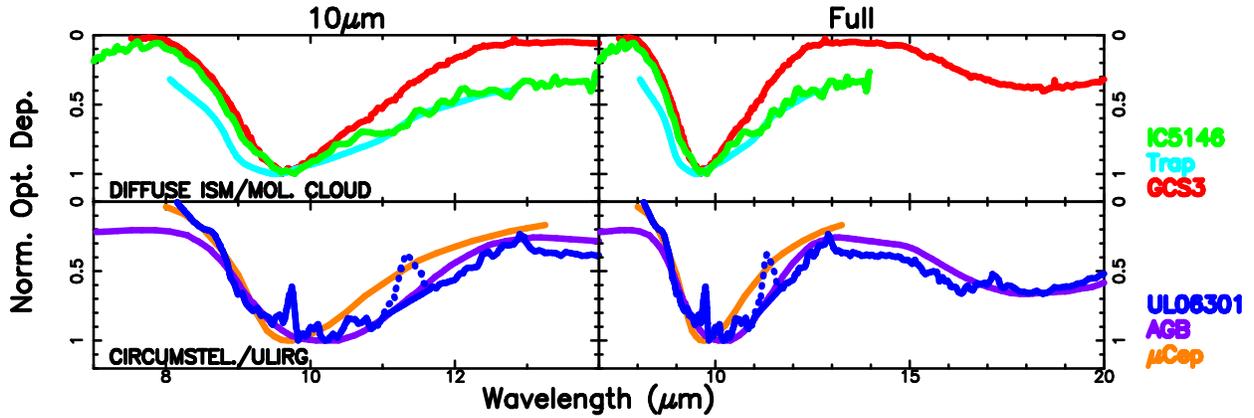}
\caption{
Optical depth template profiles for observed (astrophysical) sources drawn from the literature, as detailed in Table~\ref{templatesOBS}, split into two rows for clarity.
The left column depicts the profile over the 10~$\micron$ silicate feature. The right column
depicts the profile over the extended fitting region including the 18~$\micron$ silicate feature, if data is available.}\label{profilesOBS}
\end{figure}

\begin{figure}
\epsscale{1.}
\plotone{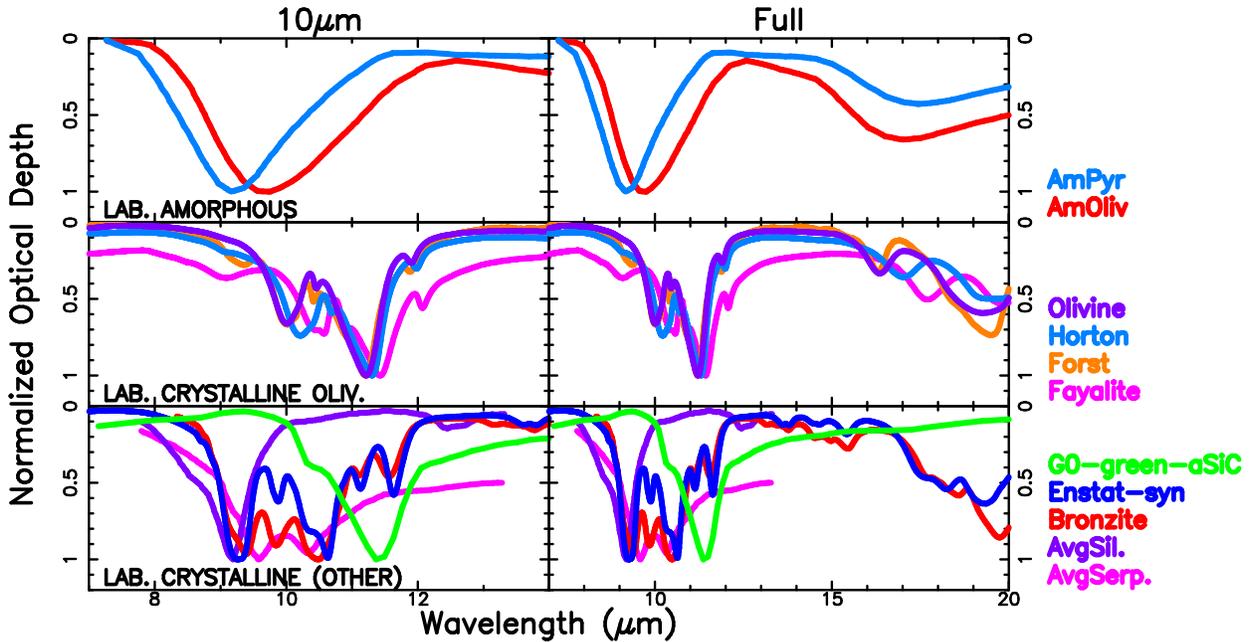}
\caption{
Similar to Figure~\ref{profilesOBS}, but for the laboratory profiles (Table~\ref{templatesLAB}) drawn from the literature. Note the complex structure for the crystalline silicate profiles, in contrast with the the relatively featureless
amorphous profiles. Small variations in chemical composition can produce notable differences in the profile, as illustrated by the 4 crystalline olivine templates in the middle panel.}\label{profilesLAB}
\end{figure}

To constrain the composition of the material producing the 10~$\micron$ silicate absorption in the z=0.886 absorber, we fit
more than 100 unique optical depth profiles, derived from astrophysically observed and laboratory sources in the literature, as detailed in Tables~\ref{templatesOBS}-\ref{templatesLAB}. 
These templates probe a wide range of 
laboratory profiles for natural and synthetic terrestrial minerals, and environments including the Solar System (comets),
circumstellar material, the Galactic diffuse interstellar medium (ISM), dense molecular clouds, and  
extragalactic objects (e.g. UltraLuminous InfraRed Galaxies; ULIRGs). 
For brevity, we include a subset of 18 of the most representative profiles in the bulk of this paper, as detailed in Table~\ref{templatesOBS} (observational templates) and Table~\ref{templatesLAB} (laboratory templates). 
However, in order to more fully explore the possible minerals comprising the
detected silicate dust, we additionally examine an expanded subset of the templates, including 44 laboratory and 18 observational profiles, including comets, in Appendix~\ref{expandedAP}.
The full range of profiles are illustrated in Figure~\ref{profilesOBS} (observed) and Figure~\ref{profilesLAB} (laboratory). 
This complete set of templates spans a range of chemical compositions and degrees of silicate crystallinity, ranging from the purely amorphous [diffuse ISM \citep{Kemper04}, amorphous laboratory templates) to 
10-15\% crystallinity [AGB star \citep{Sylvester99, Kemper01,Molster}, ULIRG \citep{Spoon,Kemper10}], to purely crystalline laboratory silicates.

\begin{figure}
\epsscale{1.}
\plotone{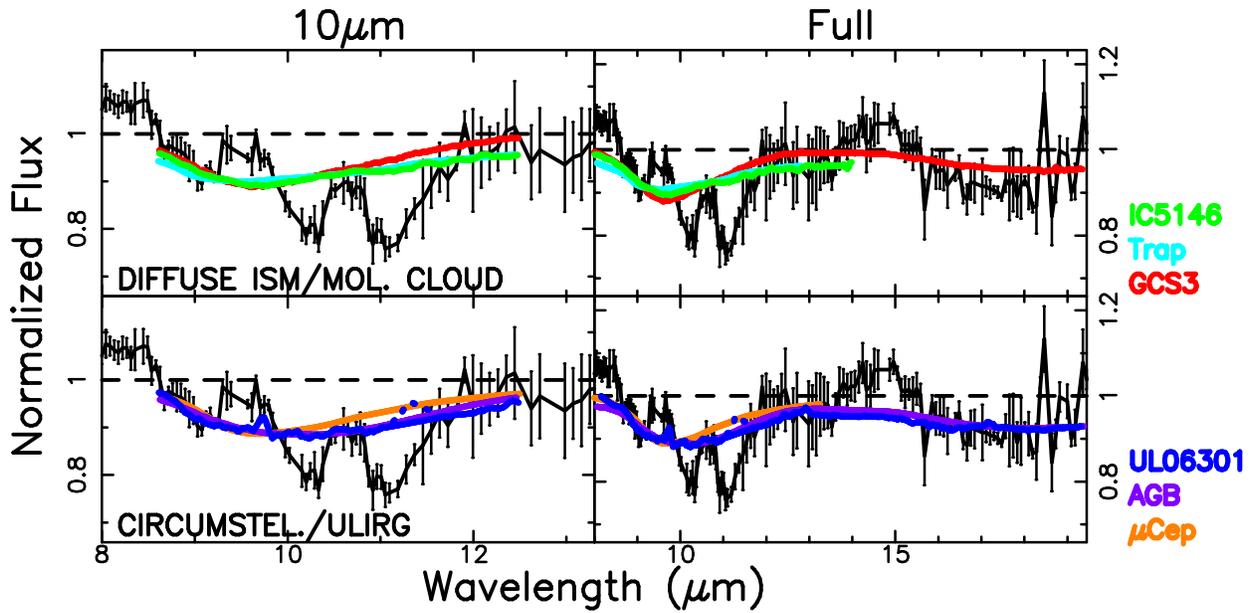}
\caption{
Continuum-normalized spectrum of PKS 1830-211, in the rest frame of the z=0.886 absorber,
overlaid with fits using the observed optical depth template profiles. The left panels show fits performed solely over the 10~$\micron$ feature
(8.6-12.5~$\micron$). The right panels show the fits to the expanded range (8.0-19.45~$\micron$), when available; otherwise a more limited fit is depicted. 
Abbreviations for the source names are as in Table~\ref{templatesOBS}.
None of the astrophysically observed optical depth profiles replicate the observed structure in the PKS 1830-211 absorber 10~$\micron$ region.}\label{fitsfigureOBS}
\end{figure}

\begin{figure}
\epsscale{1.}
\plotone{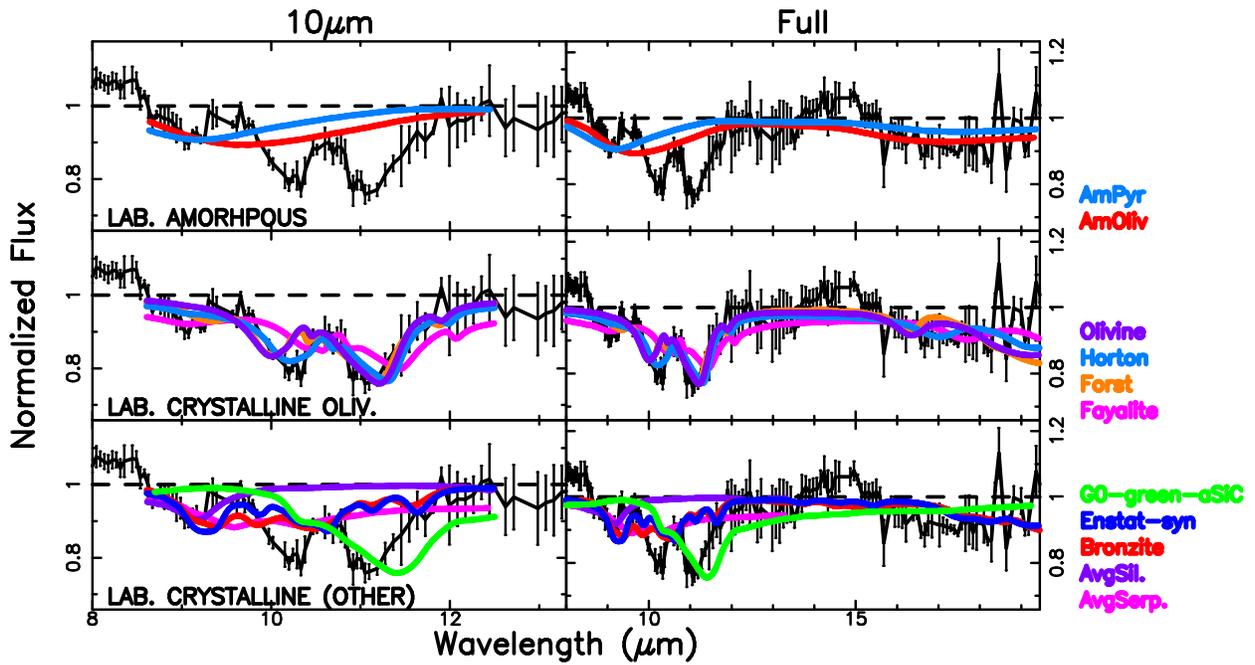}
\caption{
Similar to Figure~\ref{fitsfigureOBS}, but for the laboratory optical depth templates. Abbreviations for the source names are as in Table~\ref{templatesLAB}.
The rough agreement between the 10~$\micron$ and combined 10+18~$\micron$ template fits suggests that the continuum normalization constraints applied at long wavelengths are reasonable.
This is addressed further in Appendix~\ref{normAP}.
The only profiles which reproduce the strong double-peaked profile within the 10~$\micron$ silicate feature region are the crystalline olivines.}\label{fitsfigureLAB}
\end{figure}

The optical depth fits obtained using  the 18 templates are presented in Table~\ref{fits}.
We list the peak optical depth for fitting both solely over the 10~$\micron$ spectral region (8.6-12.5~$\micron$) and over the 8.0-19.45~$\micron$ fitting range.  
In Figures~\ref{fitsfigureOBS} and \ref{fitsfigureLAB} we over-plot the fitted profiles on top of our PKS 1830-211 spectrum, using a template profile sampling of $\Delta\lambda$=0.01~$\micron$ for uniformity.
Considering our complete subset of 18 profiles,
we find 0.11$\leq\tau_{10}\leq0.13$ for the observationally-based profiles, and 0.10$\leq\tau_{10}\leq0.28$ for the laboratory-based profiles. Most of these are visibly poor fits, as discussed in the following sections.
Our best fit estimate of the peak optical depth is $\tau_{10}$=0.27$\pm$0.05 for the crystalline olivine hortonolite template.
This best fit was determined
from both a visual inspection of the fits and based on the $\chi_r^2$, and is at the highest end of the optical depth range derived using our template profiles. 

In order to ensure sampling uniformity for all of the computed fits, we have (cubic-spline) interpolated the values in each of the template profiles at the wavelengths at which our PKS 1830-211 spectrum has measured data points. 
By sampling every template at the wavelengths of our PKS 1830-211 data points, we ensure consistency when comparing the derived fits. This is essential since some of the laboratory profiles have been finely sampled with significantly
sub-micron measurements, while the observationally derived profiles are more coarsely sampled. A consequence of sampling at our observed spectral dispersion scale is that we may potentially miss extremely narrow lines,
but studies examining unresolved atomic/molecular transitions in similar IRS spectra, e.g. star-forming galaxies \citep{Smith}, suggest that this is not a significant problem.

Although the template profiles are sampled at a range of resolutions, since observationally-based templates cannot achieve the high resolutions of laboratory measurements, 
we do not believe this is an issue for our analysis; the locations of the peaks play a more significant role than their shape. In order to test the effects of resolution variations, we have convolved three of the most-structured,
crystalline olivine template profiles with two different Gaussians representative of the Spitzer IRS instrumental resolution; one of which represents the lowest achieved resolution typical of LL1, and one of which is more representative of
an average spectral resolution across the wavelength range covering the 10~$\micron$ feature. We find that for the more average spectral resolution there is a negligible change in the derived optical depth, 
with $\Delta\tau$=0.01. For the worst-case scenario, the optical 
depth is at most lower by $\Delta\tau$=0.04-0.05, which is consistent within our 1-$\sigma$ stated uncertainties. Since the instrumental resolution varies with wavelength over each IRS spectral module, it is difficult to correct
for the instrumental profile. Given that our tests indicate no significant impact on the derived optical depths, no significant improvements in the $\chi_r^2$ for our best-fitting profiles, and no visual improvement in the quality of the fit, we have
made no corrections for the resolution variations between our template profiles and the PKS 1830-211 spectrum.

\subsection{Fits with Observational Templates}\label{OBSTEMPfits}

While the laboratory-based profiles are arguably more precisely measured, and do not include the same observational uncertainties as astrophysical sources, 
we also include templates derived from astrophysical sources in this analysis for the following reasons: (i) they may contain compounds which are not commonly
found terrestrially or synthesized in a laboratory; (ii) they may trace minerals in temperature and density environments more physically similar to our quasar absorber than can be achieved in ground-based laboratories; and (iii)
in order to establish the similarity or dissimilarity of higher-redshift extragalactic dust with that locally, it is essential to directly compare Galactic and other extragalactic dust sources with our sample. 

\subsubsection{Fits with Galactic Source Templates}\label{GalSrc}

We have first considered optical depth profiles derived from astrophysical sources within the Milky Way, including the diffuse ISM, dense molecular clouds, and circumstellar material. 
If extragalactic dust is similar to that in local systems, as is commonly implicitly assumed in extragalactic studies, then at least some of these sources should provide a good match to the 10~$\micron$ (and 18~$\micron$)
silicate absorption in the PKS 1830-211 z=0.886 quasar absorber. Furthermore, \citet{Kulkarni2} have previously found
that those quasar absorption systems in which the 10~$\micron$ silicate feature is not fit by laboratory amorphous olivine, are instead well-fit by Galactic interstellar clouds, in which the dust is known to be composed of primarily
amorphous material \citep{Kemper04}. They have not found any instances wherein dense molecular clouds provide the best fit. 

In our analysis of the PKS 1830-211 absorber, however, \textit{none} of these Galactic template profiles adequately fit the 10~$\micron$ feature. 
In general, these profiles exhibit three flaws: (i) the peak wavelength of the silicate absorption is shifted to shorter wavelengths relative to our absorber; 
(ii) they exhibit a single peak while we observe three distinct substructure peaks in the 10~$\micron$ absorption region; and (iii) the breadth of our combined feature is relatively larger than can be
reproduced by the Galactic templates. 
The best fit amongst the Galactic profiles is produced by a profile characteristic of an AGB star, perhaps the richest in crystalline silicates of our considered templates, but this is still a relatively poor fit when compared with the laboratory
templates in the following sections. 
Neither the molecular clouds, nor the diffuse ISM profiles, nor the supergiant stellar environments produce a particularly good fit. It is remarkable that the Galactic molecular cloud templates do not fit the molecule-rich
PKS 1830-211 absorption system; this may suggest that the molecular ISM in the PKS 1830-211 absorber differs from that in our own Galaxy.

\subsubsection{Fits with Extragalactic Source Templates}\label{ULIRG}

We have also considered a silicate optical depth profile for an extragalactic source: a ULIRG (06301-7934, z=0.156), which may plausibly be more similar to our extragalactic z=0.886 absorber, despite the fact that it exhibits more
robust star-formation signatures than are associated with our system. This ULIRG was selected from a set of 12 such objects in \citet{Spoon} because it produced the best match to our absorption system; coincidentally at 13\% 
silicate crystallinity \citep{Kemper10}, it is also among the most crystalline-enriched ULIRGs from \citet{Spoon}. We find that while the template profile based on this ULIRG produced a better fit than any of the Galactic sources, it still suffers from the 
same flaws in terms of an offset in the peak wavelength for the profile, and being unable to match either the observed substructure or the breadth of the feature. Removing the prominent 11.3~$\micron$ PAH complex emission feature
from the ULIRG template, which is not seen in our system, improves upon the fit slightly but still does not allow us to reproduce our observed spectrum.

\subsection{Fits with Laboratory Templates}\label{LABTEMPfits}

\subsubsection{Fits with Amorphous Silicate Templates}\label{AmSils}

We have next considered laboratory amorphous silicate profiles. 
As addressed in \citet{Bowey02}, the majority of analyses in both the local Galaxy, and at higher
redshift [e.g., \citet{Spoon}], have considered the 10~$\micron$ silicate feature to be produced primarily by an amorphous silicate, with residual structural features attributed to either crystalline silicates or to non-silicate emission/absorption lines.
The analyses by \citet{Kulkarni1,Kulkarni2} found that in most previously studied quasar absorption systems, the detected 10~$\micron$ silicate absorption can
be  reasonably reproduced by amorphous olivine. We have thus considered profiles representing both amorphous olivine ($Mg_{2x}Fe_{2-2x}SiO_4$) and pyroxene ($Mg_{x}Fe_{1-x}SiO_3$)
compositions, where $0<x<1$\footnote{Throughout our analysis, we note that amorphous olivine/pyroxene refers to amorphous silicates of an olivine/pyroxene chemical composition, since
intrinsically all olivines and pyroxenes are crystalline in structure \citep{Henning10}.}.
We find that amorphous olivine provides a better fit than amorphous pyroxene, or than intermediate (not illustrated) amorphous silicate profiles, 
such as an amorphous version of the 21 Ferraro blend and an amorphous blend from \citet{Bowey02}. 
However, the amorphous silicate fits are still poorer than those produced by the crystalline silicates discussed below, or even by the more crystalline-rich observational templates. 
As with the observational template profiles, the laboratory amorphous silicate profiles are peaking at a lower wavelength, exhibit a shallower/narrower shape, and cannot reproduce the 
distinctive substructure we see in the PKS 1830-211 spectrum. If amorphous silicates are invoked to explain the 10~$\micron$ feature in the system, we would also require significant superposed absorption and emission from other mechanisms.
This possibility is addressed in \S\ref{CONCL}, but leads to a poorer fit than our best case fit for a crystalline silicate template.

We note that for both amorphous and crystalline (discussed below) olivine silicates, the values of $\tau$ which are obtained fitting over the 10~$\micron$ region are consistent with those obtained 
over the extended fitting range including the 18~$\micron$ feature. 
This consistency suggests that our long-wavelength quasar continuum 
constraints are not unreasonable, although the relatively higher noise associated with the 18~$\micron$ data points, and
the fact that we marginally cover the full extent of the 18~$\micron$ feature, places relatively weaker constraints on the derived optical depth measurements relative to the 10~$\micron$ fitting. 

\subsubsection{Fits with Laboratory Crystalline Olivine Templates}\label{CRYSoliv}

\begin{figure}
\epsscale{1.}
\plotone{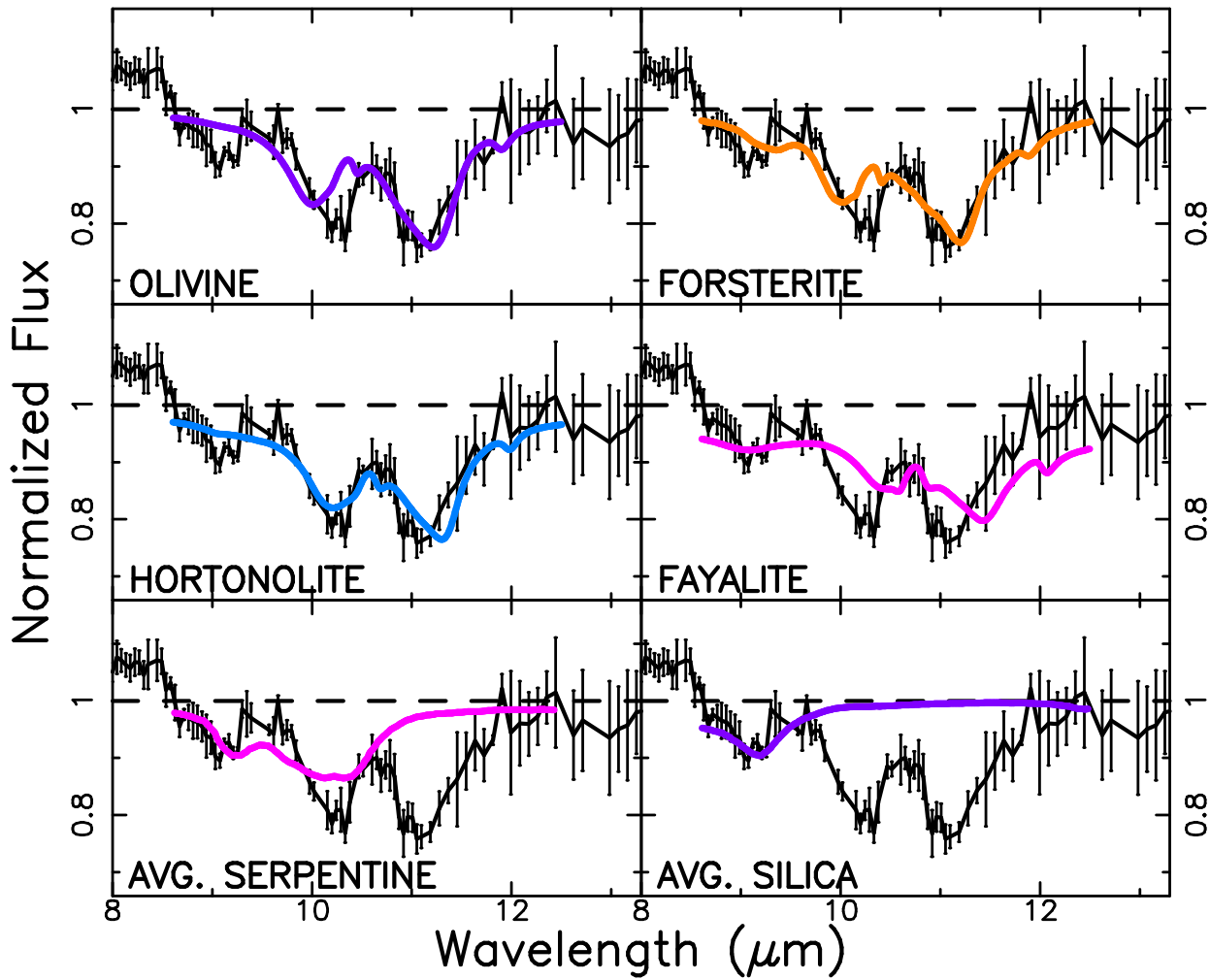}
\caption{
Six of the most viable profile fits to the 10~$\micron$ absorption feature, overlaid on the PKS 1830-211 normalized flux curve, in the rest frame of the z=0.886 absorber. 
The top two rows illustrate the crystalline olivine profile fits. Note that while the central profile peak is well-matched by hortonolite, the rightmost peak is offset to slightly shorter wavelengths relative to hortonolite.
The trend is reversed for forsterite. This suggests the need for an olivine composition which is intermediate between the two profiles, or which exhibits slight variations in the grain morphology.
Formally hortonolite provides a slightly better fit, but none of the olivine profiles reproduces the leftmost peak. This lower-wavelength peak could be fit by serpentine (which also contributes to the slightly under-fit central peak) or by silica. 
We, thus, conclude that a mixture of olivine, and either a second silicate material or ammonia absorption, as discussed in \S\ref{CRYSoliv}, is the most likely explanation for the composition of the observed silicate dust.
}\label{crystalsFIG}
\end{figure}

Given the relatively poor fits for both the amorphous silicates and for the observational sources, and the fact that the more crystalline-rich sources provide better fits than those which are crystalline-poor,
we have next considered crystalline olivine silicates. We have examined fits for four such olivine templates (Mg$_{2x}Fe_{2-2x}SiO_4$) spanning a range of Mg/Fe ratios: 
synthetic forsterite (x=1.0), natural olivine (x=0.94), natural hortonolite (x=0.55), and synthetic fayalite (x=0.0). 
As illustrated in Figure~\ref{crystalsFIG}, these minerals are able to explain the bulk of our profile shape and substructure, including the offset in the peak wavelength,
the breadth of the feature, and the majority of the substructure peaks, without invoking secondary absorption and emission mechanisms such as would be required for amorphous silicates. The best fit is produced by 
hortonolite, both over the 10~$\micron$ fitting region, and when considering the expanded fitting region covering both the 10~$\micron$ and 18~$\micron$ absorption features, as illustrated in Figure~\ref{fitsfigureLAB}. 
As is evident in Figure~\ref{crystalsFIG}, hortonolite well-reproduces the general shape of the profile, but cannot fully account for the leftmost peak ($\sim$9~$\micron$). Additionally, the rightmost peak ($\sim$11~$\micron$)
is offset to slightly longer wavelengths relative to the PKS 1830-211 absorber spectrum. Comparatively forsterite, which produces the second-best fit, is able to match the shape and location of the rightmost ($\sim$11~$\micron$)
substructure peak, but finds the middle peak ($\sim$10~$\micron$) and the intermediate ridge between these two features offset to slightly lower wavelengths. This strongly suggests that a crystalline olivine with an intermediate
Mg/Fe ratio could simultaneously fit both peaks. While natural olivine is such an intermediate profile, it provides a poorer fit than either forsterite or hortonolite, fitting neither peak perfectly, suggesting an unexplored ratio between 0.55 and 1.0 may
provide the best match. In order to determine the exact Mg/Fe ratio, longer wavelength data would be helpful as the similar profile shapes for the four olivine compositions are significantly different longward of 20~$\micron$,
as discussed by \citet{Jaeger}.  

\subsubsection{Fits with Other Laboratory Templates}\label{LABsils}

Lastly, in order to explore options for a compound which could account for the leftmost ($\sim$9~$\micron$) substructure absorption feature, or which could perhaps explain all three substructure peaks simultaneously, 
we have considered a range of other crystalline materials including an average serpentine, an average silica mixture, and two crystalline pyroxenes. We have also considered a SiC terrestrial/laboratory material, 
since although SiC has not been found in abundance in the ISM, it is found in meteorites and in evolved stars and stellar products, including carbon stars, planetary nebulae, and some supernova remnants, 
e.g., \citet{Whittet90,Speck,Stanghellini,Zijlstra,Yang,Gruendl,Bernard,Rho}. We find that the best fitting of these alternative compositions is synthetic enstatite, a crystalline pyroxene, but that while it is an improvement over the
other considered materials, and even over the laboratory amorphous silicate templates, it is not as good as any of the crystalline olivine templates. We also find that silica and serpentine, as illustrated in Figure~\ref{crystalsFIG},
are successful at accounting for the leftmost ($\sim$9~$\micron$) substructure feature in our spectrum. Serpentine not only fits the $\sim$9~$\micron$ feature, but also contributes some absorption in the 10~$\micron$ region which is under-fit by some of the crystalline olivine profiles, such as the natural olivine and the fayalite.

\subsubsection{Evidence for Multiple Minerals in Combination}\label{bivar}

\begin{figure}
\plotone{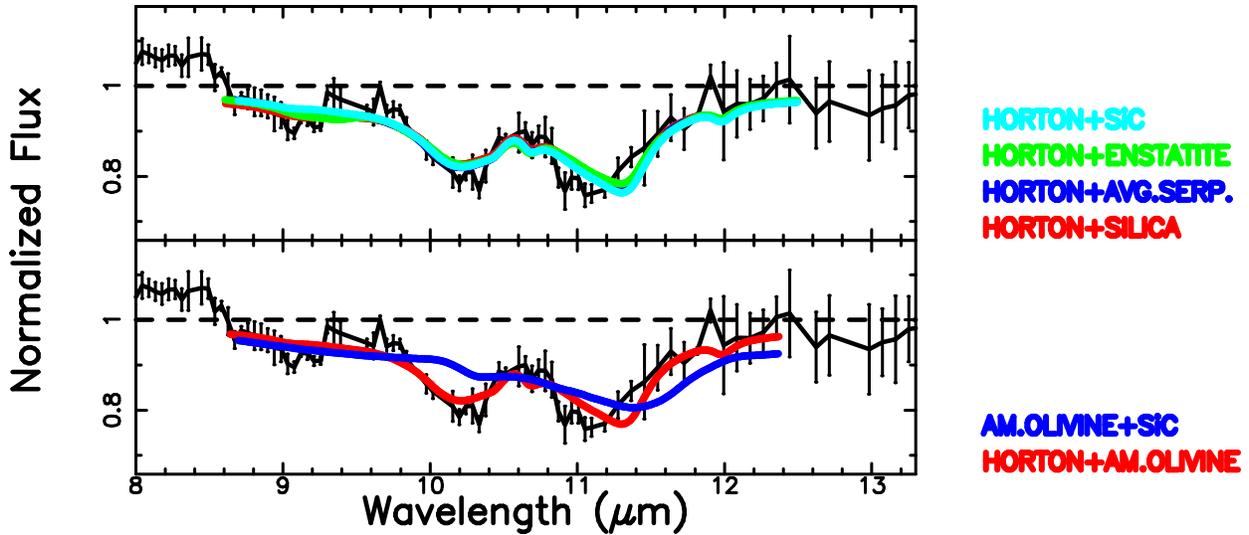}
\caption{
Bi-variate fits combining the best-fitting hortonolite profile with other templates, overlaid on top of the normalized flux curve for PKS 1830-211, in the rest-frame of the z=0.886 absorber. 
The best fit is produced by hortonolite+silica, although in \textit{every} case hortonolite is the dominant component. The minimal relative contributions from the secondary constituents produces the visual similarity between the 4 different fits
in the top panel. We also illustrate the combination of hortonolite+amorphous olivine and amorphous olivine+SiC, for comparison; in both of these cases amorphous olivine provides the more minor contribution to the fit.
}\label{multivarFIGS}
\end{figure}

As our template fits have revealed that hortonolite produces the best fit, but that it cannot explain the $\sim$9~$\micron$ absorption substructure, we have finally considered whether a combination of minerals could fully explain 
the absorption feature observed in the PKS 1830-211 spectrum. To test this we have considered a two-mineral mixture within the dusty quasar absorption system, combining hortonolite and a second mineral, such that 
$I/I_0 = \exp[{-(a_{n}\tau_{1,norm}+b_{n}\tau_{2,norm})}]$. 
The optimal values for $a_n$ and $b_n$ are obtained simultaneously, with the uncertainties conservatively determined by ascertaining parameter values which produce $\chi_r^2-\chi_{r,min}^2=1$
within a given parameter, while holding the other parameter fixed.
The resultant bi-variate fits are presented in Table~\ref{multivarTAB}, and depicted in Figure~\ref{multivarFIGS}.

\begin{figure}
\plotone{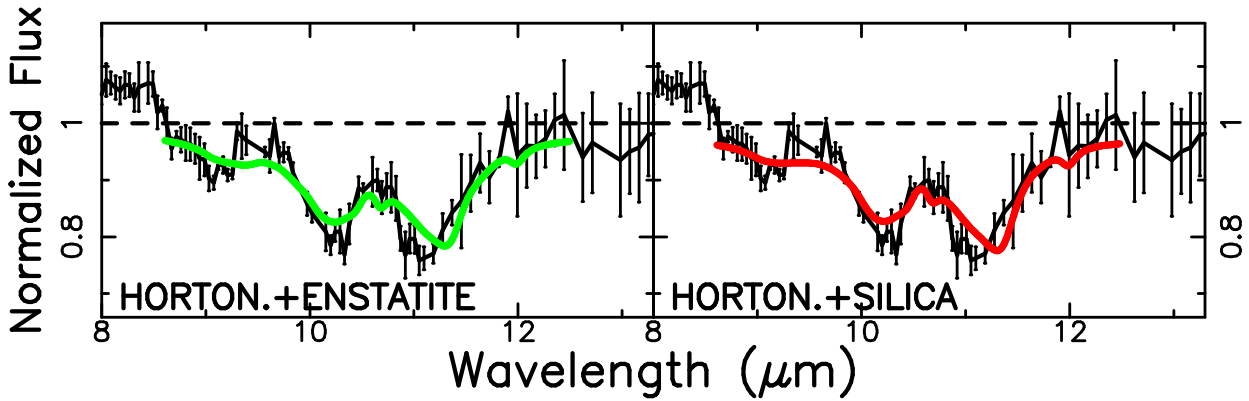}
\caption{
Bi-variate fits combining the hortonolite and enstatite (left) and silica (right) optical depth profiles, overlaid on the PKS 1830-211 normalized flux curve, in the rest-frame of the z=0.886 absorber. 
These two combinations formally produce the best fits; although the improvement over the pure hortonolite profile is slight.
}\label{bestmultivarFIGS}
\end{figure}

We find that the combination of hortonolite with average silica, average serpentine, or enstatite all reduce the $\chi_r^2$ for the fit over the 10~$\micron$ feature, and none 
of these profiles have significant features, for which would we need to account, in the 12.5-17~$\micron$ region. 
We have isolated the two best combinations in Figure~\ref{bestmultivarFIGS}, in which it is evident that the strongest contribution is from hortonolite. 
The largest difference between these two fits is in the region between the 9~$\micron$ and 10~$\micron$ features where the enstatite combination exhibits an additional, very small, inflection. 
None of these fits quite match the height of the region intermediate between the 9~$\micron$ and 10~$\micron$ features, though, and while 
overall the fit is improved by the addition of the second mineral, there is only a weak contribution from this secondary material. For instance, in the combination of enstatite with hortonolite,
enstatite is only contributing 10\% of the column density relative to hortonolite. 
By adding these minerals we slightly reduce the hortonolite optical depth
from 0.27$\pm$0.05 to 0.24 at minimum (for enstatite). The optical depth of the second material is consistent with zero in every combination. We have also considered the combination of SiC with hortonolite, but find this is not an improvement.

Alternatively, we note that it is possible that the $\sim$9~$\micron$ absorption feature is produced by ammonia (NH$_3$) absorption, which is seen in some ice-rich Galactic environments, 
such as W33A, a dust-embedded young stellar object \citep{Gibb,Gibb2004}. As discussed in \citet{Gibb2004}, NH$_3$ has an umbrella mode transition
at 9.3~$\micron$ which, when NH$_3$ is a small component of the ice, shifts to 9.0~$\micron$. We have compared the modeled H$_2$O:NH$_3$ (100:9) ice at 50K for W33A from \citet{Gibb} with our spectrum and find that both the overall shape and location of that absorption feature is well-matched by our 9~$\micron$ absorption. Furthermore, ortho-NH$_3$ has previously been detected in the PKS 1830-211 z=0.886 quasar absorption system by \citet{Menten}, with a total column density of 
$\sim$10$^{14}$ cm$^{-2}$. 

In addition to considering combinations of hortonolite with a second material, we have also considered amorphous olivine in combination with either hortonolite or with SiC. We find that in the latter fit, SiC dominates the fit, and that while an improvement over 
pure amorphous olivine, is still a relatively poor fit in comparison with crystalline olivine, since the combined profile is unable to explain the shorter wavelength substructure features. 
The hortonolite and amorphous olivine fit is dominated completely by hortonolite, with little to no contribution from amorphous olivine. The amorphous olivine
contribution produces a $\tau\leq$0.005, the lower limit considered in the combination, leading us to conclude that the contribution from crystalline silicates (1-N$_{am}$/N$_{cry}$) is $\geq$95\%, following \citet{Spoon}.

\section{DISCUSSION \& CONCLUSIONS}\label{CONCL}

In conclusion, based on our examination of a range of both observed and laboratory template profiles characterizing the 10~$\micron$ silicate absorption feature in the  
obscuring z=0.886 absorber towards PKS 1830-211, we have determined that our data are best fit by
a laboratory crystalline olivine profile, possibly in combination with secondary silicate materials or ammonia. This finding is exceedingly unusual in the context of what is known about silicates in other astrophysical sources.
The 10~$\micron$ and 18~$\micron$ features in Galactic interstellar matter are usually produced by the Si-O stretching and O-Si-O bending modes, respectively, of primarily amorphous silicate material.
In the following section, we begin with a discussion of the plausibility of finding pure crystalline silicates in the PKS 1830-211 absorber, and then discuss several alternative scenarios which could potentially produce some of the
observed structural features in the 10~$\micron$ region. 
\subsection{Crystalline Silicates}
Crystalline silicates have been observed in a multitude of Galactic and extragalactic sources, but they generally contribute $<$15\% of the silicate mass. 
\citet{Bowey02} have examined whether the broad 10~$\micron$ silicate feature could instead represent a superposition of numerous crystalline silicate features, but \citet{Molster}
deem this scenario unlikely given that the corresponding expected crystalline silicate resonances at longer wavelengths have not been observed.
We discuss in the following subsections a comparison of the PKS 1830-211 quasar absorber crystallinity with other Galactic and extragalactic sources, and with other quasar absorption systems, and finally
we discuss the physical environments for crystalline silicates. 

\subsubsection{Comparison With Other Astrophysical Sources}

While such high crystallinity as we tentatively observe towards PKS 1830-211 is very rare, objects with some degree of crystalline silicates are ubiquitous both in the local universe, and at higher redshifts. 
Galactic objects with noted crystalline silicates include comets (7-90\% estimated crystallinity, depending on grain properties), interplanetary dust particles (IDPs), 
primitive meteoritic materials, young stellar environments including pre-main sequence stars such as Herbig Ae/Be stars and T-Tau stars, and evolved stars, particularly those with 
oxygen- and dust-rich outflows such as (high mass loss) AGB stars (10-15\% crystallinity), post-AGB stars (20-60\% crystallinity in some post-AGB binary systems surrounded by a dusty circumbinary disc),
and planetary nebulae \citep{Bouwman03,Gielen,Honda03,Kemper01,Meeus03,Molster,Uchida04,Wooden99}. 
Although a few rare stellar objects, such as the circumstellar disk around the peculiar carbon star IRAS 09425-6040, exhibit up to 75\% small-grain crystallinity \citep{Molster01}, the material around the more populous low mass stars is 
generally less crystalline-enriched; the winds of low mass post-MS stars exhibit $\leq$10\% crystallinity \citep{Molster,Kemper01}. 
The diffuse ISM, at the lowest end of the spectrum, exhibits $\leq$ 2-5\% crystalline silicates \citep{Kemper04,Li07}. 
Among extragalactic sources, ULIRGs have been 
measured with at most 13\% silicate crystallinity \citep{Spoon,Kemper10}, and the crystalline features are most pronounced longward of the 10~$\micron$ region. These ULIRG templates provide a better match to our spectrum 
than any of the other observational
templates. Galactic templates rich in crystalline silicates, such as the AGB OH/IR template, produce a better fit to our data than crystalline-poor templates, such as those representing the diffuse ISM. 
This all suggests that the PKS 1830-211 quasar absorption system at z=0.886 may indeed be unusually crystalline-rich.

\subsubsection{Comparison With Other Quasar Absorbers}

Previous studies by our group of the silicate dust in other quasar absorbers did not find evidence for crystalline structure, but no crystalline-rich templates were utilized.
Closer examinations of the fit residuals suggest additional unaccounted for substructure within the absorption features.
\citet{Kulkarni1,Kulkarni2} examined 5 quasar absorbers and found that the 10~$\micron$ feature
was best reproduced by templates from either laboratory amorphous silicates of an olivine composition, or by the primarily amorphous diffuse Galactic ISM template. 
We have re-examined these fits in Aller et al. in preparation, using our expanded template library.
We find that while in some of the systems, such as Q0235+164, the dominant component may be amorphous olivine, there are suggestions of additional crystalline silicate
material in every system. In the case of the Q0937+5628 absorber we, in fact, find evidence that a high degree of silicate crystallinity may be present, similar to PKS 1830-211. Likewise, in 3C196 we find suggestions of significant crystallinity. 
These results lend further credence to our conclusion that the silicate dust in the PKS 1830-211 z=0.886 absorber may exhibit substantial crystallinity. 

Furthermore, it is plausible that the temperature of the absorber may impact the crystallinity.
Measurements of the spin temperatures in some of these systems which exhibit the weakest signatures of crystallinity, e.g. Q0852+3435 [T$_s<$536~K; \citet{Srianand}] and Q0235+164 [T$_s$=210~K; \citet{KC03}], 
may be lower than that within the PKS 1830-211 system [T$_s \sim$1000 K in some of the H~I material \citet{Subrah}]. 
The primarily amorphous silicates in these other quasar absorption systems could be linked to their lower spin temperatures. 
(We note, however, that the spin temperatures in quasar absorbers are somewhat uncertain, as there is debate about the fraction of the quasar radio emission covered by the foreground absorber,
e.g., \citet{Curran,Kanekar}).

\subsubsection{Physical Conditions}

If the features in our spectrum are indeed produced by crystalline silicates, it would be interesting to understand both the stimulus and physical environment producing these crystallines and whether they are similar
to the crystalline silicates detected in more local systems.
Considering the relatively large value of the total to selective extinction [R$_V$=6.34$\pm$0.16; \citet{Falco99}] estimated for the PKS 1830-211 absorber, which is suggestive of large grains associated with relatively dense regions, it is possible that the environment in the galaxy
may be conducive to grain growth which could be linked with the apparent build-up of crystalline silicates detected. One should, however, note cautions by \citet{McGough} about the interpretation of R$_V$ values in such lensed systems.
In light of the measurements of crystalline silicates in other astrophysical sources, and in the laboratory, there persist a number of puzzling questions for the z=0.886 absorption towards PKS 1830-211, if crystalline silicates are invoked to explain the spectral structures.  

First, it is unclear why there is no evidence of amorphous silicates in the 10~$\micron$ region of the PKS 1830-211 absorption system.
Galactic sources, even those noted to be relatively crystalline-rich, are spectrally-dominated by the contribution from the amorphous silicates in the 10~$\micron$ region, as is well-illustrated by the collage in Figure 4 of \citet{Bouwman01}
and in the analysis of \citet{Bouwman03}. In our spectrum, however, we see significant potential crystalline structure, which could imply a deficit of the amorphous silicate material.
It is not clear what mechanism would produce such a surfeit of crystalline material in the z=0.886 absorber towards PKS 1830-211, or alternatively, such a suppression of the amorphous silicates in the 10~$\micron$ region. 
In the 18~$\micron$ region, where crystalline silicates might be expected to also be prominently detected,  
our data are consistent with a broad amorphous-type structure with no distinctive crystalline features; however, we note that our visibly noisier data in this region could possibly obscure narrow crystalline signatures. 

The second question which remains unclear is the temperature of the media producing and hosting the crystalline silicates. 
Galactic crystalline silicates are found in a range of different temperature environments and may be formed through differing physical processes.
Temperature estimates within the z=0.886 absorber vary significantly according to \citet{Henkel}, who estimate kinetic temperatures of $\sim$80~K for 80-90\% for the system, 
with warmer temperatures for the remaining 10-20\% of the material, some of which is $\gtrsim$600~K and generally concentrated in a spiral arm of the absorber galaxy.
Detailed studies of laboratory crystallines, in combination
with observations of crystallines in Galactic sources, indicate that efficient production of crystallines requires temperatures in excess of $\sim$1000~K, 
and that if the material cools too rapidly after formation then the material becomes quickly amorphized.
However, since we are observing the silicate profile in \textit{absorption} rather than emission, this suggests that the material may be relatively cool, 
although we note that, in principle, some warm material could also be detected in absorption.
The detection of crystalline silicates in a cooler medium is not completely unusual; for instance meteorites exhibit significant quantities of crystalline silicates.
The presence of such cool crystalline silicates in an extragalactic environment would suggest that the crystalline material may have been transported in, e.g., a cooling outflow from 
an initially hotter medium; 
in ULIRGs, the cooling outflows stemming from evolved stars within the star-bursts are invoked to explain the temperature discrepancy between the observed crystalline silicate absorption and the
required high temperature formation medium. 
Alternatively, \citet{Molster99} have found evidence that the cooler, lower-density, disk-like material in some evolved stellar systems exhibits a high abundance of crystalline silicates at temperatures below
the annealing temperatures, and have attributed their existence to an unidentified low-temperature crystallization process. Laboratory experiments have also indicated that partial crystallization can be achieved
at room temperature through electron irradiation, although it is unclear whether this process is effective in astrophysical environments \citep{Carrez01}. 

The third aspect which remains unclear is the iron content in the material. In addition to constraining the species and derived expected behavior of the quasar absorber dust material, 
it would be interesting to compare the iron content in silicate dust with the Fe/Si ratios inferred indirectly from gas phase depletions in other systems. (Although we note that some iron may be locked up
in non-silicate dust, such as metallic or metallic oxide grains.)
For instance, \citet{Welty01} have illustrated that along one sight line in the SMC, Fe exhibits severe depletion while Si can exhibit much milder depletions than would be expected from the depletions of 
other refractory species, in contrast with what is observed for the Milky Way.
Galactic crystalline silicates are generally Mg-rich and Fe-poor, i.e. to be much closer to forsterite than to fayalite,
although some large ($>$1~$\micron$ diameter) crystalline silicates with Fe are found in the Solar System \citep{Molster02}. However, many DLAs [e.g., \citet{Prochaska07,Kulkarni10}] 
show at least some Fe-depletion, i.e. [Fe/Zn]$\sim$-0.62 for z$<$1.5 \citep{Meiring}, so their dust grains are unlikely to be completely Fe-poor. Longer wavelength IR spectral data are required to establish the relative Fe-enrichment
in this source, as described in \S\ref{SUMMARY}. 

Finally, there remain a number of alternatives for the formation mechanism of the crystalline silicates within the PKS 1830-211 absorption system.
In general Galactic crystalline silicates are believed to form via one of three primary mechanisms:
(i) condensation at or above the glass temperature, which is thought to be the primary mechanism around evolved stars; 
(ii) annealing (crystallization via heating), or formation following melting, of amorphous grains in a high temperature environment, which is the primary mechanism invoked in the accretion disks of young stars; or
(iii) a by-product of planet-formation mechanisms, as invoked for comets, such as through the collisional cascade of asteroid-sized objects due to gravitational effects, or through flash-heating by shocks stimulated by protoplanet-disk
tidal interactions \citep{Molster,Molster02,Bouwman03,Harker+Desch02}. 
Although these mechanisms can successfully produce sizable quantities of crystalline silicates, these crystallines are rapidly amorphized or otherwise destroyed by a range of mechanisms including sputtering, evaporation, cosmic-rays,
grain-grain collisions, supernova shock waves and adsorption of Fe \citep{Molster,Molster02,Tielens98}. In the ISM, crystalline silicates may be amorphized within $\sim$40-90 Myr, by, e.g., heavy ion cosmic rays \citep{Kemper04,Kemper04e,Bringa}. 

In the case of ULIRGs, the relatively high ISM crystallinity may be explained by enhanced ongoing star-formation. 
\citet{Spoon} ascribe the high percentage of crystallines to recent merger-triggered star-formation, and postulate that the amorphization process lags the injection of dust which is driven by the intense star-formation. 
They posit that this crystalline-rich dust is originating in evolved massive stars, such as red supergiants, luminous blue variables,
and Type II supernovae. Numerical simulations in \citet{Kemper10}, however, suggest that even under the assumption of intense star-formation rates (1000~M$_{\sun}$~yr$^{-1}$) and highly efficient dust production by supernovae, 
they are barely able to explain the 6.5-13\% crystallinity observed in ULIRGs, and with more realistic, lower, values for these rates, the addition of a secondary heating source, such as an AGN, is required to fully explain the observed crystallinity. 

The scenarios which appear the most probable to explain the observed crystalline silicates in the PKS 1830-211 absorption system run a wide gamut.
(i) The first scenario is an extreme burst of star-formation within the SW spiral arm, which is heavily obscured by both dust and molecular material. 
Alternatively, one of the PKS 1830-211 lines of sight could trace a path directly through an isolated, high stellar mass cluster in which strong outflows have transported the crystalline silicates into a slightly cooler region, 
wherein the crystalline material has not yet amorphized. While intense star-formation has been invoked to explain the crystallinity in ULIRGs,
 there is no evidence for intense star-formation within the spiral arms of the PKS 1830-211 absorber galaxy. Studies have previously indicated that the z=0.886 absorber host is a typical late-type spiral, 
based on its position on the Tully-Fisher relationship [see, e.g., \citet{winn}]. 
While $R_V$ is relatively high \citep{Falco99}, there is also evidence that the dust-to-gas ratio in this system is consistent with the Galactic value \citep{Dai}.
Furthermore, the SW arm is noted to be significantly rich in molecules and even displays unusual isotopic ratios \citep{Muller06,Muller11}.
(ii) The second scenario is intense, recent, AGN activity within the absorber galaxy which produced outflows to export crystalline silicates from the central torus, 
or a strong flux of UV radiation to anneal the amorphous silicates, as has been invoked by \citet{Kemper10} to help explain the ULIRG crystallinity. 
However, there is no published evidence of a strong AGN in the absorber galaxy. 
(iii) The crystalline material could have been abundantly produced in young stellar disks. However, in this scenario,
the material should be hotter, and potentially detected in emission, rather than absorption. Furthermore, this mechanism would again require a massive (undetected) 
starburst to produce adequate quantities of crystalline silicates to dominate the profile.
(iv) The remaining mechanism which could be invoked is that of shock-heated material, perhaps from a localized, extreme shock resulting from a transient event. 
Shock-heating mechanisms have been used to explain the crystalline silicates in comets \citep{Harker+Desch02}, 
although we note that very strong shocks could destroy not only the crystalline structure but also the dust grains themselves. 
The origin of the shocking mechanism in the absorbing galaxy remains unclear, however, and detailed simulations would be required to determine whether a large scale shock could produce 
sufficient temperatures to form the crystalline structures without destroying the dust grains.  We regard scenario (i) as the most likely given the high molecular content and large R$_V$ value of the absorber.

\subsection{Amorphous Silicate Absorption Combined with Atomic/Molecular Features}
While we believe that crystalline silicates provide the simplest explanation for
the observed substructure within the 10~$\micron$ absorption feature, we now explore a variety of scenarios to assess whether the observed features could be reproduced without crystalline silicates. 
The first alternative explanation which we explore is the possibility that the observed structure is the result of a combination of a broad absorption feature, 
produced by an amorphous silicate of olivine composition, and a series of other additional atomic and molecular absorption/emission features within the quasar absorber material. 
This possibility is worth exploring because 
(i) the 10~$\micron$ amorphous silicate feature is present in many Galactic and extragalactic sources; 
(ii) this absorption system is exceedingly rich in molecular material \citep{WiklindN,Wiklind,Muller11};
(iii) narrow absorption and emission features have been observed in ULIRGs \citep{Spoon}; and 
(iv) previous detections of crystalline silicates in astrophysical sources, e.g. ULIRGs, have been much weaker than in our profile.

\begin{figure}
\epsscale{1.}
\plotone{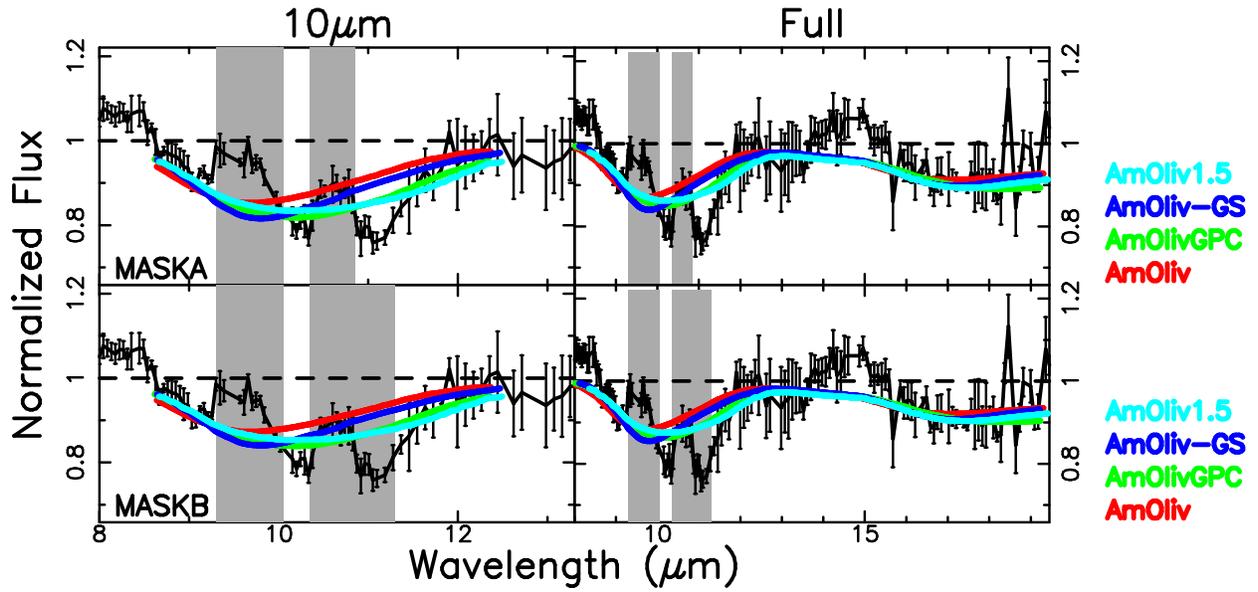}
\caption{
Illustration of a test to see whether a version of the PKS 1830-211 spectrum in which the substructure features are masked (and attributed to alternative absorption/emission sources) could be well-fit by amorphous olivine. We depict these masked regions,
which were excluded from the fit, with shaded grey areas. The top panel masks two `emission' features (mask A), while the bottom additionally masks an `absorption' feature (mask B). None of the four amorphous olivine templates
adequately matches the absorption feature, even with masking, because the amorphous olivine profiles are both too shallow and peak at lower wavelengths than our observed broad feature. As in previous figures, 
the left panel depicts the 10~$\micron$ fit, and the right the full fit; wavelengths are in the rest frame of the z=0.886 absorber.}
\label{maskFIG}
\end{figure}

The first test of this scenario which we have implemented is to manually mask potential absorption and emission features within the broader 10~$\micron$ absorption feature. We have then attempted to fit the remaining points 
with a series of four amorphous olivine profiles which were selected as providing the best range of olivine templates (see Appendix Table~\ref{APtemplatesLAB} for template details). 
We have implemented these fits applying two different masking combinations: mask A which 
attributes the 9.3-10.05~$\micron$ and 10.35-10.85~$\micron$ regions to unidentified, superposed emission features, and mask B, which includes both of these purported emission features, as well as an absorption 
feature at 10.85-11.3~$\micron$.  (We have also considered two other masks which exclude the 8.65-9.25~$\micron$ spectral region, in which the enstatite, silica, or serpentine material, or ammonia, absorption, may be present, 
instead of the 9.3-10.05~$\micron$ region; but these two masks do not fit the data as well as mask A or mask B, and so we do not illustrate these fits.)
We have adopted the same fitting procedure as utilized in \S\ref{fitting}. These fits are illustrated in Figure~\ref{maskFIG} and parameterized in Table~\ref{maskshiftT}. We present the $\chi_r^2$ values in this table in order to facilitate
comparisons between the different templates and masking schemes, but we note that the bulk of the structure in the 10~$\micron$ region is masked, and as a result relatively few points in the masked profiles are being considered in the $\chi_r^2$. 
For the less aggressive mask A, the amorphous olivine silicate fits remain too narrow/shallow relative to the PKS 1830-211 profile and peak at shorter wavelengths. 

\begin{figure}
\epsscale{1.}
\plotone{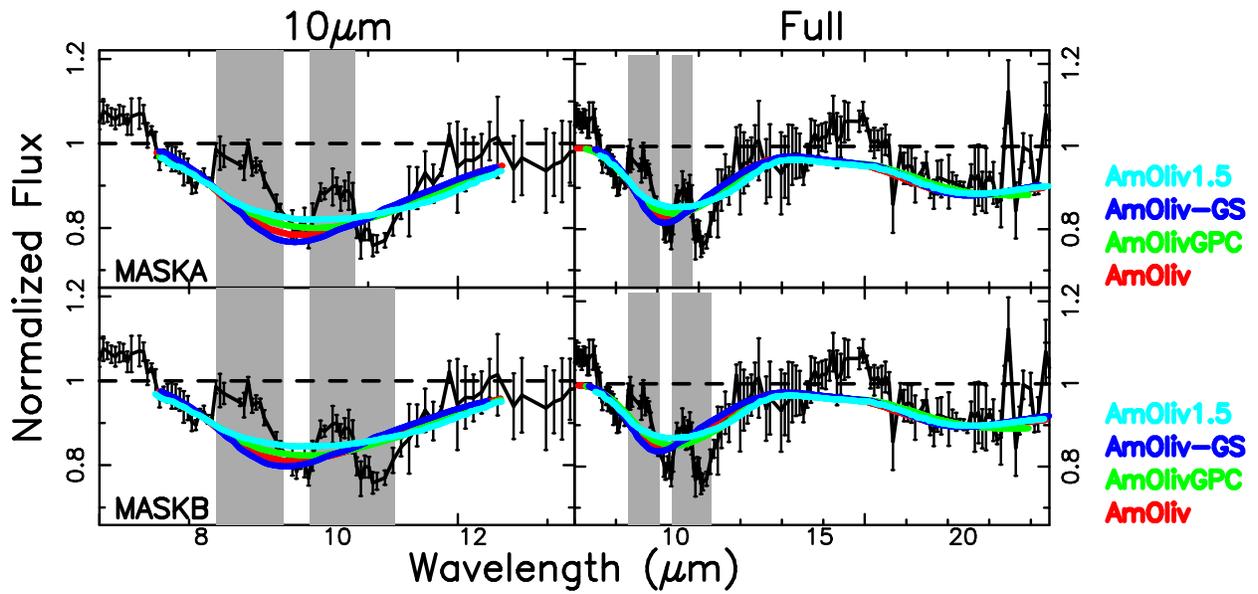}
\caption{
Similar to Figure~\ref{maskFIG}, but illustrating the fits when both masking of PKS 1830-211 and wavelength-shifting of the template profile are implemented.
These fits, particularly those for mask B, are an improvement over the pure masking. However, while the fit is able to reproduce the underlying broad shape of the absorption feature, it requires significant masking and consequently
emission/absorption produced by alternative materials, in addition to a shift of $\sim$0.1-0.5~$\micron$, perhaps attributable to grain variations. The combination of these phenomena seems unlikely in comparison with the crystalline
silicate explanation for the substructure.}
\label{maskshiftFIG}
\end{figure}

As a second test we have employed the same masking of the PKS 1830-211 absorption spectrum, and we have additionally allowed the amorphous olivine templates to be arbitrarily shifted in wavelength-space to produce the best fit. 
We permit this wavelength-shifting because 
(i) as discussed in Appendix~\ref{expandedAP}, variations in silicate grain morphology can produce marked offsets in the peak of the optical depth profile; and
(ii) recent work by R. Nikutta (private communication) suggests that sizable wavelength shifts in the 10~$\micron$ peak can be produced around quasars with a clumpy dust torus model which combines simultaneous silicate detections in absorption and emission, in conjunction with the rising quasar continuum. 
Although we are not looking at dust within a quasar torus, the work of Nikutta raises the possibility that a combination of absorption and emission by clumpy dust 
could contribute to the observed wavelength offsets. 
The results of this fitting are shown in Figure~\ref{maskshiftFIG} and are tabulated in Table~\ref{maskshiftT}. We find that using mask B with shifting of +0.1 to +0.5~$\micron$, we can achieve relatively low $\chi_r^2$ values. However,
as previously noted, this is based on substantial masking of the absorption feature, and there are few points which are being directly fitted within the deepest part of the optical depth profile; the bulk of the fitting leverage is being 
applied by the wings. Given the reduced number of sample points being fit, we do not consider the masked fits to be as reliable as those in our primary analysis. 

\begin{figure}
\epsscale{0.8}
\plotone{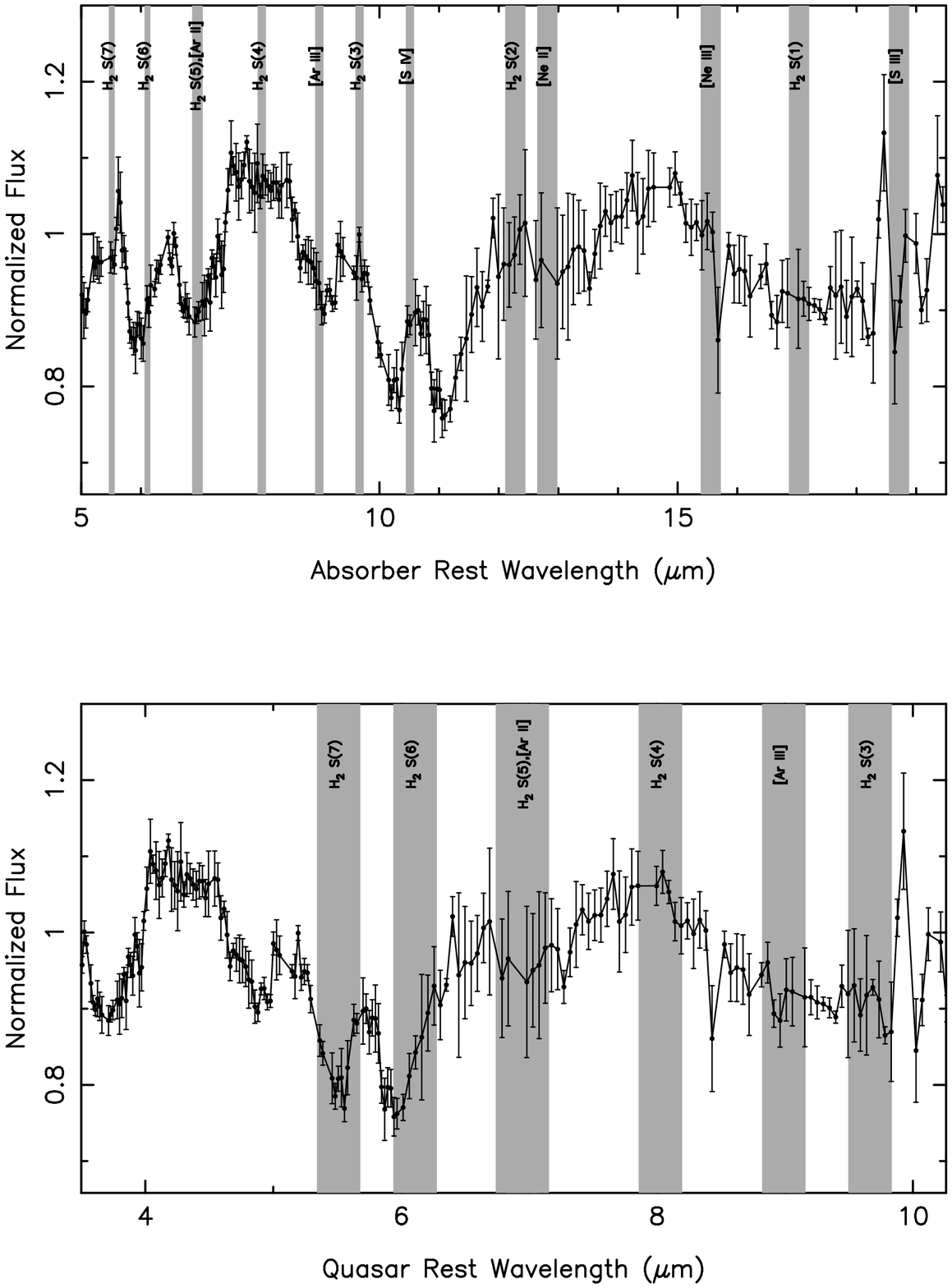}
\caption{
Examination of whether the substructure features in the PKS 1830-211 spectrum (depicted in both the z=0.886 absorber rest frame and the quasar rest frame) could originate from atomic/molecular transitions which are found in 
star-forming galaxies \citep{Smith}, ULIRGs \citep{Spoon}, or active galaxies/AGN \citep{Hao}. We show in grey these lines, taken from \citet{Smith}, with the breadth of the feature determined from the unresolved FWHM in the four
IRS spectral regions, following that paper. While some of the smaller/weaker features in the PKS 1830-211 spectrum could be attributed to the depicted transitions, such as the small peak in the absorber rest frame at the location
of the H$_2$ S(3) feature, which is prominent in some ULIRGs \citep{Spoon,Higdon}, these transitions in combination with broad amorphous silicate absorption cannot explain the broader substructure features.}
\label{lines}
\end{figure}

Furthermore, the masking scenario is contingent upon all of the identified emission and absorption features being physically explicable by molecular and atomic transitions. In order to determine whether there are plausible
atomic and molecular lines which could produce these features, we have considered prominent lines which are seen in other extragalactic and Galactic sources (see Figure~\ref{lines}).
As is evident from this figure, none of the commonly observed features are broad enough to explain the significant substructure within our absorption feature. We note that H$_2$ S(3) emission, which is commonly
seen in ULIRGs, may be present in our z=0.886 absorber, producing the small hump in the larger substructure peak. This emission, however, cannot explain the entire substructure peak, as the line is not only relatively narrow in comparison,
based on the line widths in star-forming galaxies \citep{Smith} and ULIRGs \citep{Spoon}, but it is also not centered on the broad substructure peak seen in our data. We also do not see prominent H$_2$ S(2) or H$_2$ S(1) emission at longer
wavelengths, although these features could be lost in the noise. The emission feature from 10.35-10.85~$\micron$ could be associated with the [S IV] 10.51~$\micron$ feature, which is again seen in some ULIRGs \citep{Spoon}
and star-forming galaxies \citep{Smith}, but our feature appears broader, offset to longer wavelengths, and more prominent than the unresolved line seen in these other systems. The best candidate to explain the absorption feature from 
10.85-11.3~$\micron$ remains crystalline silicates, based on the work of \citep{Spoon}; however, we note that if we invoke a combination of amorphous silicates with molecular/atomic transitions to explain the bulk of the substructure
in the profile, we require a much smaller contribution from crystalline silicates than claimed elsewhere in this paper.

In general, these superposed emission features in conjunction with broad amorphous absorption appear inadequate to explain the detected substructure. 
Also, there is no strong evidence that all of these particular features can be physically produced and detected. 
Emission lines of H$_2$ and [S IV] are detected in ULIRGs, but these objects are undergoing extreme star-formation,
and can be detected in emission. By contrast, the PKS 1830-211 quasar absorber host is a regular spiral which is barely detected in HST images, with no noted pronounced star-formation. It seems unlikely that strong emission lines 
of these species would be produced in such an environment, and certainly not in the required line strengths/breadths to explain our substructure peaks. 

Lastly, we address the possibility that the substructure we attribute to crystalline silicates is, in fact, a by-product of PAH emission overlaid on a broad silicate absorption feature. Given the location of the expected PAH emission 
features in the rest frame of the z=0.886 absorber, illustrated in Appendix Figure~\ref{PAHS}, we deem this unlikely. While the 10.68~$\micron$ PAH emission feature in the rest frame of the absorber is consistent in location with the peak between the 
10~$\micron$ and 11~$\micron$ absorption sub-features, this would still require a shift in the amorphous olivine template and an additional absorption mechanism to explain the $\sim$9~$\micron$ feature. 
The sub-peak between 10 and 11~$\micron$ (in the z=0.886 absorber rest-frame) aligns with the 5.7~$\micron$ PAH feature in the quasar rest-frame, but this feature is quite weak in most quasars \citep{Hao}.
We deem it more likely that our substructure originates from crystalline silicates than from a wavelength-shifted amorphous olivine profile in combination with superposed PAH emission, although
we concede that this sub-peak may be enhanced by some PAH contributions.
\subsection{Foreground Galaxy Absorption}
An alternative scenario to explain the structure in the PKS 1830-211 absorber spectrum is that we are detecting superposed amorphous silicate features from two absorption sources at different redshifts which serendipitously combine
to produce the appearance of multi-peaked structure. 
This is plausible in that the NE quasar sightline is passing within 10 kpc of the center of a z=0.19 foreground galaxy \citep{Lovell1996}, and so our spectrum is also probing the material in the outskirts of this z=0.19 foreground galaxy, 
which may contain silicate dust. If we perform a bi-variate fit in which we consider an amorphous olivine template at the rest-frame of our z=0.886 quasar-absorption-host, combined with an amorphous olivine template at the rest-frame of the 
z=0.19 galaxy, as in Appendix~\ref{foreground}, we do not find this to be viable, however. At the wavelength at which we would expect 10~$\micron$ silicate absorption in the z=0.19 foreground galaxy, 
we instead detect an emission-type feature. We, therefore, conclude that the structure in the 10~$\micron$ region of the PKS 1830-211 absorber spectrum is not the superposition of 10~$\micron$ silicate absorption from the absorber host galaxy
and 18~$\micron$ silicate absorption from the z=0.19 foreground galaxy. We also note that while such a superposition might account for two of the substructure peaks, it would not account for three, and would require a third absorption
system to explain the $\sim$9~$\micron$ absorption feature, or once again invoke additional mineral/molecular absorption in this region. Furthermore, while in principle two absorption features could combine to produce the observed
`W' shape formed by the 10-11~$\micron$ substructures, individual amorphous olivine silicate profiles are too intrinsically broad to produce a `W' of the required shape.

We have also repeated the bi-variate fitting both allowing for a foreground galaxy at any arbitrary redshift, in combination with the z=0.886 absorber, and allowing for two systems at arbitrary redshifts out to z=1, with possibly no absorption from
the z=0.886 system. These scenarios cover the possibility in which a dark object which is rich in silicate material was missed in previous studies, as well
as the possibility that the z=0.19 galaxy might in fact be at a higher redshift closer to the z=0.886 galaxy, as suggested by \citet{Lehar}, or that the z=0.886 galaxy might in fact be multiple galaxies, as proposed by
\citet{Courbin}. However, we do not find any such ``hidden" object responsible for producing superposed silicate absorption features in our spectrum. 

Finally, we have examined the possibility that we are detecting a combination of crystalline silicate absorption at z=0.19 with amorphous olivine absorption at z=0.886. While we do see
structure near 6-7~$\micron$, which we have attributed in Appendix~\ref{residbump} to H$_2$O and other molecular absorption, we note that these features occur in the z=0.19 rest-frame
10~$\micron$ spectral region. We, thus, tested whether this structure might be attributable to crystalline silicates at z=0.19. However, when we fit the hortonolite spectrum to the z=0.19 rest-frame absorption 
spectrum, we find that the hortonolite features are too narrowly spaced to explain the $\sim$6-7~$\micron$ features, and that the structure at longer wavelengths is not well-reproduced by 18~$\micron$ crystalline silicates in the rest-frame
of the z=0.19 galaxy. Thus, we discount this scenario.
\subsection{Amorphous Silicates with Superposed Emission}
Lastly, we consider the possibility that the PKS 1830-211 spectrum could be produced by a broad amorphous olivine feature associated with the z=0.886 absorber in combination with superposed foreground or coeval emission. 
For instance, if the presumed z=0.19 galaxy is instead a massive galaxy located at z=0.886, as proposed by \citet{Lehar}, or the z=0.886 host galaxy is instead comprised of two neighboring galaxies, as proposed
by \citet{Courbin}, then we might expect a blend of silicate absorption with either some limited emission or absorption reflecting a differing chemical composition, from the second galaxy. 
This scenario has largely been excluded by the bi-variate fitting tests in \S\ref{bivar}, in which we did not find compelling evidence arguing for two distinct chemical compositions, although we cannot exclude a weak contribution from a second source.
Alternatively, we could be detecting broad silicate absorption along the SW quasar line of sight probing the molecule-rich material, while simultaneously detecting emission from, e.g. crystalline silicates, along the other quasar line of sight.
However, the broad amorphous silicate absorption would still need to be offset in wavelength to fit our spectrum, as would the crystalline silicate emission template; and  
no single offset or explored crystalline silicate could simultaneously reproduce all of the observed peaks/substructure.
Furthermore, as in the case of invoking atomic and molecular emission lines, it is not clear that such emission should be detectable from the z=0.886 host galaxy, or any companion galaxies. 

Alternatively, some of the foreground emission could be produced by a transient object within our own Galaxy or Solar System, perhaps connected to the scattered light feature discussed in Appendix~\ref{scatlight}. 
The flux from a foreground object could be probing a longer rest-frame wavelength region than the z=0.886 absorber, where crystalline silicate resonances would be more abundant. For instance,
\citet{Bouwman10} find abundant crystalline silicates around a young, low-mass M4 star with a proto-planetary disk, and as discussed in \S\ref{foreground}, there is an M4 dwarf within our slit. 
However, we do not believe that the flux from this particular star contributes significantly to our PKS 1830-211 absorption spectrum, as detailed in \S\ref{foreground}, 
nor is it believed to be a particularly young stellar system, and so any crystalline silicates produced during its youth are expected to have dissipated.
Furthermore, the likelihood of a transient source providing significant crystalline structure seems unlikely, and as discussed in Appendix~\ref{scatlight}, 
we do not expect contamination from the scattered light feature to be strong enough to produce structure at the level we observe in the PKS 1830-211 spectrum.
\section{SUMMARY}\label{SUMMARY}
In summary, based on an infrared spectrum of PKS 1830-211 obtained with the Spitzer IRS, we have detected the presence of a 10~$\micron$ silicate absorption feature in the z=0.886 absorber at a $>$10$-\sigma$ significance level, as well as the presence
of an 18~$\micron$ silicate absorption feature. This reinforces previous findings from \citet{Kulkarni1,Kulkarni2} that silicate material is a component of dust in quasar absorption systems. We have found that the 10~$\micron$ silicate feature
is best fit by crystalline olivine, possibly in combination with a second material such as enstatite, silica, or serpentine, or with superposed ammonia absorption. Given the rarity of such highly crystalline silicate dust, 
we have explored a variety of other scenarios invoking atomic and molecular absorption/emission lines, and a superposition of silicate dust in two quasar absorption systems; but none of these alternative
scenarios satisfactorily reproduce the observed structure in the PKS 1830-211 spectrum. 

In order to confirm the presence of the postulated crystalline silicates, and to eliminate some of these suggested formation scenarios, supplemental data are required. 
In particular, moderately high resolution data of the PKS 1830-211 absorption system at $>$20~$\micron$, in the z=0.886 absorber rest frame, 
are needed to provide evidence for the presence of longer wavelength crystalline resonances and to constrain 
the relative amount of Fe present in the system. 
As illustrated in \citet{Jaeger}, at $>$20~$\micron$ the crystalline olivine profiles vary significantly with the ratio of Mg/Fe, 
with peaks shifting to longer wavelengths, by as much as several $\micron$, as more Fe is introduced to the mineral. As summarized in \citet{Molster},
these differences have been used to provide evidence of the high Mg/Fe ratio in, e.g., evolved stars, using the 69~$\micron$ band. Furthermore, as discussed in \citet{Koike06}, variations in the temperature of the material will produce offsets
in the peak wavelengths of the spectral features; these offsets become increasingly sizable at longer wavelengths. 
For example, \citet{BoweyTemp} have exploited these offsets to place constraints on the temperatures of a range of stellar objects using the 69~$\micron$ forsterite band.
Astronomical observations spanning a broader spectral range could also provide evidence to support/rule out other atomic, PAH, and molecular transitions in both the quasar and the absorber. 
Higher resolution spectra covering the absorber rest-frame 10~$\micron$ and 18~$\micron$ spectral features for the PKS 1830-211 z=0.886 absorber could also be of great use for ascertaining better estimates of the 
structure features and line widths, which in turn could help discriminate between
the emission/absorption scenario and the crystalline silicate scenario, as well as helping to determine whether we we are seeing broad emission
line features, or emission features which are a blend of multiple narrower emission lines.  
These data would also help rule out any contributions to the spectra from the anomalous scattered light feature (Appendix~\ref{scatlight}), as it is believed to
result from a transient source, and so should be absent in follow-up observations. 
Finally, spectroscopic or photometric observational constraints on the amount of star-formation within the obscured SW spiral arm of the z=0.886 absorbing galaxy could help to further constrain the formation mechanisms of the silicate dust. 

\acknowledgments

We thank the anonymous referee for his/her suggestions to improve this paper, and C. J\"{a}ger and Th. Henning for their assistance in helping us obtain the laboratory-based optical depth profiles for our analysis.
Additionally, we thank the Spitzer Science Center help desk for their assistance with the IRS data analysis and interpretation of the scattered light feature. 
We also acknowledge use of several subroutines taken from \citet{Press} in our analysis.
This work is based on observations made with the Spitzer Space Telescope, and has made use of the NASA/ IPAC Infrared Science Archive, which are operated by the Jet Propulsion Laboratory, California Institute of Technology under a contract with the National Aeronautics and Space Administration (NASA). Support for this work is provided by NASA through an award issued by JPL/Caltech: Spitzer grant  PID-50783 (PI V. P. Kulkarni). Additional support comes from National Science Foundation grant AST-0908890 to the Univ. of South Carolina (PI V. P. Kulkarni).

{\it Facilities:} \facility{Spitzer (IRS)}.
\appendix
\section{PKS 1830-211 Quasar-Continuum Normalization}\label{normAP}
\subsection{Spectral Joining Methodology}\label{joinspect}

\begin{figure}
\epsscale{1.0}
\plottwo{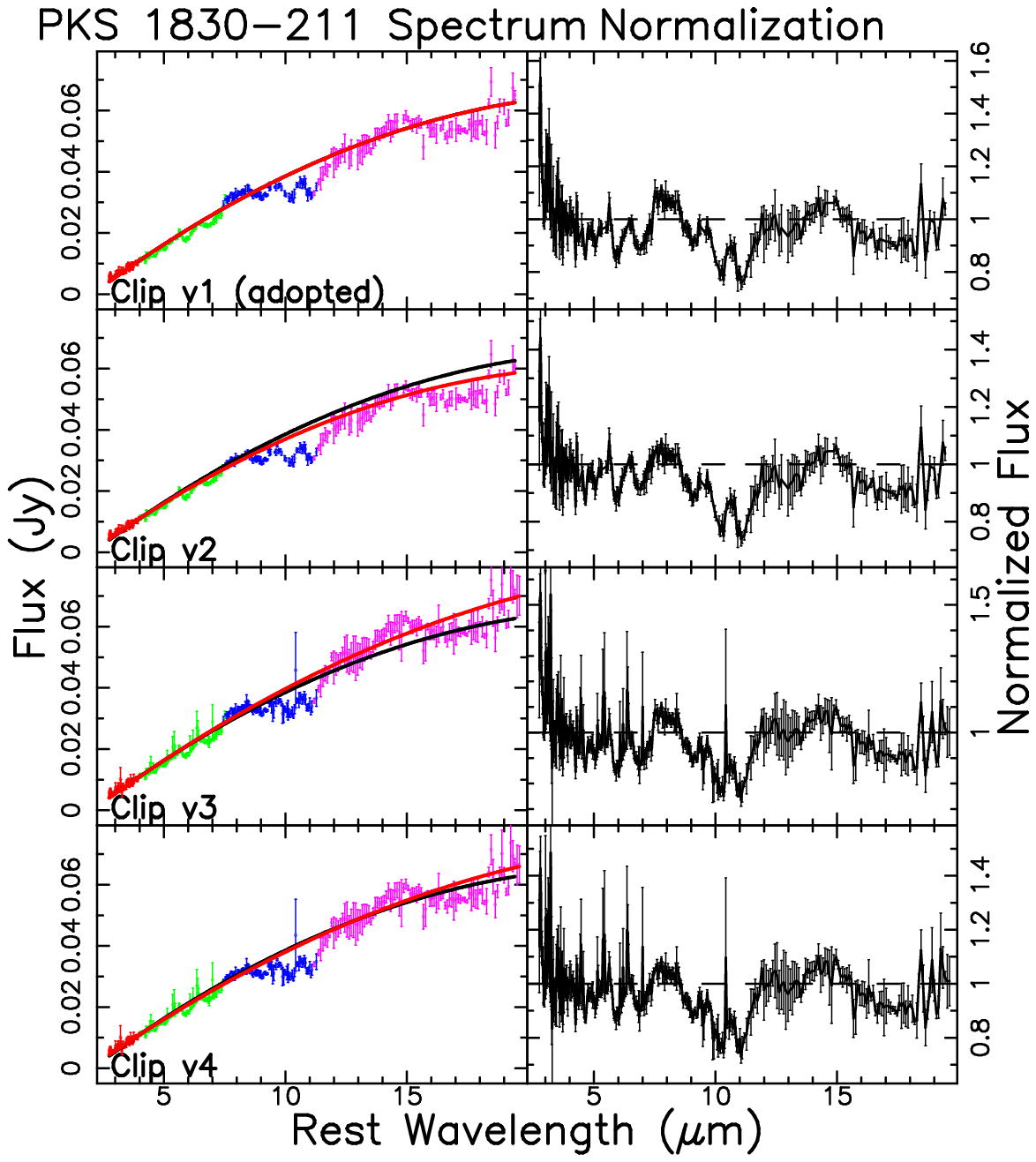}{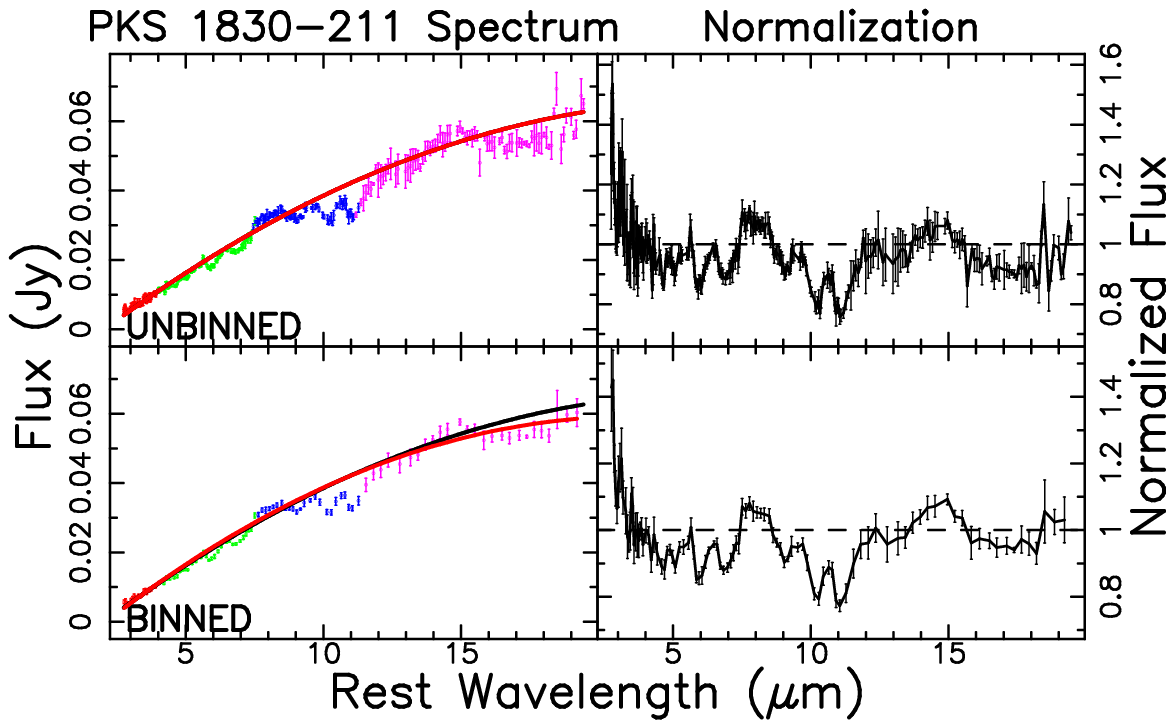}
\caption{
Illustration of the impact of variations in the adopted clipping and binning on the quasar-continuum normalizations,
and resultant normalized flux profiles, for PKS 1830-211, in the z=0.886 absorber rest-frame. 
Left: Comparison of the derived profiles when applying additional or reduced clipping of data points, with the adopted clipping shown in the top row.
Right: Depiction of the impact of binning (bottom panels) versus not binning (top panels). 
Within each panel, the left plot illustrates the PKS 1830-211 spectrum, with data point colors as described in Figure~\ref{spectrum}. The 
black line depicts our fiducial adopted clipping, with no binning, combination, while the red line shows the alternative normalization for comparison. The right plot shows
the spectrum normalized by this alternative normalization. 
None of these explored variations significantly impact the derived fits. These variations are discussed in Appendix~\ref{normAP}.
}\label{clipbinvar}
\end{figure}

As addressed in \S\ref{spice}, we have joined the four individual nod-combined SL and LL spectra in order to form a single spectrum for PKS 1830-211, following a manual trimming of the data to eliminate lower quality data points, 
i.e., those points which appear visually discrepant or noisy. As discussed further in \S\ref{vardepth}, the exclusion of these data points does not significantly impact our derived fits and conclusions. 
The majority of the removed points constitute isolated data points where a bad pixel was not removed through our conservative spectral cleaning described in \S\ref{clean}. The remaining points occur in regions where the two nods showed discrepant spectral features, indicative of bad-pixel contamination in one position, or occur beyond 38~$\micron$ where the noise notably increases. 
Lastly, we have eliminated points at the edges of each spectral order which do not connect smoothly with the adjoining orders. As in \S\ref{clean}, we have been as 
conservative as possible when making these cuts, so as not to eliminate spectral features which may be potentially physical in origin. We illustrate the spectrum which would result when excluding the bulk of this clipping in the left, bottom two 
panels of Figure~\ref{clipbinvar} (clip v3 and v4). As is evident in this figure, there are several additional data points with large error bars, which are absent in our fiducial clipping (v1), but the overall shape of the spectrum is unaltered.

Following this manual clipping of the data, we have normalized and stitched together the data in the SL2, SL1, LL2, and LL1 spectral orders, to form a single SL spectrum and a single LL spectrum for the PKS 1830-211 system. Within
the SL and LL orders we have normalized the first and second order data to the bonus (third) order. The individual data points from the bonus orders have been excluded from our final spectrum, 
as they provide no additional information since they align with the data in the retained
spectral orders and exhibit similar features within the overlapping spectral regions; furthermore, their inclusion would lead to uneven sampling over the spectral regions spanned by the bonus order relative to the full spectrum. We note
that the joining of the LL1 and LL2 spectral orders occurs within the 10~$\micron$ feature of our z=0.886 absorber, which is the focus of our analysis, and so could potentially impact the shape of this feature. However, as the orders align well, have 
overlapping data points at the edges of the orders, and match the bonus order spectrum, we do not believe that the observed substructure illustrated in Figure~\ref{spectrum}, and attributed in \S\ref{CONCL} to physical mechanisms, is affected by this joining. 

We have generated the final PKS 1830-211 spectrum by scaling down the LL spectrum and joining it with the SL spectrum. An examination of the well-separated longer (radio) and short (X-ray) wavelength data for PKS 1830-211 in the literature does not convincingly argue for either scaling up the SL data or scaling down the LL data. Since the SL data generally have smaller associated formal uncertainties, we have given that order preference in setting the absolute-flux-normalization. This has resulted
in final normalization-scaling factors of 0.97, 0.99, 0.91, and 0.87 relative to the original extracted data for the SL2, SL1, LL2, and LL1 data respectively (including the shifts of the first and second order data to match the corresponding bonus order data), 
with a slight overlap in SL and LL data points at the location of the joining. 
These scaling factors are consistent with scalings employed in other analyses [e.g., \citet{Spoon,Smith}]. As our analysis is based on a quasar-continuum-normalized spectrum, any offsets in the absolute flux normalization will not impact our conclusions.
We have also considered a slightly modified version of the joining in which we employed additional clipping at the long-wavelength edge of the SL spectrum,
and then shifted the LL spectrum down by a larger factor. This alternative joining is illustrated in the clip v2 (and clip v4 for the less aggressively clipped data) panels of Figure~\ref{clipbinvar}, and does not significantly impact our conclusions. 

We have, lastly, experimented with binning the data by a factor of three, using an unweighted average. While, as illustrated in the right panel of Figure~\ref{clipbinvar} this can reduce the uncertainty associated with each data point, and produce a cleaner
looking spectrum, it can also obscure the finer structure in the absorption feature. Given that we find indications of significant substructure in our analysis, we have opted against binning. As discussed in \S\ref{vardepth}, however, had we chosen
binning it would not have altered our conclusions that crystalline olivine provides the best match to the 10~$\micron$ data feature. 

\subsection{Variations in Spectral Joining and Continuum Normalization}\label{normspect}

In order to fit the silicate template profiles in \S\ref{fitting} to the PKS 1830-211 absorber silicate features (in the rest frame of the z=0.886 absorber), we have normalized the observed spectrum by a fit to the underlying continuum.
We have first examined the data at both longer (radio) and shorter (X-ray) wavelengths to ascertain whether a single quasar continuum could be fit through an extended frequency range. However, given the discrepancies in the literature,
the known temporal variability across a wide spectral range of the quasar, and uncertainty in the shape of the quasar spectral energy distribution in the IR region \citep{SED}, as well as the large wavelength separation between our data and these
supplementary data, we decided to select a continuum which fits in detail solely our own IR spectral data. 

We then explored a range of more than 20 fits to the quasar continuum. In the analysis of \citet{Kulkarni1,Kulkarni2}, the underlying quasar continuum, in the quasar absorption systems, was well-represented by a polynomial or power-law. 
Thus, in this
analysis we considered not only a linear polynomial and a power-law normalization, but also 3rd-5th order Chebyshev polynomials, 3rd order Legendre polynomials, and single component cubic spline fits. 
The power-law and linear polynomial fits were determined
using a least-squares algorithm, while the higher order polynomial fits were obtained using the IRAF task \textit{continuum} with unweighted fitting in linear space, and no averaging of the sample points. A subset of the most representative
of these fits is tabulated in Table~\ref{normvarTBL} and illustrated in Figure~\ref{normvar}. Unless specifically noted, each of these fits was obtained using masking to exclude the four prominent absorption features in the spectrum:
5.5-6.2~$\micron$, 6.5-7.1~$\micron$, 8.5-12.2~$\micron$, and 14.8-18.3~$\micron$ in the z=0.886 absorber rest-frame. We also forced the IRAF-fitted curves to pass through points longward of 18.5~$\micron$ in order to constrain the curvature
of the continuum normalization; the implications of this constraint are addressed below. 
Based on the relative contributions of the continua to the observed and laboratory fits over solely the 10~$\micron$ feature and over the combined 10 and 18~$\micron$ features, 
as discussed in \S\ref{fitting}, and a visual inspection of the resultant normalizations, we have determined that the 3rd order Chebyshev polynomial provides the optimal fit. As addressed in \S\ref{vardepth}, however, adopting an alternative normalization
would not have altered our conclusions in this paper.

Lastly, we have experimented with forcing the 3rd order Chebyshev polynomial to pass through/near certain critical parts of the spectrum. Since the continuum fitting is unweighted, this prevents anomalous, high uncertainty points, like those at the longest
wavelengths, from skewing the fit. In Figure~\ref{normvarX} we illustrate five alternative assignments of these regions. The fiducial fitting solely forces the curve to pass through points at the longest wavelengths ($\lambda\gtrsim$18.5~$\micron$),
termed `A' in the figure. We have also examined forcing the spectrum through points between the 10 and 18~$\micron$ features (termed `B' in the figure), points just shortward of the 10~$\micron$ absorption feature (termed `C'), 
points just shortward of the 6~$\micron$ H$_2$O absorption feature (termed `D'), and points in the substructure peak near 9.6~$\micron$ (termed `E'). 
These variations are illustrated by the shaded grey regions in Figure~\ref{normvarX}, and listed in Table~\ref{normvarTBL}, with the resultant continuum variations
shown by the red lines in the left side of Figure~\ref{normvarX}. These variations can impact the relative depth of our absorption feature, as addressed in \S\ref{vardepth}, and illustrated in the right side of Figure~\ref{normvarX}, but they do \textit{not}
alter our conclusion that the absorption feature in the PKS 1830-211 spectrum is best-represented by crystalline olivine.

\subsection{Potential Residual `Emission/Absorption Bumps in Normalized Spectrum}\label{residbump}

An examination of our final normalized spectrum, shown in the right side of Figure~\ref{spectrum}, illustrates that in addition to the prominent 10~$\micron$ and 18~$\micron$ absorption features, which are the focus of this analysis, there are several
additional emission and absorption features, which we briefly address here. First, there are two absorption features located near 6~$\micron$ and 7~$\micron$, respectively. The first of these we attribute to water ice (H$_2$O). Galactic
observations of, e.g., ice-rich embedded young stellar objects (YSOs), have also found this feature, and a direct comparison of these Galactic profiles with our absorption feature finds good agreement in both the breadth and shape of the H$_2$O
feature \citep{Gibb,Chiar11}. The slightly longer wavelength feature is also seen prominently in some Galactic environments, alongside the H$_2$O feature, e.g., in the W33A spectrum \citep{Gibb}. Although its origin is not completely
understood in the literature, it is likely attributable to other molecular transitions associated with the ices. Given that the PKS 1830-211 z=0.886 absorber is well-documented to be rich in molecules, we find that these associations are plausible.

We also note the presence of a visible emission-type feature just shortward of the 10~$\micron$ absorption feature, near 8~$\micron$, which we contend may be evidence of PAH emission. As illustrated in Figure~\ref{PAHS}, the location of the feature
is consistent with the 7.7~$\micron$ complex and 8.3-8.8~$\micron$ PAH emission features in the rest-frame of the z=0.886 absorber. Following  the classifications of \citet{Peeters} and \citet{vanDied}, the shape of the feature is 
similar to a blend of perhaps Type A and Type C PAH emission, perhaps originating in a range of stellar sources.
We note that if this feature does originate in PAH emission, it is unusual that we do not see evidence of the usually prominent 11.3~$\micron$ complex emission, although the relative strengths
of the PAH features can vary, for example, due to variations in the ionization levels [see, e.g., \citet{Tielensrev,Smith}]. These PAH ratio variations are illustrated in Figure~\ref{PAHSil} comparing the 
PAH features for two star-forming galaxies, NGC 5866 and NGC 3198, rescaled and overlaid on top of our PKS 1830-211 spectrum. Furthermore, it is possible
that our absorption feature could be hiding some weak PAH emission, i.e., that the rightmost ($\sim$11~$\micron$) substructure peak may indeed be deeper or broader than our spectrum indicates. This would not alter our assessment that the primary
constituent of the silicate dust is crystalline olivine, but might alter our ascribed species of crystalline olivine or the requisite secondary materials (\S\ref{ANALYSIS}). Finally, we note that in the Q0852+3435 quasar absorption spectrum examined
by \citet{Kulkarni2}, there is also evidence of not only the two absorption features near 6 and 7~$\micron$, but also the emission features shortward and longward of the 10~$\micron$ absorption; thus, these emission features may be physically connected
to an enriched environment.

Finally, we address the `emission' feature longward of our 10~$\micron$ absorption. The physical origin of this feature is less clear. It may originate from one or a combination of mechanisms including weaker PAH emission features in the rest frame
of the absorber, emission from the 7.7~$\micron$ PAH complex emission in the frame of the quasar host galaxy, H$_2$ S(2) emission in the frame of the z=0.886 absorber (which would be consistent with evidence for the H$_2$ S(3) line at shorter
wavelengths), or it may be related to the scattered light feature addressed in Appendix~\ref{scatlight}, which provides the greatest contamination in this region, as is evidenced by the larger associated error bars. Given this range of plausible mechanisms, 
and the added complication of the contaminating scattered light, we do not attribute this feature to an artifact of the quasar continuum normalization. We note, however, that treating this emission as an artifact, and eliminating it through an altered continuum shape,
is encompassed by our tests explored in Figure~\ref{normvarX} and Table~\ref{normvarTBL}, and we emphasize that these normalizations would not impact our qualitative conclusions in this paper.

\subsection{Impact of Data Treatment on Derived Peak Optical Depth}\label{vardepth}
\begin{figure}
\epsscale{0.7}
\plotone{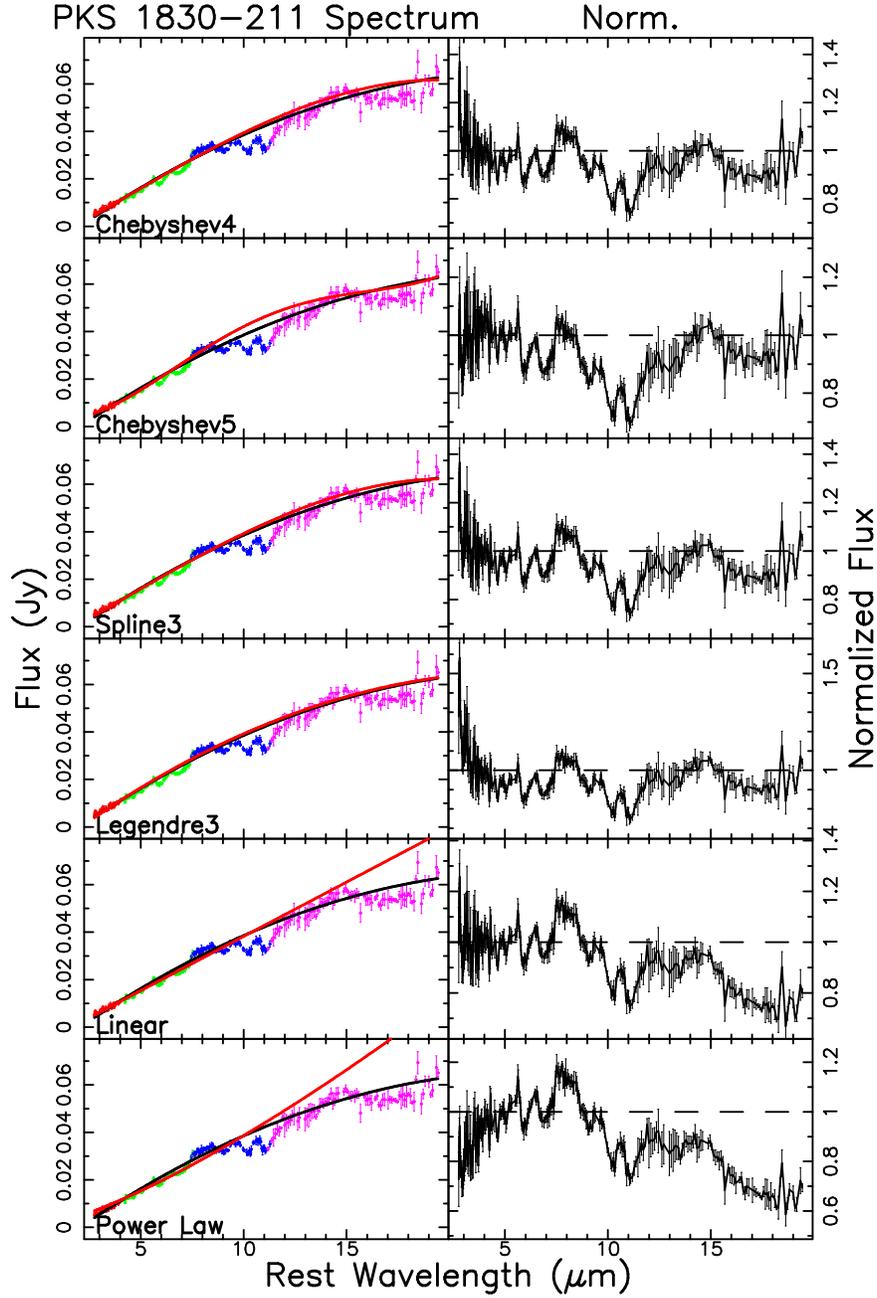}
\caption{
Illustration of explored quasar-continuum normalizations, including higher order Chebyshev polynomials, a cubic spline, a 3rd order Legendre polynomial, a linear fit, and a power law fit, 
and the resultant normalized flux profiles, for PKS 1830-211, in the z=0.886 absorber rest-frame, with panels as in Figure~\ref{clipbinvar}. 
None of these explored variations significantly impact the derived fits, as discussed in Appendix~\ref{normAP}.
}\label{normvar}
\end{figure}

\begin{figure}
\epsscale{0.7}
\plotone{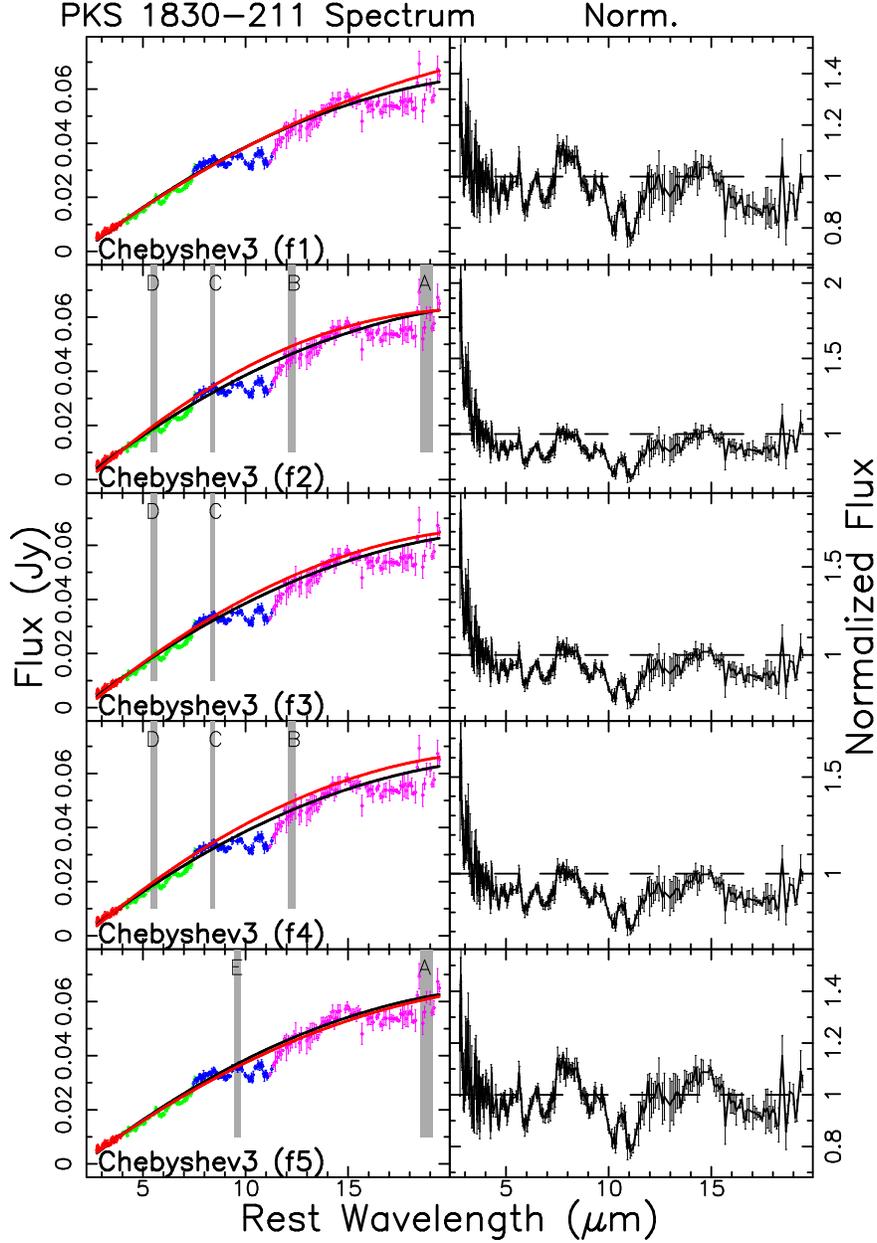}
\caption{Similar to Figure~\ref{normvar}, but illustrating variations in the quasar-continuum normalization shape when forcing the 3rd order Chebyshev polynomial to pass through certain regions of the spectrum, as depicted
by the grey shaded regions. Our fiducial fit employes only forcing passage through points at the longest wavelengths, as shown by region A in the figures. While, as illustrated on the right, these variations can impact the shape of the normalized
quasar spectrum and the depth of the absorption feature, they do not qualitatively alter our conclusions about crystalline olivine silicates providing the origin for the feature. 
}\label{normvarX}
\end{figure}

Lastly, in order to ensure that our data processing choices regarding clipping, binning, and joining (Figure~\ref{clipbinvar}), the quasar continuum polynomial (Figure~\ref{normvar}), and the continuum shape (Figure~\ref{normvarX})
have not impacted our conclusions, we have repeated our template fitting described in \S\ref{fitting} using the alternative normalizations in Table~\ref{normvarTBL}. We find that, as detailed below, while the continuum variations can
have some impact on the optical depth, even the most extreme plausible normalizations do not alter our conclusions that the observed structure is best represented by crystalline olivine. 

We find that neither the possible clipping, joining, and binning options, nor the variations in the shape of the adopted polynomial normalization, significantly impact our peak optical depth measurements. All 
explored clipping, joining, and binning options increased the derived optical depth by values within one-sigma, and over the 10~$\micron$ region hortonolite remains the best fit. 
The linear and power-law normalizations produce optical depth measurements which are consistent over the 10~$\micron$ range, but we discount the fitting over the full
fitting range because of their visibly poor fits at the longest and shortest wavelengths. We also discount the 5th order Chebyshev, as we believe that it has too much curvature and is over-fitting
the continuum, and thus removing some of the residual substructure. The remaining continuum normalization fits are all consistent within one-sigma of the fiducial fit, and also consistently find that hortonolite produces the best fit. 

Finally, we examine the impact of the different shape constraints on the third order polynomial. We find that the more aggressive forcing, which eliminates the emission peaks preceding and following the 10~$\micron$ absorption feature,
can raise the optical depth to at most 0.42$\pm$0.05 using the hortonolite template, or to a maximum of 0.37 using the crystalline olivine template composition which provides the best fit. The $\Delta\tau$=0.10 is nearly consistent within the one-sigma uncertainties, relative to the fiducial fit. We also note that imposing these constraints we consistently find that crystalline olivine provides the best fit, although the precise species of olivine varies. We prefer our fiducial normalization to these alternatives, because it is able
to better fit the lowest wavelength data in our spectrum, which is under-fit in these more aggressively-forced polynomials. If we employ less aggressive forcing, as explored in test `f1' (see Table~\ref{normvar}), 
the 10~$\micron$ optical depth ($\tau_{10}$) decreases slightly to 0.26, while
the fit including longer wavelengths ($\tau_{full}$) increases to 0.34; it is at this long wavelength end that the continuum normalization differences are the most significant. We believe that our adopted polynomial, however, provides a more reasonable 
fit to the overall continuum, and is less influenced by noisy points at the longest wavelengths. Lastly, using a polynomial shape which passes through the $\sim$9.6~$\micron$ substructure peak, we find that the optical depth is decreased
to $\tau_{10}$=0.20 and $\tau_{full}$=0.24, both of which are consistent within one-sigma of our fiducial fits, and again, argue for crystalline olivines as the origin of the substructure.
\section{Minor Caveats}\label{caveat}

We note the presence of two distinct spectral contaminants, one physical and one spurious, which impact our data but which we believe do not significantly alter our conclusions.

\subsection{Scattered Light Feature}\label{scatlight}
First, we have identified an anomalous emission feature in our 2-D LL spectra which we associate with scattered light. The relatively broad feature is adjacent to our quasar spectrum at approximately
$(\alpha,\delta)_{J2000}$=(18:33:36, -21:03:41.3) in each of our LL1, LL2, and LL3 spectra. The centroid of this feature is approximately 8 pixels away from the quasar spectrum at the wavelength relevant to our 10~$\micron$ absorption feature.
This prominent anomalous feature exhibits a spatially narrower distribution at low LL wavelengths, which broadens progressively as the wavelength increases; 
at the longest LL1 wavelengths it appears to be bifurcated with the light generally concentrated in two spatially distinct prongs, and a diffuse contribution in-between. 
Thus, while at the shortest LL wavelengths we do not believe this feature substantially contributes to our spectrum, at the longest wavelengths it visibly crosses through our narrow PKS 1830-211 spectrum. 

We have determined that the origin of this feature is most likely scattered light, from a bright off-field object. A visual inspection of both ground-based and HST visible and near-IR data reveals no significant objects at the spatial location
of the feature. Furthermore, we have utilized \textit{SPICE} to extract the spectrum of this feature, and compared it with a catalog of known stellar spectra observed with the Spitzer IRS \citep{IRSspectCAT}, and
with quasar spectra; the feature does not match any of these spectral shapes. The Spitzer Science Center help desk has confirmed that this feature is in all probability a very rare scattered light feature, of unidentified
origin, although the most likely candidates are either an IR bright source near the galactic plane, or a transitory object from our own solar system. Neither we, nor the help desk, could find any bright optical or IR sources in the 
immediate vicinity of our target. We have also ruled out detector features, such as latents or bad dark corrections, as the origin of this feature.

As this scattered light feature is spatially broad, and visibly crosses through the quasar spectrum at the longest wavelengths, it may be the origin of the observed discrepancies between the two LL1 nod positions in the
22-26~$\micron$ spectral region. This location corresponds to the 11.7-13.8~$\micron$ region in the rest frame of the z=0.886 absorber.
As discussed in \S\ref{spice}, a discrepancy on the order of 15-20\%, in which one nod position exhibits a broad emission-type feature while the other exhibits a broad absorption-type feature, is 
present over this spectral range. In order to identify the origin of this discrepancy, which is larger than those observed in the SL and LL2 spectral orders, we have first examined the position of the slit-centering to determine whether perhaps
an offset in centering is causing a foreground/background object to fall within the slit for one of the nod positions, but not for the other. However, the header information indicates that there are variations of at most 
($\Delta\alpha$,$\Delta\delta$)=(0.\arcsec018, 0.\arcsec009)  and (0.\arcsec020, 0.\arcsec014) within the two nod positions, and no significant offsets (within one pixel) in-between the nod positions; 
IRS exhibits pointing uncertainties on the order of 1\arcsec (which is smaller than the pixel scale). 
There is no evidence to suggest that we have a slight pointing mismatch, which could result in the inclusion of extra flux from a foreground/background object in only some of the exposures, nods, or orders.

We have also explored whether the difference between the two nod positions could stem from the background subtraction. As we were unable to subtract the background by differencing the two nod positions because of the 
anomalous scattered light feature, we have instead constructed a background generated from the combined off-source LL2 images. It is plausible that a background gradient, independent from the scattered light feature, might
be contaminating one nod background or region of the chip but not the other. In order to check this, we have utilized \textit{SPICE} to extract a background spectrum from multiple locations across our LL2 background image, 
as well as across a region of each of the LL1 spectra not covered by the quasar or the scattered light feature. 
We do not detect any substantial variations between the intrinsic backgrounds in the two nod positions, nor do we detect any significant spatial variations in the subtracted background, indicative of either a gradient or of a region of 
slightly bad pixels underlying one of the nod positions. We have also confirmed that the feature is not a result of any treatment of the background by the extraction algorithm, by replacing the pixels within the scattered light feature 
of the 2D spectrum with 
representative background pixels, and repeating the extractions; the extracted structural features remain unchanged. An additional test using the IRAF task \textit{apall} confirms that treatment of the background in our extraction algorithm is
not the origin of the derived structure. 

Additionally, we have explored whether the difference in the 22-26~$\micron$ spectral region could stem from a localized event along one of the spectra, such as an isolated cosmic ray burst or transitory event. In order to 
test this, we have extracted the combined 20-exposure spectrum, for each nod, as well as four subset spectra with 5 exposures each. We find that all of the spectral features in each nod position, 
including the discrepancy, are virtually identical between these five extractions. Therefore, it is unlikely that an isolated transitory event produced the discrepancy.

Thus, the most likely explanation for the discrepancy in the 22-26~$\micron$ region is contamination from the scattered light feature, 
which peaks in intensity over this 4~$\micron$ region, and more strongly affects one of the nod positions than the other. 
An inspection of the 2-dimensional spectrum shows that the scattered light feature is both the most narrow and has the highest surface brightness at the short wavelength end of LL1, 
where it is well-separated ($\sim$8 pixels) from the quasar
spectrum. As the wavelength increases, the breadth of this feature also increases.
However, the potential impact of this increased contamination at the longest wavelengths is mitigated by the decrease in surface brightness of the feature with increasing wavelength. 
In order to determine whether a narrower extraction window might minimize this contribution, 
we have compared a \textit{SPICE} extraction obtained using the default point-source extraction template, which is of variable spatial width becoming
broader at longer wavelengths, with two manual template extractions, one of which is 3 pixels wide, and one of which is 5 pixels wide. Modifying the extraction template impacts the absolute calibration of the extracted spectrum,
but as we are only utilizing the manual extractions to compare features within the spectrum, this is not an issue. While we do find that the continuum level is slightly different between the extractions, as expected, we find
no significant differences in the shape of the PKS 1830-211 spectrum or in any of the spectral features, along either nod position. Thus, employing a narrower extraction template would not change our analysis.
We also have no method by which to separate any possible
contamination occurring directly over the location of the quasar spectrum from the quasar spectrum itself, as this scattered light feature does not exhibit any standard spectral type. We, thus, rely on the combination of the two nod positions to effectively
cancel out the contamination pattern within this 22-26~$\micron$ region, with resultant larger uncertainties associated with the data points. 
\subsection{Contamination from Foreground Objects}\label{foreground}

The second contaminant of our quasar absorption system spectrum comes from several foreground objects which are also present within the slit. 
However, none of these objects seem to be bright enough to affect our analysis, as we describe below.
As discussed briefly in \S\ref{clean}, in addition to the two images of the lensed quasar and 
the z=0.886 quasar absorber host galaxy within the slit, a foreground star and a lower-redshift foreground galaxy are also encompassed by the IRS slit in all spectral orders. 
The brighter of the two quasar sight lines, termed the NE or A component in the literature, is located at 18h33m39.931s and -21$\degr$03$\arcmin$39.$\arcsec$75 (J2000), while the dimmer, and more dust-obscured quasar
sightline to the SW is located at $(\Delta\alpha, \Delta\delta)=(-0.\arcsec64, -0.\arcsec73)$ from the NE component\citep{Subrah90,Jin,winn}. The host galaxy for the absorber, which has been identified as an Sb/Sc spiral, is separated by 
 $(\Delta\alpha, \Delta\delta)=(-0.\arcsec33, -0.\arcsec49)$ from the NE quasar component \citep{winn}. The absorber subtends a physical scale of 7.6 kpc/$\arcsec$, assuming H$_o$=73 km~s$^{-1}$~Mpc$^{-1}$ and a standard
 $\Lambda$CDM cosmology.
 Our requested central slit position of 18h33m39.89s, -21d03m40.4s places it close to the center of this quasar absorption system. 

The foreground star which is also present within the slit, termed S1 in the literature, dominates the combined light distribution at optical wavelengths, but has a minimal impact at the mid-IR wavelengths under consideration in this analysis. 
This star, which is located at $(\Delta\alpha, \Delta\delta)=(0.\arcsec09, 0.\arcsec53)$ relative to the NE quasar component, has been identified as a likely M4 dwarf star by \citet{Courbin} and \citet{Djorgovski}. 
Star S1 dominates the total light at visible wavelengths, contributing between 54-74\% of the total light at F555W (V) and 68-83\% of the total light at F814W (I), based on the magnitudes for the star and other objects within the slit provided
by \citet{winn} and \citet{Courbin}; differences in the total contribution by S1 reflect variations in separating the light contributed by different objects by these authors. In the near-infrared, however, S1 contributes a much smaller percentage
of the total light: 32-44\% at F160W (H) and 17-23\% at F205W (K) \citep{Lehar,Courbin}. 
Extrapolating these rapidly decreasing relative contributions to the spectral region covered by the IRS, we would expect $\sim$1\% of the total flux to be contributed
by S1 at the shortest (5$\micron$) wavelengths covered, and $\lesssim$0.01\% at the longest wavelengths covered. These findings of a negligible contribution to the total flux are reconfirmed by rescaling the IRS spectra for three brighter
early M dwarf stars (HD180617, GJ687, and GJ849) provided in the compilation of IRS stellar spectra by \citet{IRSspectCAT}; we find $\approx$0.5\% contribution to the total flux near 5~$\micron$, and 0.001\% contribution
to the total flux near 35~$\micron$. In the region where the z=0.886 quasar absorber 10~$\micron$ silicate feature is located, we find only an 0.01\% contribution to the spectrum by these rescaled, similar stars from the Spectral Atlas. The 
small contribution of the M4 dwarf to our combined spectrum is a result of the rapidly decreasing strength of the stellar emission at mid-IR wavelengths, with the increasing strength of the underlying quasar continuum flux density. 
Furthermore, we have verified through an inspection of the HD180617 stellar spectrum that we do not expect any significant structural features to be produced in our spectrum from an M4 dwarf star. We, thus, exclude
the possibility that the spectral features we see in the PKS 1830-211 spectrum are produced by this foreground star.  

The second potential foreground contaminant to our quasar spectrum is the z=0.19 \citep{Lovell1996,Wiklind,Courbin,winn} galaxy, termed G2 in the literature, which also lies within the slit, 
and is separated by $(\Delta\alpha, \Delta\delta)=(0.\arcsec24, -2.\arcsec49)$ from the NE quasar component. We do not believe that this galaxy will produce a significant contribution for three reasons. 
First, it is substantially ($\Delta m_{F205W} = 3.35$ mag) fainter than the NE quasar sightline in the near-infrared, and is less reddened than the quasar \citep{Lehar}. Hence, any mid-IR emission produced by the galaxy should be negligible.
Second, any spectral features contributed by the galaxy will be well-separated spectrally from those of the 0.886 quasar absorber host galaxy. 
Third, the central regions of the foreground galaxy are not being directly illuminated by either of the quasar sight lines, and so material from the foreground galaxy should not be producing significant absorption in the quasar spectrum.
However, the closer, NE quasar line of sight passes within 10 kpc of the center of this foreground galaxy (assuming a standard $\Lambda$CDM cosmology with H$_o$=73 km~s$^{-1}$~Mpc$^{-1}$) and so may be probing the outer regions
of this z=0.19 foreground galaxy. Thus, in order to confirm that none of the prominent features which we observe coincide with silicate absorption 
in a z=0.19 galaxy, we have shifted the intermediate amorphous olivine profile (AmOliv) into the rest frame of the z=0.19 galaxy and repeated our fitting procedure.
As illustrated in Figure~\ref{FGND}, we find that while the purported 10~$\micron$ feature could be explained by 18~$\micron$ silicate absorption in the z=0.19 foreground galaxy, there is no corresponding 10~$\micron$ absorption feature
for this foreground galaxy. At the wavelength at which we would expect to detect 10~$\micron$ silicate absorption in the z=0.19 foreground galaxy there is a weak emission-type feature detected instead. We, therefore, conclude
that the structure we see in the PKS 1830-211 spectrum does not originate from a superposition of 10~$\micron$ absorption from the quasar absorber and 18~$\micron$ absorption from the z=0.19 foreground galaxy. 

The deblending of the objects in this field, based on their HST WFPC2 and NICMOS images, remains uncertain, and it is possible that there are further additional objects contained within the slit. 
\citet{Courbin}, \citet{Meylan} and \citet{Lehar} have identified a second foreground star, termed star P, located at $(\Delta\alpha, \Delta\delta)=(-0.\arcsec3, -0.\arcsec5)$ relative to the NE quasar component; \citet{winn} have instead
argued that this object is the bulge of the z=0.886 galaxy, and the current photometry is unable to distinguish between the two scenarios. 
Given that this object is 0.02-0.05 times fainter than star S1 at F205W, its emission would not have a significant impact on the derived profile should it indeed be a foreground star. 
Furthermore, \citet{Courbin} argue that the object considered to be a single z=0.886 galaxy by \citet{winn}, may in fact be comprised of two galaxies near z=0.89, one of which exhibits significant spiral features. 
The impact of two companion galaxies is addressed briefly in \S\ref{CONCL}, in which we have searched for simultaneous absorption, or simultaneous emission and absorption, at arbitrary redshifts 
in order to explain the observed spectral features.
However, this explanation does not adequately account for the spectral structure which we observe, which indicates that if two galaxies are contributing to the observed silicate absorption profile, they either have a similar chemical
composition and redshift, or one of the objects is dominating the absorption profile. 
\section{Expanded Selection of Template Fits}\label{expandedAP}

In this appendix, we have considered an expanded set of observationally-derived and laboratory optical depth templates. While this expanded analysis does not alter our fundamental conclusion that
crystalline olivine provides the most viable explanation for the structure in the 10~$\micron$ silicate absorption feature, it allows us to explore how variations in environment, temperature, and silicate
grain shape, size, and metallicity can impact our fits. We have considered this large selection of template profiles to encompass the broadest range of both the physical environments in which silicates are detected, and of the observed silicate
chemical compositions. The degree of crystallinity in these silicates varies widely amongst different astrophysical sources, as discussed in detail in \S\ref{CONCL}, ranging
from nearly 90\% for some comets to nearly 0\% for the ISM. Thus, our broad range of template profiles implicitly allows us to probe a range of crystalline:amorphous silicate ratios, 
although we do not expect that the formation mechanisms producing
the silicate absorption towards PKS 1830-211 are necessarily similar to those in, e.g., a comet. Reflecting the range of chemical compositions and grain properties producing the putative 10~$\micron$ silicate feature,
the observed profiles span a range of values for the peak  feature wavelength from 9.37~$\micron$ for a YSO embedded in the Taurus molecular cloud to 10.20~$\micron$ for a ULIRG.
The laboratory silicate optical depth profiles show an even broader range of peaks. 
We note that in our PKS 1830-211 spectrum, as discussed in the following sections, the peak wavelength of the absorption feature is longward of 9.7~$\micron$. 
Laboratory experiments have found that variations in grain morphology, 
including size, shape, porosity, agglomeration, SiO$_4$ polymerization,
and metallicity, can substantially alter both the shape of the optical depth profile, and its peak wavelength [e.g., \citet{Henning05,Henning10}].

\subsection{Expanded Subset of Observational Templates}\label{expOBS}

We first examine an expanded set of observationally-based templates for astrophysical sources ranging from comets to star-forming galaxies. Although these profiles are inherently subject to more sources
of uncertainty than the laboratory profiles, which are the product of controlled experiments, they can provide valuable insight into the astrophysical environment which might be lacking should we solely use laboratory templates. 
We list our expanded set of observational templates, separated by environment, in Table~\ref{APtemplatesOBS}, with the derived fits listed in Table~\ref{OBSfits}, and illustrated in Figure~\ref{AP-OBSFITTING}. It is
apparent both visually, and from examining the $\chi_r^2$ of the fits, that none of these objects matches the profile in our system as well as crystalline olivine. This is significant because it suggests that our system
is dissimilar from the Galactic ISM and from the ISM in several star-forming galaxies. 

The profiles which best match our system are the two comets and the ULIRG 060301-7934 (when excluding the absent 11.3~$\micron$ PAH feature). What unites these systems is that they are all crystalline-rich. 
While we do not have a precise estimate of the crystallinity for the Bradfield and Levy comets, comets, in general, are found to be up to 90\% crystalline. The ULIRG 060301-7934 has a 13\% crystallinity and provides a better fit
than either ULIRG 18443+7433 or ULIRG 00397-1312 \citep{Spoon}, both of which exhibit slightly lower levels of crystallinity. These three ULIRGS were selected as the best matches from those in the \citet{Spoon} study; the ULIRGs
with 6.5-9\% crystallinity generally provided even poorer fits. 

The Galactic and stellar environments generally provide relatively poor fits. The best fit is from the AGB star, and since AGB outflows have been found to be crystalline rich (10-15\%), this is consistent
with our need for silicate crystallinity. The diffuse ISM sources, which were adequate for reasonably fitting the silicate absorption features in the \citet{Kulkarni1,Kulkarni2} quasar absorbers, are a poor fit here,
being shallower and offset to lower wavelengths relative to our feature. The circumstellar environments, dense molecular clouds, and stars embedded within molecular clouds all suffer from this same general failing.
Some of the molecular clouds, such as $\rho$ Oph, 
characterized as a disturbed environment \citep{Bowey03}, and the visually similar (not depicted) Serpens profile, produce significantly inferior fits 
when compared with, e.g., profiles for sources probing the the Taurus molecular cloud.
Interestingly, profiles for sources embedded within the dense molecular clouds tend to produce slightly better fits than those illuminating the dark molecular clouds, which may stem from the fact that
young sources surrounded by circumstellar material can be rich in crystalline silicates [e.g., see review in \citet{Molster}]. 

\subsection{Variations in Laboratory Amorphous Olivine Templates}\label{amOLIVtemps}
We also explore, in the context of the amorphous olivine silicate profiles, the range in derived optical depths which can occur as a result of variations in the shapes and sizes of the dust particle grains (Table~\ref{APtemplatesLAB}), using
profiles from a variety of literature sources, as noted in that table. 
The reviews by \citet{Henning05} and \citet{Molster} emphasize that variations in grain size, shape, and metallicity can have an impact on the shape and peak-wavelength of the silicate profiles. This is well-illustrated
 in Figures 6 and 7 of \citet{Chiar06} for variations in the dust particles' shapes, including solid versus porous particles, and solid glass versus a continuous distribution of ellipsoids.
These differences are most pronounced in the 18~$\micron$ region. 

We first explore the impact of the variations in grain shape by fitting three separate amorphous olivine profiles to our PKS 1830-211 absorption features: one which is
typical of solid spheres, one which is typical of a porous continuous distribution of ellipsoids, and one which is intermediate between these two extremes.
Figure~\ref{AP-LABAMFITTING} illustrates the visible differences in the fits which can result from these variations. 
As detailed in Table~\ref{LABfits}, the best fit results from the profile representing the continuous distribution of ellipsoids with a porous structure, although none of the profiles adequately fit our substructure, as
discussed elsewhere, and all have relatively poor $\chi_r^2$ values. Furthermore, as detailed in Table~\ref{LABfits}, while these variations impact the quality of the fit, 
they have relatively little effect on the derived optical depth which only varies from 0.11$\leq\tau_{10}\leq$0.13. 

We next consider the impact of variations in grain size, considering three similar particle distributions with particle sizes of 2~$\micron$, 1.5~$\micron$, and those in the Rayleigh limit, all taken from \citet{Dorschner95}. 
We find that these size variations can have an even more significant impact both visually (see Figure~\ref{AP-LABAMFITTING}) and in terms of the $\chi_r^2$ (Table~\ref{LABfits}), compared with the explored particle
shape variations. The best fit over the 10~$\micron$ range is produced by the 2.0~$\micron$ particles, while over the full fitting range the 1.5~$\micron$ particles provide a better fit. However, despite the range in relative quality of 
of the fits, all poor in comparison with our crystalline olivine fits, the optical depths are consistently in the 0.11-0.13 range. 

\subsection{Variations in Laboratory Crystalline Olivine Templates}\label{cryOLIVtemps}
Given our results in \S\ref{ANALYSIS} that the PKS 1830-211 z=0.886 absorber silicate profile is best represented by crystalline olivine, we explore in this section the effects of varying the chemical composition (i.e. the Mg/Fe ratio), particle shape,
and material temperature on the derived fits. To implement these tests we examined a total of 19 different varieties of crystalline olivines, as detailed in Table~\ref{APtemplatesLAB}. The derived fits using these profiles are illustrated
in Figure~\ref{AP-LABCRYFITTING}, where it is evident that both particle shape and size variations can significantly alter the fit; temperature has a reduced effect over this region of the spectrum, but would play a larger role at longer
wavelengths. 

We first consider the variations in composition by exploring six distinct olivine compositions, as well as two different hortonolite profiles, since there can be variations between different laboratory samples.
These profiles include both naturally occurring (terrestrially found) compounds, as well as more synthetic blends, as described in the references given in Table~\ref{APtemplatesLAB}. As discussed in \S\ref{CRYSoliv}, we see that
as the Mg/Fe ratio is varied, the leftmost ($\sim$11~$\micron$) and central ($\sim$10~$\micron$) substructure features are increasingly well or poorly fit, and despite exploring 7 olivine varieties, we do not find any that perfectly
fit our data. However, the wide range of fits which we do find, lead us to believe that a ratio of Mg/Fe exists that will fully, simultaneously fit these two features. The best fits are obtained using the two hortonolite profiles,
an intermediate Mg/Fe ratio, with the profile from \citet{Jaeger} providing slightly superior results. The next best fit is provided by forsterite, the Fe-free end of the olivine compositional spectrum. As with the amorphous olivine, 
we find similar fits if we fit over solely the 10~$\micron$ range, or if we expand fitting to include the full range. The best fit produces an optical depth of $\tau_{10}$=0.27$\pm$0.05, and the full set of profiles ranges from 
0.23$\leq\tau_{10}\leq$0.28, well-within the one-sigma limits. Since these templates span the full range of Mg/Fe ratios, we anticipate that the true Mg/Fe ratio will likewise produce
an optical depth within this range.

We next explore the effects of variations in particle shape using four chemically consistent compounds (Mg$_{1.9}$Fe$_{0.1}$SiO$_4$) close to forsterite, from \citet{Fabian01}. These variations, which include considering
a naturally occurring particle distribution, as well as spherical particles, and two variations of a continuous distribution of ellipsoidal (CDE) particles produce even more visually and statistically significant variations in the quality
of the fit than the chemical variations discussed above. The best fit is produced from the powdered olivine, and it is apparent that the CDE distributions are not good fits. However, again despite significant variations in fit quality, the derived
optical depth measurements are self-consistent within one-sigma across both shape variations and fitting regions. 

Lastly, we consider the effects of variations in temperature. As addressed, in detail, by \citet{Koike06}, variations in temperature can induce visible shifts in the profile features, particularly at long wavelengths. To investigate
whether these can impact our fits, we have considered both fayalite (the pure Fe olivine) and forsterite (the pure Mg olivine) over a range of temperatures from 10-300K. The variations for forsterite are minimal in terms of optical depth, 
although $\Delta\chi_{r,10}^2$=0.9. For fayalite, $\Delta\tau_{10,full}$=0.02, which is consistent within one-sigma, and the $\Delta\chi_{r,10}^2$=0.24, which is somewhat smaller. From these tests, we conclude that particle shape, size,
and composition may significantly impact the quality ($\chi_r^2$) of our derived fits, albeit not the optical depth, while in this regime temperature variations are largely insignificant. 

\subsection{Expanded Subset of Laboratory Templates for  Other Minerals}\label{expLAB}
Lastly, we explore an expanded set of laboratory-derived non-olivine silicate profiles including 3 amorphous silicate minerals, 5 crystalline pyroxenes, 4 phyllosilicates, 3 crystalline silicate blends, and 4 types of SiC (Table~\ref{APtemplatesLAB},
with included references for template origins).
These substances span a wide range of chemical compositions, some of which have been proposed as potentially significant contributors to Galactic dust sources, such as the phyllosilicates \citep{Dorschner78}, and others of which are
known to be prominent in specific environments, such as SiC which is found in meteorites and carbon stars. 
This large variety of compounds are illustrated in Figure~\ref{AP-LABFITTING}. However, we find that none of the materials fit the silicate absorption feature as well as virtually all of the crystalline olivine silicates, although as discussed in \S\ref{ANALYSIS},
some of the materials such as silica, synthetic enstatite, serpentine, and SiC can fit a subset of the structural features. 
In conclusion, while in principle a combination of several of these minerals might be constructed to produce a plausible fit to the substructure, and certainly several of them are candidates to explain the $\sim$9~$\micron$ feature
in combination with crystalline olivine, they do not provide a compelling alternative, by themselves, to crystalline olivine when explaining the origin of the detected absorption feature.

\begin{deluxetable}{lrrll}
\tabletypesize{\scriptsize}
\tablecaption{PKS 1830-211 Spitzer IRS Observations \label{obs}}
\tablewidth{0pt}
\tablehead{
\colhead{Order} & \colhead{$\lambda$-Coverage} & \colhead{t$_{exp}$(s)}& \colhead{N$_{exp}$} & \colhead{t$_{obs} (m)$}}
\startdata
SL2 & 5.13-7.60 & 60 & 19 & 38 \\
SL1 & 7.46-14.29 & 60 & 20 & 40\\
LL2 & 13.90-21.27 & 120 & 10 &40 \\
LL1 & 19.91-39.90 & 120 & 20 & 80\\
\enddata
\tablecomments{Summary of Spitzer IRS observations. In the first column we list the spectral order, followed by the wavelength coverage (in $\micron$), the exposure time per frame (in seconds), 
the number of exposures, and the total on-target exposure (in minutes) for each nod position in the given spectral order. This total time is double $t_{exp}N_{exp}$, because of the two nod positions.}
\end{deluxetable}
\begin{deluxetable}{lcrll}
\tabletypesize{\scriptsize}
\tablecaption{Template Summary: Observational Profiles \label{templatesOBS}}
\tablewidth{0pt}
\tablehead{
\colhead{Profile}  & \colhead{$\lambda_{peak}$} & \colhead{$\lambda_{range}$} & \colhead{Ref} & \colhead{Description}}
\startdata
GCS3 & 9.60 & 7.52-34.94 & 1 & Galactic center source GCS-3 (Galactic diffuse ISM) \\
Trap & 9.53 & 8.06-12.83 & 2 & Trapezium region of Orion nebula (dense molecular cloud) \\
IC5146 & 9.78 & 5.54-13.97 & 3 & Q21-6 G8.5 IIIa Fe-0.5 field star behind IC5146 (quiescent dense cloud complex) \\
$\mu$Cep  & 9.69 & 8.00-13.50 & 4 & $\mu$ Cephei red supergiant (stellar env.) \\
AGB & 10.08 & 7.04-29.96 & 5 & model profile for O-rich AGB star OH/IR 127.8+0.0 (stellar env.) \\
UL06301 & 10.20 & 8.16-33.55 & 6 & ULIRG 06301-7934 (extragalactic) \\
UL06301t & 10.20 & 8.16-33.55 & 6 & ULIRG 06301-7934, masking 11.3 PAH complex (extragalactic) \\
\enddata
\tablecomments{
Summary of  optical depth profile templates extracted from the literature for observed, astrophysical, sources. In the first column we list our abbreviated name, followed by 
the peak wavelength for the 10~$\micron$ silicate feature, the range of wavelengths spanned by the profile, 
the literature reference, and finally a description of the profile. All wavelengths are specified in~$\micron$. For broad profiles, with significant sub-structure, 
the wavelength at which the deepest optical depth is achieved is listed as the peak. 
REFS: (1) \citet{Spoon} based on data from \citet{Chiar06}; (2) \citet{Bowey01} based on data from \citet{Forrest}; (3) \citet{Chiar11}; (4) \citet{mucep};
(5) \citet{Chiar06} based on data from \citet{Kemper02}; \& (6) \citet{Spoon}.}
\end{deluxetable}
\begin{deluxetable}{lcrll}
\tabletypesize{\scriptsize}
\tablecaption{ Template Summary: Laboratory Profiles \label{templatesLAB}}
\tablewidth{0pt}
\tablehead{
\colhead{Profile} & \colhead{$\lambda_{peak}$} & \colhead{$\lambda_{range}$} & \colhead{Ref} & \colhead{Description}}
\startdata
\cutinhead{AMORPHOUS}
AmOliv & 9.76 & 7.28-33.87 & 1 & amorphous olivine \\
AmPyr & 9.16 & 7.26-33.81 & 1 & amorphous pyroxene \\
\cutinhead{CRYSTALLINE OLIVINE - Mg$_{2x}$Fe$_{2-2x}$SiO$_4$}
Fayalite & 11.43 & 2.00-199.44 & 2 & synthetic fayalite (x=0.0) \\
Forst & 11.21 & 1.67-199.44 & 2 & synthetic forsterite (x=1.0)  \\
Horton & 11.30 & 1.67-669.09 & 2 & natural hortonolite (x=0.55) \\
Olivine & 11.21 & 2.00-199.44 & 2 & natural olivine (x=0.94)  \\
\cutinhead{OTHER SILICATE MATERIALS}
AvgSerp. & 10.31 & 7.80-13.28 & 3 & avg. serpentine (amalgam antigorite, chrysotile, serpentine) \\
AvgSil. & 9.19 & 7.79-13.31 & 3 & avg. silica (amalgam quartz, fumed silica, hydrated silica, opal, micronized am. silica) \\
Bronzite & 10.49 & 8.34-505.90 & 2 & natural orthobronzite (pyroxene) \\
Enstat-syn & 9.27 & 1.67-199.44 & 2 & synthetic clinoenstatite (pyroxene) \\
G0-green-aSiC & 11.36 & 6.67-31.25 & 4 & green $\alpha$-SiC (G0 sample)\\
\enddata
\tablecomments{
As in Table~\ref{templatesOBS}, but listing optical depth profiles for laboratory sources. For profiles with multiple peaks over the spanned region, the deepest peak is adopted as $\lambda_{peak}$. 
REFS: (1) \citet{Spoon} based on data from \citet{Fabian01}; (2) \citet{Jaeger} with tabulated data provided by C. J\"{a}ger; (3) \citet{Bowey02} based on data from \citet{Ferraro};  \& (4) \citet{Friedemann}}
\end{deluxetable}
\begin{deluxetable}{lrrrrrr}
\tabletypesize{\scriptsize}
\tablecaption{Optical Depth Fits for PKS 1830-211 z=0.886 Absorber \label{fits}}
\tablewidth{0pt}
\tablehead{
\colhead{Profile} & \colhead{$\tau_{10}$} & \colhead{$\chi_{r,10}^2$} & \colhead{Fit-$\lambda_{range}$}& \colhead{$\tau_{full}$} & \colhead{$\chi_{r,full}^2$} & \colhead{Fit-$\lambda_{range}$}}
\startdata
 \cutinhead{OBSERVATIONALLY DERIVED TEMPLATES} 
GCS3 & 0.12$\pm$0.03 & 15.59 & 8.60-12.50 & 0.13$\pm$0.04 & 10.46 & 8.00-19.45 \\
Trap & 0.11$\pm$0.02 & 13.30 & 8.60-12.50 & 0.10$\pm$0.02 & 14.30 & 8.06-12.83 \\
IC5146 & 0.12$^{+0.03}_{-0.02}$ & 13.09 & 8.60-12.50 & 0.11$\pm$0.03 & 11.32 & 8.00-13.97 \\
$\mu$Cep & 0.12$\pm$0.03 & 14.07 & 8.60-12.50 & 0.12$\pm$0.03 & 12.75 & 8.00-13.25 \\
AGB & 0.12$^{+0.03}_{-0.02}$ & 10.62 & 8.60-12.50 & 0.12$\pm$0.03 & 7.94 & 8.00-19.45 \\
UL06301 & 0.13$\pm$0.03 & 10.27 & 8.60-12.50 & 0.13$\pm$0.03 & 7.57 & 8.16-19.45 \\
UL06301t & 0.13$\pm$0.03 & 9.42 & 8.60-12.50 & 0.13$\pm$0.03 & 7.20 & 8.16-19.45 \\
\cutinhead{AMORPHOUS SILICATE TEMPLATES (LABORATORY)}
AmOliv & 0.11$^{+0.03}_{-0.02}$ & 14.79 & 8.60-12.50 & 0.11$\pm$0.03 & 9.60 & 8.00-19.45 \\
AmPyr & 0.10$\pm$0.03 & 21.60 & 8.60-12.50 & 0.10$^{+0.04}_{-0.03}$ & 14.19 & 8.00-19.45 \\
\cutinhead{CRYSTALLINE OLIVINE SILICATE TEMPLATES (LABORATORY)}
 Fayalite & 0.23$\pm$0.04 & 6.33 & 8.60-12.50 & 0.20$\pm$0.05 & 6.74 & 8.00-19.45 \\
Forst & 0.27$\pm$0.05 & 5.09 & 8.60-12.50 & 0.25$\pm$0.07 & 6.04 & 8.00-19.45 \\
Horton & 0.27$\pm$0.05 & 3.73 & 8.60-12.50 & 0.26$^{+0.07}_{-0.06}$ & 4.78 & 8.00-19.45 \\
Olivine & 0.28$^{+0.06}_{-0.05}$ & 7.32 & 8.60-12.50 & 0.26$\pm$0.07 & 6.52 & 8.00-19.45 \\
\cutinhead{OTHER SILICATE TEMPLATES (LABORATORY)}
AvgSerp. & 0.15$\pm$0.03 & 14.82 & 8.60-12.50 & 0.14$\pm$0.04 & 13.23 & 8.00-13.28 \\
AvgSil. & 0.10$\pm$0.03 & 27.66 & 8.60-12.50 & 0.09$\pm$0.04 & 23.97 & 8.00-13.31 \\
Bronzite & 0.13$\pm$0.03 & 13.55 & 8.60-12.50 & 0.13$\pm$0.04 & 9.38 & 8.34-19.45 \\
Enstat-syn & 0.14$\pm$0.03 & 13.39 & 8.60-12.50 & 0.14$\pm$0.04 & 8.92 & 8.00-19.45 \\
G0-green-aSiC & 0.28$\pm$0.06 & 14.30 & 8.60-12.50 & 0.28$\pm$0.09 & 10.83 & 8.00-19.45 \\
\enddata
\tablecomments{
Best fits to PKS 1830-211 absorption spectrum using our optical depth template profiles. In the first column we list the profile name, followed by the peak optical depth
normalization factor (a$_n$, termed $\tau$ here and throughout the paper), the reduced chi-squared, and the wavelength range over which the fit was performed. All wavelengths are in~$\micron$.
Columns 2-4 correspond to fitting performed solely over the (z=0.886 absorber rest-frame) 10~$\micron$ feature (8.6-12.5~$\micron$). Columns 5-7 correspond to fits extended 
to cover both the 10~$\micron$ and 18~$\micron$ features (spanning 8.0-19.45~$\micron$), if adequate profile data are available; otherwise a more limited fitting range is utilized, based on limits derived from the template profile.
The best fit is produced by hortonolite (crystalline olivine), both when considering the narrower fitting range and when fitting over the full profile. All of the crystalline olivine templates fit better than the alternative laboratory and
observationally derived profiles.}
\end{deluxetable}
\begin{deluxetable}{llrrrr}
\tabletypesize{\scriptsize}
\tablecaption{Two-Mineral Fitting of the 10~$\micron$ Feature \label{multivarTAB}}
\tablewidth{0pt}
\tablehead{
\colhead{Profile1+Profile2} &\colhead{$\tau_1$} & \colhead{$\tau_2$} & \colhead{$\chi_r^2$} & \colhead{Fit-$\lambda_{range}$} & \colhead{$\tau_2/\tau_{1}$} }
\startdata
\cutinhead{HORTONOLITE + SECOND SILICATE}
HORTON+SILICA & 0.25$\pm$0.05 & 0.02$^{+0.03}_{-0.02}$\tablenotemark{a} & 3.37 & 8.60-12.50 & 0.09 \\
HORTON+AVG.SERP. & 0.26$\pm$0.05 & 0.01$^{+0.03}_{-0.01}$\tablenotemark{a} & 3.72 & 8.60-12.50 & 0.02 \\
HORTON+ENSTATITE & 0.24$\pm$0.05 & 0.02$^{+0.03}_{-0.02}$\tablenotemark{a} & 3.44 & 8.60-12.50 & 0.10 \\
HORTON+SiC\tablenotemark{b} & 0.27$\pm$0.05 & 0.01$^{+0.06}_{-0.01}$\tablenotemark{a} & 3.74 & 8.60-12.50 & $\leq$0.02 \\
\cutinhead{AMORPHOUS OLIVINE + SECOND SILICATE}
HORTON+AM.OLIVINE\tablenotemark{b} & 0.26$\pm$0.05 & 0.01$^{+0.03}_{-0.01}$\tablenotemark{a} & 3.75 & 8.60-12.50 & $\leq$0.02 \\
AM.OLIVINE+SiC & 0.07$\pm$0.02 & 0.19$\pm$0.06 & 6.74 & 8.60-12.50 & 2.49 \\
\enddata
\tablenotetext{a}{1-$\sigma$ limit formally extends below 0.0; lower limit is considered to be effectively 0.}
\tablenotetext{b}{$\tau_2$ value is at the lowest boundary explored in the calculation, $\tau_2=0.005$, and is consistent with zero. We, thus, discount this fit.}
\tablecomments{ 
Fits resulting from bi-variate profile combinations, i.e., $\exp[{-(a_{n}\tau_{1,norm}+b_{n}\tau_{2,norm})]}$. 
In the first column, we list the two combined minerals, followed by the peak optical depth for the first and second minerals, the reduced chi-squared of the fit, the wavelength range over which the fit was performed,
and the ratio of the derived optical depths for mineral 2 and mineral 1; the nomenclature is as detailed in Table~\ref{fits}. All wavelengths are in~$\micron$.
Hortonolite, which produced the lowest $\chi_r^2$ in Table~\ref{fits} is used for the first profile in every combination but the last. The best fit is the hortonolite + silica ($\chi_r^2$=3.37) although
it is only a marginal improvement over the pure hortonolite fit ($\chi_r^2$=3.73).}
\end{deluxetable}
\begin{deluxetable}{lrrrrrrrr}
\tabletypesize{\scriptsize}
\tablecaption{Amorphous Olivine Profiles with PKS 1830-211 Masking}
\tablewidth{0pt}
\tablehead{
\colhead{Profile} & \colhead{$\tau_{10}$} & \colhead{$\chi_{r,10}^2$} & \colhead{Fit-$\lambda_{range}$}& \colhead{$\Delta\lambda$}& \colhead{$\tau_{full}$} & \colhead{$\chi_{r,full}^2$} & \colhead{Fit-$\lambda_{range}$}& \colhead{$\Delta\lambda_{shift}$}}
\startdata
\cutinhead{LABORATORY - AMORPHOUS - AmOliv}      
fiducial & 0.11$^{+0.03}_{-0.02}$ & 14.79 & 8.60-12.50 & \nodata & 0.11$\pm$0.03 & 9.60 & 8.00-19.45 & \nodata \\
MASKA & 0.16$\pm$0.03 & 12.79 & 8.60-12.50 & \nodata & 0.14$\pm$0.04 & 8.36 & 8.00-19.45 & \nodata \\
MASKB & 0.14$\pm$0.03 & 6.74 & 8.60-12.50 & \nodata & 0.13$\pm$0.04 & 5.97 & 8.00-19.45 & \nodata \\
SHIFTMASKA & 0.24$\pm$0.04 & 2.56 & 8.60-12.50 & 0.55 & 0.19$\pm$0.05 & 5.00 & 8.00-19.45 & 0.55 \\
SHIFTMASKB & 0.21$\pm$0.04 & 1.41 & 8.60-12.50 & 0.48 & 0.16$\pm$0.05 & 4.10 & 8.00-19.45 & 0.45 \\
\cutinhead{LABORATORY - AMORPHOUS - AmOlivGPC}     
fiducial & 0.13$\pm$0.03 & 10.46 & 8.60-12.50 & \nodata & 0.13$\pm$0.03 & 7.19 & 8.02-19.45 & \nodata \\
MASKA & 0.20$\pm$0.03 & 4.55 & 8.60-12.50 & \nodata & 0.16$\pm$0.04 & 5.53 & 8.02-19.45 & \nodata \\
MASKB & 0.17$\pm$0.03 & 2.39 & 8.60-12.50 & \nodata & 0.15$\pm$0.04 & 4.39 & 8.02-19.45 & \nodata \\
SHIFTMASKA & 0.22$\pm$0.04 & 2.78 & 8.60-12.50 & 0.23 & 0.18$\pm$0.04 & 4.64 & 8.27-19.45 & 0.25 \\
SHIFTMASKB & 0.19$\pm$0.04 & 1.83 & 8.60-12.50 & 0.15 & 0.16$^{+0.05}_{-0.04}$ & 3.98 & 8.27-19.45 & 0.25 \\
\cutinhead{LABORATORY - AMORPHOUS - AmOlivGS}     
fiducial & 0.13$\pm$0.03 & 14.29 & 8.60-12.50 & \nodata & 0.13$\pm$0.04 & 8.84 & 8.12-19.45 & \nodata \\
MASKA & 0.20$\pm$0.04 & 8.85 & 8.60-12.50 & \nodata & 0.17$\pm$0.05 & 6.46 & 8.12-19.45 & \nodata \\
MASKB & 0.17$\pm$0.03 & 4.07 & 8.60-12.50 & \nodata & 0.16$^{+0.05}_{-0.04}$ & 4.47 & 8.12-19.45 & \nodata \\
SHIFTMASKA & 0.27$^{+0.05}_{-0.04}$ & 2.77 & 8.60-12.50 & 0.38 & 0.20$\pm$0.05 & 4.50 & 8.49-19.45 & 0.38 \\
SHIFTMASKB & 0.23$\pm$0.04 & 1.27 & 8.60-12.50 & 0.28 & 0.18$\pm$0.05 & 3.45 & 8.34-19.45 & 0.23 \\
\cutinhead{LABORATORY - AMORPHOUS - AMOLIV1.5}     
fiducial & 0.13$^{+0.03}_{-0.02}$ & 9.85 & 8.60-12.50 & \nodata & 0.12$\pm$0.03 & 6.81 & 8.36-19.45 & \nodata \\
MASKA & 0.18$\pm$0.03 & 5.03 & 8.60-12.50 & \nodata & 0.15$\pm$0.04 & 5.38 & 8.36-19.45 & \nodata \\
MASKB & 0.16$\pm$0.03 & 3.15 & 8.60-12.50 & \nodata & 0.13$\pm$0.04 & 4.23 & 8.36-19.45 & \nodata \\
SHIFTMASKA & 0.20$\pm$0.03 & 4.00 & 8.60-12.50 & 0.20 & 0.16$\pm$0.04 & 4.89 & 8.61-19.45 & 0.25 \\
SHIFTMASKB & 0.17$\pm$0.03 & 2.94 & 8.60-12.50 & 0.10 & 0.14$\pm$0.04 & 4.05 & 8.50-19.45 & 0.15 \\
\enddata
\tablecomments{
Optical depth profile fits for four amorphous olivine templates (for template details, see Tables~\ref{templatesLAB} and ~\ref{APtemplatesLAB}), listing the fiducial fits (from Table~\ref{fits}), 
as well as the fits derived when employing masking of presumed absorption/emission
features in the PKS 1830-211 spectrum, and simultaneous feature masking and wavelength-shifting of the template profile, in order to obtain the best fit. Two broad fitting regions are tabulated:
the 10~$\micron$ region (8.6-12.5~$\micron$) and the full fitting region (8.0-19.45~$\micron$), with the fitting region adjusted as appropriate to ensure template coverage. The nomenclature is 
as in Table~\ref{fits}, with the addition of the requisite shift applied in the fifth and last columns for the shift+mask fits, and all wavelengths in~$\micron$. 
The masks are applied to cover the following spectral regions: A (9.3-10.05~$\micron$; 10.35-10.85~$\micron$) and  B (9.3-10.05~$\micron$; 10.35-11.3~$\micron$). 
The formally best fits result when using mask B and wavelength-shifting of the templates, but these fits are also weakly constrained in the 10~$\micron$ region and 
assume that plausible physical mechanisms are present which can explain the requisite emission/absorption features. The lowest overall $\chi_r^2$ is obtained with the AmOlivGS template, employing
a shift of 0.23-0.28~$\micron$, and maximal masking.} \label{maskshiftT}
\end{deluxetable}
\clearpage
\begin{deluxetable}{llllllllllll}
\tabletypesize{\scriptsize}
\rotate
\tablecaption{PKS 1830-211 Spectral Normalization/Processing Variations \label{normvarTBL}}
\tablewidth{0pt}
\tablehead{
\colhead{Name} & \colhead{Type} &  \colhead{Order} & \colhead{Norm.} & \colhead{H $\tau_{10}$} & \colhead{H $\chi_{r,10}^2$} & 
\colhead{B(10~$\micron$)} & \colhead{B $\tau_{10}$} & \colhead{B $\chi_{r,10}^2$} & 
\colhead{B(full)} & \colhead{B $\tau_{full}$} & \colhead{B $\chi_{r,full}^2$}}
\startdata
\cutinhead{Adopted Normalization}
Fiducial &  Chebyshev poly.  & 3 & A & 0.27$\pm$0.05 & 3.73 & Horton & 0.27$\pm$0.05 & 3.73 & Horton & 0.26$^{+0.07}_{-0.06}$ & 4.78 \\
\cutinhead{Clipping Variations}
Clip v2   &  Chebyshev poly.  & 3 & A & 0.33$\pm$0.05 & 4.58 & Horton & 0.33$\pm$0.05 & 4.58 & Horton & 0.32$\pm$0.07 & 4.45 \\  
Clip v3  &  Chebyshev poly.  & 3 & A  & 0.30$^{+0.06}_{-0.05}$ & 4.71 & Horton & 0.30$^{+0.06}_{-0.05}$ & 4.71 & Horton & 0.29$\pm$0.07 & 4.90 \\                               
Clip v4   &  Chebyshev poly.  & 3 & A & 0.32$\pm$0.06 & 5.25 & Horton & 0.32$\pm$0.06 & 5.25 & Horton & 0.30$\pm$0.07 & 4.94 \\  
\cutinhead{Binning Variations}
Binned  &  Chebyshev poly.  & 3 & A & 0.29$\pm$0.05 & 2.48 & Horton & 0.29$\pm$0.05 & 2.48 & Forst & 0.27$\pm$0.07 & 5.38 \\
\cutinhead{Normalization Polynomial Shape Variations}
Chebyshev4   &  Chebyshev poly.  & 4 & A & 0.32$\pm$0.05 & 3.70 & Horton & 0.32$\pm$0.05 & 3.70 & Horton & 0.31$\pm$0.07 & 4.66 \\ 
Chebyshev5  &  Chebyshev poly.  & 5 & A & 0.46$^{+0.07}_{-0.05}$ & 8.06 & Fayalite & 0.40$\pm$0.04 & 6.31 & Horton & 0.41$\pm$0.07 & 6.98 \\
Spline3  &  cubic-spline  &  \nodata  & A & 0.31$\pm$0.05 & 3.64 & Horton & 0.31$\pm$0.05 & 3.64 & Horton & 0.31$\pm$0.07 & 4.73 \\ 
Legendre3   &  Legendre poly.  & 3 & A & 0.31$\pm$0.05 & 4.01 & Horton & 0.31$\pm$0.05 & 4.01 & Horton & 0.30$\pm$0.07 & 4.44 \\                                         
Linear  &  linear  & 2 & \nodata & 0.26$\pm$0.05 & 4.37 & HortonB & 0.25$\pm$0.05 & 3.86 & FayalT300 & 0.51$^{a}_{-0.06}$ & 20.97 \\ 
Power Law  &  power-law  &  \nodata  & \nodata & 0.29$\pm$0.05 & 5.36 & HortonB & 0.29$\pm$0.05 & 4.00 & AmOliv2.0 & 0.28$\pm$0.03 & 42.20 \\   
\cutinhead{Third-Order Chebyshev Normalization Shape Variations}
Chebyshev3 (f1)  &  Chebyshev poly.  & 3 & \nodata & 0.26$\pm$0.05 & 3.65 & Horton & 0.26$\pm$0.05 & 3.65 & FayalT300 & 0.34$\pm$0.08 & 5.65 \\                                    
Chebyshev3 (f2)  &  Chebyshev poly.  & 3 & A,B,C,D & 0.42$\pm$0.05 & 7.76 & Fayalite & 0.37$\pm$0.04 & 6.22 & Horton & 0.40$\pm$0.07 & 5.70 \\ 
Chebyshev3 (f3)  &  Chebyshev poly.  & 3 & C,D  & 0.37$\pm$0.05 & 5.22 & Horton & 0.37$\pm$0.05 & 5.22 & Horton & 0.37$\pm$0.07 & 4.59 \\      
Chebyshev3 (f4)  &  Chebyshev poly.  & 3 & B,C,D  & 0.42$\pm$0.05 & 7.19 & Fayalite & 0.37$\pm$0.04 & 5.91 & Fayalite & 0.35$^{+0.06}_{-0.05}$ & 5.13 \\ 
Chebyshev3 (f5)\tablenotemark{b}  &  Chebyshev poly.  & 3 & A,E& 0.20$\pm$0.05 & 4.38 & Horton & 0.20$\pm$0.05 & 4.38 & FayalT10 & 0.24$\pm$0.08 & 5.64 \\   
\enddata
\tablecomments{Comparison of the explored binning, clipping, and normalization variations applied to the PKS 1830-211 spectrum. 
We list in col. 1 the fitting variation name, in col. 2 the shape of the fit applied, in col. 3 the order of the fitted polynomial, in col. 4 the code for the regions through which the normalization was required to pass (see below),
in cols. 5-6 we list the peak optical depth and reduced chi-squared using the given normalization and the hortonolite template, in cols. 7-9 we list the best fitting optical depth template over the 10~$\micron$ region and the
associated peak optical depth and reduced chi-squared, and in cols. 10-12 we list the same for the expanded fitting range. 
When fitting the quasar continuum profile, the standard applied masking excluded the 10~$\micron$ silicate absorption feature, the 18~$\micron$ silicate absorption feature, the $\sim$6.0~$\micron$ H$_2$O feature, and the
$\sim$6.8~$\micron$ feature, as they appear in the z=0.886 absorber rest-frame. 
For the primary analysis in this paper the adopted combination consists of un-binned, maximally-clipped data with a third-order Chebyshev normalization 
with fitting forced to pass (see note-A below) through data points near 19~$\micron$, and standard masking.
The quasar continuum fit was required to pass through specific regions in the z=0.886 absorber rest-frame, denoted in col. 4 as follows:
(A) points at the longest wavelengths ($\gtrsim$ 18.5~$\micron$; 
(B) points between the 10 and 18~$\micron$ absorption features;
(C) points just shortward of the 10~$\micron$ absorption feature;
(D) points just shortward of the 6~$\micron$ H$_2$O absorption feature; \&
(E) points in the sub-structure peak near 9.6~$\micron$.}
\tablenotetext{a}{Unconstrained within plausible $\tau$ values explored.}
\tablenotetext{b}{The $\sim$6.0 and 6.8~$\micron$ absorption features were not explicitly masked in the quasar-continuum fit.}
\end{deluxetable}
\begin{deluxetable}{lcrll}
\tabletypesize{\scriptsize}
\tablecaption{Expanded Template Summary: Observational Profiles \label{APtemplatesOBS}}
\tablewidth{0pt}
\tablehead{
\colhead{Profile}  & \colhead{$\lambda_{peak}$} & \colhead{$\lambda_{range}$} & \colhead{Ref} & \colhead{Description}}
\startdata
 \cutinhead{EXTRAGALACTIC SOURCES: ULIRGS} 
UL00397\tablenotemark{a}& 9.80 & 7.85-30.19 & 1 & ULIRG 00397-1312 \\
UL06301\tablenotemark{a} & 10.20 & 8.16-33.55 & 1 & ULIRG 06301-7934 \\
U18443\tablenotemark{a} & 9.97 & 8.38-34.17 & 1 & ULIRG 18443+7433 \\
\cutinhead{GALACTIC DIFFUSE ISM}
GCS3 & 9.60 & 7.52-34.94 & 2 & Galactic center source GCS-3  \\
WR118 & 9.81 & 7.10-27.61 & 3 & diffuse ISM illum by WC Wolf-Rayet star WR118 \\
\cutinhead{DENSE MOLECULAR CLOUD/CLOUD COMPLEX}
Trap & 9.53 & 8.06-12.83 & 4 & Trapezium region of Orion nebula  \\
Tau16 & 9.55 & 8.07-12.82 & 5 & Taurus molecular cloud illum. by field star Elias 16  \\
IC5146 & 9.78 & 5.54-13.97 & 6 & Q21-6 G8.5 IIIa Fe-0.5 field star behind IC5146  \\
$\rho$Oph & 9.81 & 7.51-12.29 & 7 & $\rho$ Ophiuci illum. by SSTc2d\_J163346.2-242753 \\
\cutinhead{T-TAURI STAR EMBEDDED IN MOLECULAR CLOUD}
emb-Tau & 9.37 & 8.08-12.83 & 8 & T-Tauri star Elias 7N embedded in Taurus mol. cloud complex \\
emb-$\rho$Oph & 9.47 & 8.35-13.29 & 9 & T-Tauri stsr Elias 29 embedded in disrupted molecular cloud $\rho$Oph \\
\cutinhead{STELLAR MATERIAL}
$\mu$Cep & 9.69 & 8.00-13.50 & 10 & $\mu$ Cephei red supergiant \\
AGB & 10.08 & 7.04-29.96 & 11 & model profile for O-rich AGB star OH/IR 127.8+0.0  \\
\cutinhead{COMETS}
Bradfield & 10.09 & 7.67-13.04 & 12 & Bradfield 1987 XXIX comet (r=0.99AU), norm. by 335K blackbody continuum  \\
Levy & 9.87 & 7.68-13.43 & 13 & Levy 1990 XX comet, norm. by 266K blackbody continuum  \\
\enddata
\tablecomments{
As in Table~\ref{templatesOBS}, but listing the extended set of optical depth profile templates based on observed, astrophysical sources. 
These two comets were selected from a larger sample of 9 such objects, because they provided the best fits to our system. 
For ease in comparison, the profiles in Table~\ref{templatesOBS} are repeated here.
REFS: (1) \citet{Spoon}; (2) \citet{Spoon} based on data from \citet{Chiar06}; (3) \citet{Chiar06}; (4) \citet{Bowey01} based on data from \citet{Forrest};
(5) \citet{Bowey02} based on data from \citet{Bowey98}; (6) \citet{Chiar11}; (7) \citet{vanBreemen}; (8) \citet{Bowey03} based on data from \citet{Bowey01};
(9) \citet{Bowey03} based on data from \citet{Hanner}; (10) \citet{mucep}; (11) \citet{Chiar06} based on data from \citet{Kemper02}; 
(12) \citet{Hanner94} based on data from \citet{Lynch89}; \& (13) \citet{Hanner94} based on data from \citet{Lynch92}.}
\tablenotetext{a}{A nearly identical profile, but with the 11.3 PAH complex masked, has also been utilized in fitting; these additional profiles have a ``t" appended to the profile name.}
\end{deluxetable}
\begin{deluxetable}{lrrrrrr}
\tabletypesize{\scriptsize}
\tablecaption{Expanded Observational Optical Depth Fits for PKS 1830-211 Absorber \label{OBSfits}}
\tablewidth{0pt}
\tablehead{
\colhead{Profile} & \colhead{$\tau_{10}$} & \colhead{$\chi_{r,10}^2$} & \colhead{Fit-$\lambda_{range}$}& \colhead{$\tau_{full}$} & \colhead{$\chi_{r,full}^2$} & \colhead{Fit-$\lambda_{range}$}}
\startdata
 \cutinhead{EXTRAGALACTIC SOURCES: ULIRGS} 
UL00397 & 0.11$\pm$0.02 & 12.00 & 8.60-12.50 & 0.11$\pm$0.03 & 9.02 & 8.00-19.45 \\
UL00397t & 0.11$\pm$0.02 & 11.53 & 8.60-12.50 & 0.11$\pm$0.03 & 8.81 & 8.00-19.45 \\
UL06301 & 0.13$\pm$0.03 & 10.27 & 8.60-12.50 & 0.13$\pm$0.03 & 7.57 & 8.16-19.45 \\
UL06301t & 0.13$\pm$0.03 & 9.42 & 8.60-12.50 & 0.13$\pm$0.03 & 7.20 & 8.16-19.45 \\
U18443 & 0.13$\pm$0.03 & 11.73 & 8.60-12.50 & 0.14$\pm$0.04 & 8.15 & 8.38-19.45 \\
U18443t & 0.13$\pm$0.03 & 11.45 & 8.60-12.50 & 0.14$\pm$0.04 & 8.01 & 8.38-19.45 \\
\cutinhead{ DIFFUSE ISM}
GCS3 & 0.12$\pm$0.03 & 15.59 & 8.60-12.50 & 0.13$\pm$0.04 & 10.46 & 8.00-19.45 \\
WR118 & 0.12$\pm$0.03 & 14.92 & 8.60-12.50 & 0.13$\pm$0.04 & 9.67 & 8.00-19.45 \\
\cutinhead{DENSE MOLECULAR CLOUD/CLOUD COMPLEX}
Trap & 0.11$\pm$0.02 & 13.30 & 8.60-12.50 & 0.10$\pm$0.02 & 14.30 & 8.06-12.83 \\
Tau16 & 0.11$\pm$0.02 & 13.62 & 8.60-12.50 & 0.10$\pm$0.02 & 14.40 & 8.07-12.82 \\
IC5146 & 0.12$^{+0.03}_{-0.02}$ & 13.09 & 8.60-12.50 & 0.11$\pm$0.03 & 11.32 & 8.00-13.97 \\
$\rho$Oph & 0.11$^{+0.03}_{-0.02}$ & 16.28 & 8.60-12.29 & 0.10$\pm$0.03 & 16.69 & 8.00-12.29 \\
\cutinhead{T-TAURI STAR EMBEDDED IN MOLECULAR CLOUD}
emb-Tau & 0.11$\pm$0.02 & 13.35 & 8.60-12.50 & 0.10$^{+0.03}_{-0.02}$ & 14.30 & 8.08-12.83 \\
emb-$\rho$Oph & 0.12$\pm$0.03 & 14.49 & 8.60-12.50 & 0.12$\pm$0.03 & 13.32 & 8.35-13.29 \\
\cutinhead{STELLAR MATERIAL}
$\mu$Cep & 0.12$\pm$0.03 & 14.07 & 8.60-12.50 & 0.12$\pm$0.03 & 12.75 & 8.00-13.25 \\
AGB & 0.12$^{+0.03}_{-0.02}$ & 10.62 & 8.60-12.50 & 0.12$\pm$0.03 & 7.94 & 8.00-19.45 \\
\cutinhead{COMETS}
Bradfield & 0.16$\pm$0.03 & 8.36 & 8.60-12.50 & 0.16$\pm$0.04 & 8.71 & 8.00-13.04 \\
Levy & 0.16$\pm$0.03 & 9.94 & 8.60-12.50 & 0.16$\pm$0.04 & 9.02 & 8.00-13.43 \\
\enddata
\tablecomments{Identical to Table~\ref{fits}, but for the expanded set of observationally-based optical depth profile templates.
For ease in comparison, the results for observational profiles in Table~\ref{fits} are reproduced here. The best fit is produced by the Bradfield comet in the 10~$\micron$ fitting range, and by the AGB star
and ULIRGs when considering the full fitting range. None of these fits are as good as those found for the crystalline olivine silicates.}
\end{deluxetable}
\begin{deluxetable}{lcrll}
\rotate
\tabletypesize{\scriptsize}
\tablecaption{Expanded Template Summary: Laboratory Profiles \label{APtemplatesLAB}}
\tablewidth{0pt}
\tablehead{
\colhead{Profile} & \colhead{$\lambda_{peak}$} & \colhead{$\lambda_{range}$} & \colhead{Ref} & \colhead{Description}}
\startdata
\cutinhead{AMORPHOUS OLIVINE (particle shape var.)}
AmOliv & 9.76 & 7.28-33.87 & 1 & amorphous olivine  \\
AmOlivGPC & 10.06 & 8.02-29.50 & 2 & amorphous olivine-porous, CDE particle distribution  \\
AmOliv-GS & 9.85 & 8.12-29.74 & 2 & amorphous olivine-solid, spherical particle distribution  \\
\cutinhead{AMORPHOUS OLIVINE (particle size var.)}
AmOliv1.5 & 10.15 & 8.36-39.41 & 3 & amorphous olivine-1.5~$\micron$ particle size  \\
AmOliv2.0 & 10.51 & 8.17-39.46 & 3 & amorphous olivine-2~$\micron$ particle size  \\
AmOlivRay & 9.77 & 8.16-39.73 & 3 & amorphous olivine-Rayleigh limit particle size  \\
\cutinhead{CRYSTALLINE OLIVINE (Mg$_{2x}$Fe$_{2-2x}$SiO$_4$) - composition variations}
Fayalite & 11.43 & 2.00-199.44 & 4 & synthetic fayalite (x=0.0)  \\
Forst & 11.21 & 1.67-199.44 & 4 & synthetic forsterite (x=1.0)  \\
Horton & 11.30 & 1.67-669.09 & 4 & natural hortonolite(x=0.55)  \\
Olivine & 11.21 & 2.00-199.44 & 4 & natural olivine (x=0.94)  \\
HortonB & 11.31 & 7.80-13.30 & 5 & hortonolite(x=0.55)  \\
Chrysolite & 11.15 & 7.84-13.30 & 5 & olivine-chrysolite   \\
SanCarlT100 & 11.18 & 8.00-99.72 & 6 & San Carlos obs. olivine (T=100K)  \\
\cutinhead{CRYSTALLINE OLIVINE (Mg$_{1.9}$Fe$_{0.1}$SiO$_{4}$)- particle shape variations}
PowderOliv. & 11.21 & 2.00-199.44 & 7 & ground natural olivine crystals, powder in KBr \& PE, sizes $<$1$\micron$  \\
OlivSphere & 10.67 & 2.00-10000.00 & 7 & spherical particles in Rayleigh limit   \\
OlivCDE1 & 11.19 & 2.00-10000.00 & 7 & CDE(1) particle sizes in Rayleigh limit  \\
OlivCDE2 & 10.92 & 2.00-10000.00 & 7 & CDE(2) particle sizes in Rayleigh limit  \\
\cutinhead{CRYSTALLINE FORSTERITE (Mg$_{2}$SiO$_4$) - temperature variations}
ForstT300 & 11.19 & 8.00-99.72 & 6 & Jena forsterite T=300K  \\
ForstT200 & 11.19 & 8.00-99.72 & 6 & Jena forsterite T=200K  \\
ForstT100 & 11.18 & 8.00-99.72 & 6 & Jena forsterite T=100K  \\
ForstT10 & 11.18 & 8.00-99.72 & 6 & Jena forsterite T=10K  \\
\cutinhead{CRYSTALLINE FAYALITE (Fe$_{2}$SiO$_4$ ) - temperature variations}
FayalT300 & 11.40 & 8.00-99.72 & 6 & Jena fayalite T=300K  \\
FayalT200 & 11.38 & 8.00-99.72 & 6 & Jena fayalite T=200K  \\
FayalT100 & 11.37 & 8.00-99.72 & 6 & Jena fayalite T=100K  \\
FayalT10 & 11.37 & 8.00-99.72 & 6 & Jena fayalite T=10K  \\
 \cutinhead{OTHER AMORPHOUS SILICATES}               
AmPyr-GPC & 9.50 & 7.00-29.52 & 2 & amorphous pyroxene-porous, CDE particle distribution  \\  
AmPyr & 9.16 & 7.26-33.81 & 1 & amorphous pyroxene  \\
AmSilica & 9.04 & 7.17-31.25 & 8 & amorphous SiO$_2$  \\
 \cutinhead{CRYSTALLINE PYROXENE (Mg$_x$Fe$_{1-x}$SiO$_3$)}                 
Bronzite & 10.49 & 8.34-505.90 & 4 & natural orthobronzite (x=0.88)  \\
Enstat-nat & 9.80 & 1.67-669.09 & 4 & natural orthoenstatite (x=0.96)  \\
Enstat-syn & 9.27 & 1.67-199.44 & 4 & synthetic clinoenstatite (x=1.0)   \\
Hyperst & 10.57 & 1.67-199.44 & 4 & natural orthohypersthene (x=0.65)  \\
AvgPryrox. & 9.33 & 7.78-13.32 & 5 & avg. pyroxene (amalgam enstatite, bronzite, hypersthene, anthophyllite, diopside, pyroxene-pigeonite, wollastonite, augite)  \\
 \cutinhead{OTHER SILICATES}            
Chamosite & 10.00 & 7.81-12.50 & 9 & chamosite (phyllosilicate Fe$\cdot\cdot_4$Al[Si$_3$AlO$_{10}$][OH]$_6\cdot$nH$_2$O)  \\
Chlorite & 10.10 & 8.13-12.50 & 9 & chlorite (phyllosilicate (Mg,Fe)$_{6-p}$(Al,Fe)$_{2p}$Si$_{4-p}$O$_{10}$[OH]$_8$)  \\
Montmor. & 9.52 & 7.69-12.50 & 9 & montmorillonite (phyllosilicate m\{Mg$_3$[Si$_4$O$_{10}$][OH]$_2$\}$\cdot\cdot$p\}(Al,Fe$\cdot\cdot\cdot$)$_2$[Si$_4$O$_{10}$][OH]$_2$\}$\cdot$nH$_2$O)  \\
Serp. & 10.42 & 8.13-11.91 & 9 & serpentine (phyllosilicate Mg$_6$[Si$_4$O$_{10}$][OH]$_8$)  \\
AvgSerp. & 10.31 & 7.80-13.28 & 5 & avg. serpentine (amalgam antigorite, chrysotile, serpentine)  \\ 
21Fer & 9.15 & 7.77-13.28 & 5 & 21Ferraro mixture (amalgam 21 minerals incl. olivines, pyroxenes, serpentines, clays, talcs, silicas)  \\
AvgSil. & 9.19 & 7.79-13.31 & 5 & avg. silica (amalgam quartz, fumed silica, hydrated silica, opal, micronized am. silica)  \\
 \cutinhead{SiC}          
B0-black-aSiC & 11.36 & 6.67-31.25 & 10 & black $\alpha$-SiC  (B0 sample)\\
G0-green-aSiC & 11.36 & 6.67-31.25 & 10 & green $\alpha$-SiC (G0 sample) \\
Lonza-aSiC & 11.86 & 8.01-16.00 & 11 & high temp $\alpha$-SiC product: Lonza UF-15 sample  \\
AJ-aSiC & 11.95 & 8.01-15.99 & 11 & high temp $\alpha$-SiC product: AJ-$\alpha$ sample  \\
\enddata
\tablecomments{
As in Table~\ref{templatesLAB}, but listing an extended set of optical depth profile templates based on laboratory sources.
For ease in comparison, the profiles in Table~\ref{templatesLAB} are repeated here.
REFS: (1) \citet{Spoon} based on data from \citet{Fabian01}; (2) \citet{Chiar06} based on data from \citet{Henning99}; (3) \citet{Dorschner95};
(4) \citet{Jaeger} with tabulated data provided by C. J\"{a}ger; (5) \citet{Bowey02} based on data from \citet{Ferraro}; (6) \citet{Koike06};
(7) \citet{Fabian01}; (8) \citet{Fabian00}; (9) \citet{Dorschner78}; (10) \citet{Friedemann}; \& (11) \citet{Mutschke}.}
\end{deluxetable}
\begin{deluxetable}{lrrrrrr}
\tabletypesize{\scriptsize}
\tablecaption{Expanded Laboratory Optical Depth Fits for PKS 1830-211 Absorber \label{LABfits}}
\tablewidth{0pt}
\tablehead{
\colhead{Profile} & \colhead{$\tau_{10}$} & \colhead{$\chi_{r,10}^2$} & \colhead{Fit-$\lambda_{range}$}& \colhead{$\tau_{full}$} & \colhead{$\chi_{r,full}^2$} & \colhead{Fit-$\lambda_{range}$}}
\startdata
\cutinhead{AMORPHOUS OLIVINE (particle shape var.)}
AmOliv & 0.11$^{+0.03}_{-0.02}$ & 14.79 & 8.60-12.50 & 0.11$\pm$0.03 & 9.60 & 8.00-19.45 \\
AmOlivGPC & 0.13$\pm$0.03 & 10.46 & 8.60-12.50 & 0.13$\pm$0.03 & 7.19 & 8.02-19.45 \\
AmOliv-GS & 0.13$\pm$0.03 & 14.29 & 8.60-12.50 & 0.13$\pm$0.04 & 8.84 & 8.12-19.45 \\
\cutinhead{AMORPHOUS OLIVINE (particle size var.)}
AmOliv1.5 & 0.13$^{+0.03}_{-0.02}$ & 9.85 & 8.60-12.50 & 0.12$\pm$0.03 & 6.81 & 8.36-19.45 \\
AmOliv2.0 & 0.12$\pm$0.02 & 9.35 & 8.60-12.50 & 0.11$\pm$0.03 & 7.02 & 8.17-19.45 \\
AmOlivRay & 0.13$\pm$0.03 & 14.32 & 8.60-12.50 & 0.13$\pm$0.04 & 8.81 & 8.16-19.45 \\
\cutinhead{CRYSTALLINE OLIVINE (Mg$_{2x}$Fe$_{2-2x}$SiO$_4$) - composition variations}
Fayalite & 0.23$\pm$0.04 & 6.33 & 8.60-12.50 & 0.20$\pm$0.05 & 6.74 & 8.00-19.45 \\
Forst & 0.27$\pm$0.05 & 5.09 & 8.60-12.50 & 0.25$\pm$0.07 & 6.04 & 8.00-19.45 \\
Horton & 0.27$\pm$0.05 & 3.73 & 8.60-12.50 & 0.26$^{+0.07}_{-0.06}$ & 4.78 & 8.00-19.45 \\
Olivine & 0.28$^{+0.06}_{-0.05}$ & 7.32 & 8.60-12.50 & 0.26$\pm$0.07 & 6.52 & 8.00-19.45 \\
HortonB & 0.26$\pm$0.05 & 4.61 & 8.60-12.50 & 0.25$^{+0.06}_{-0.05}$ & 5.33 & 8.00-13.31 \\
Chrysolite & 0.28$\pm$0.05 & 6.04 & 8.60-12.50 & 0.27$\pm$0.06 & 6.74 & 8.00-13.30 \\
SanCarlT100 & 0.27$^{+0.06}_{-0.05}$ & 8.50 & 8.60-12.50 & 0.26$^{+0.07}_{-0.06}$ & 7.56 & 8.00-19.45 \\
\cutinhead{CRYSTALLINE OLIVINE (Mg$_{1.9}$Fe$_{0.1}$SiO$_{4}$)- particle shape variations}
PowderOliv. & 0.28$^{+0.06}_{-0.05}$ & 7.32 & 8.60-12.50 & 0.27$^{+0.08}_{-0.07}$ & 6.85 & 8.00-19.45 \\
OlivSphere & 0.19$^{+0.08}_{-0.07}$ & 30.23 & 8.60-12.50 & 0.22$^{+0.12}_{-0.11}$ & 18.31 & 8.00-19.45 \\
OlivCDE1 & 0.19$^{+0.08}_{-0.07}$ & 30.23 & 8.60-12.50 & 0.22$^{+0.12}_{-0.11}$ & 18.31 & 8.00-19.45 \\
OlivCDE2 & 0.23$\pm$0.05 & 14.33 & 8.60-12.50 & 0.23$\pm$0.07 & 10.31 & 8.00-19.45 \\
\cutinhead{CRYSTALLINE FORSTERITE (Mg$_{2}$SiO$_4$) - temperature variations}
ForstT300 & 0.27$\pm$0.05 & 6.52 & 8.60-12.50 & 0.25$\pm$0.07 & 7.13 & 8.00-19.45 \\
ForstT200 & 0.27$\pm$0.05 & 7.21 & 8.60-12.50 & 0.26$\pm$0.07 & 7.44 & 8.00-19.45 \\
ForstT100 & 0.27$\pm$0.05 & 7.40 & 8.60-12.50 & 0.26$\pm$0.07 & 7.61 & 8.00-19.45 \\
ForstT10 & 0.27$\pm$0.05 & 7.62 & 8.60-12.50 & 0.25$\pm$0.07 & 7.88 & 8.00-19.45 \\
\cutinhead{CRYSTALLINE FAYALITE (Fe$_{2}$SiO$_4$ ) - temperature variations}
FayalT300 & 0.31$\pm$0.06 & 6.17 & 8.60-12.50 & 0.29$^{+0.07}_{-0.06}$ & 5.24 & 8.00-19.45 \\
FayalT200 & 0.32$\pm$0.06 & 6.14 & 8.60-12.50 & 0.30$\pm$0.08 & 5.29 & 8.00-19.45 \\
FayalT100 & 0.33$^{+0.07}_{-0.06}$ & 6.25 & 8.60-12.50 & 0.31$\pm$0.08 & 5.33 & 8.00-19.45 \\
FayalT10 & 0.33$^{+0.07}_{-0.06}$ & 6.38 & 8.60-12.50 & 0.31$\pm$0.08 & 5.36 & 8.00-19.45 \\
 \cutinhead{OTHER AMORPHOUS SILICATES}         
AmPyr-GPC & 0.11$\pm$0.02 & 14.42 & 8.60-12.50 & 0.11$\pm$0.03 & 10.40 & 8.00-19.45 \\
AmPyr & 0.10$\pm$0.03 & 21.60 & 8.60-12.50 & 0.10$^{+0.04}_{-0.03}$ & 14.19 & 8.00-19.45 \\
AmSilica & 0.09$\pm$0.03 & 28.02 & 8.60-12.50 & 0.08$^{+0.05}_{-0.04}$ & 19.19 & 8.00-19.45 \\
 \cutinhead{CRYSTALLINE PYROXENE (Mg$_x$Fe$_{1-x}$SiO$_3$)}       
Bronzite & 0.13$\pm$0.03 & 13.55 & 8.60-12.50 & 0.13$\pm$0.04 & 9.38 & 8.34-19.45 \\
Enstat-nat & 0.13$\pm$0.03 & 17.86 & 8.60-12.50 & 0.14$^{+0.05}_{-0.04}$ & 11.37 & 8.00-19.45 \\
Enstat-syn & 0.14$\pm$0.03 & 13.39 & 8.60-12.50 & 0.14$\pm$0.04 & 8.92 & 8.00-19.45 \\
Hyperst & 0.12$\pm$0.03 & 12.79 & 8.60-12.50 & 0.12$\pm$0.04 & 9.65 & 8.00-19.45 \\
AvgPryrox. & 0.11$\pm$0.02 & 13.24 & 8.60-12.50 & 0.10$\pm$0.03 & 13.00 & 8.00-13.32 \\
 \cutinhead{OTHER SILICATES}      
Chamosite & 0.14$\pm$0.03 & 13.71 & 8.60-12.50 & 0.14$\pm$0.03 & 12.95 & 8.00-12.50 \\
Chlorite & 0.20$\pm$0.05 & 15.83 & 8.60-12.50 & 0.20$\pm$0.05 & 14.59 & 8.13-12.50 \\
Montmor. & 0.11$\pm$0.03 & 24.42 & 8.60-12.50 & 0.10$\pm$0.03 & 22.44 & 8.00-12.50 \\
Serp. & 0.21$^{+0.04}_{-0.03}$ & 17.19 & 8.60-11.91 & 0.21$\pm$0.05 & 15.59 & 8.13-11.91 \\
AvgSerp. & 0.15$\pm$0.03 & 14.82 & 8.60-12.50 & 0.14$\pm$0.04 & 13.23 & 8.00-13.28 \\
21Fer & 0.11$^{+0.03}_{-0.02}$ & 15.80 & 8.60-12.50 & 0.10$\pm$0.03 & 15.02 & 8.00-13.28 \\
AvgSil. & 0.10$\pm$0.03 & 27.66 & 8.60-12.50 & 0.09$\pm$0.04 & 23.97 & 8.00-13.31 \\
 \cutinhead{SiC}        
B0-black-aSiC & 0.27$\pm$0.06 & 13.98 & 8.60-12.50 & 0.27$\pm$0.09 & 11.57 & 8.00-19.45 \\
G0-green-aSiC & 0.28$\pm$0.06 & 14.30 & 8.60-12.50 & 0.28$\pm$0.09 & 10.83 & 8.00-19.45 \\
Lonza-aSiC & 0.18$\pm$0.06 & 27.01 & 8.60-12.50 & 0.18$\pm$0.08 & 17.13 & 8.01-16.00 \\
AJ-aSiC & 0.19$\pm$0.06 & 26.56 & 8.60-12.50 & 0.18$\pm$0.08 & 16.95 & 8.01-15.99 \\
\enddata
\tablecomments{Identical to Table~\ref{fits}, but for the expanded set of laboratory-based optical depth profile templates.
For ease in comparison, the results for the laboratory profiles in Table~\ref{fits} are repeated here. The best fit is produced by hortonolite ($\chi_r^2=$3.73 over the 10~$\micron$ region 
and $\chi_r^2=4.78$ over the expanded fitting region.) In general the crystalline olivines produce the best fits, although slight variations in metallicity/chemical composition, temperature, and particle
shape can have a marked effect on the profiles and derived fits.}
\end{deluxetable}
\clearpage
\clearpage
\clearpage
\begin{figure}
\epsscale{0.80}
\plotone{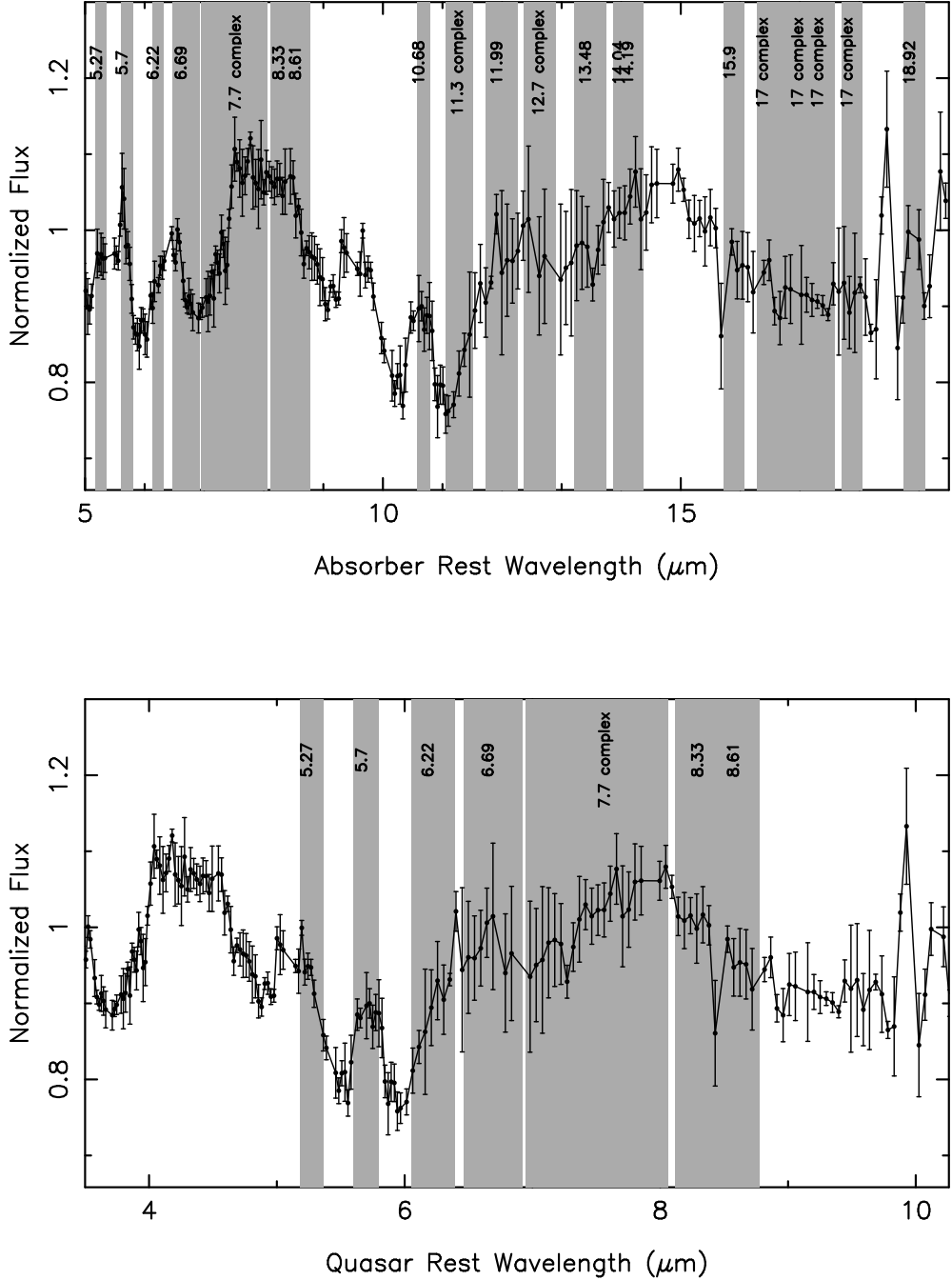}
\caption{PKS 1830-211 spectrum overlaid in grey shading with locations of the PAH features
taken from \citet{Smith}. For features below the resolution of the given IRS spectral order, the unresolved line limit FWHM was adopted from \citet{Smith}. It appears that PAH emission could explain some or all of the peak shortward of the 10~$\micron$ 
absorption feature. Likewise, it may be contributing to the rise longward of the 10~$\micron$ feature, although we note that this is the region also contaminated by the scattered light feature discussed in Appendix~\ref{scatlight}. We
note, however, that the usually prominent 11.3~$\micron$ PAH complex is absent from our spectrum, or alternatively, that the rightmost absorption substructure valley may be deeper with some superposed 11.3~$\micron$ emission.}
\label{PAHS}
\end{figure}
\begin{figure}
\epsscale{0.9}
\plottwo{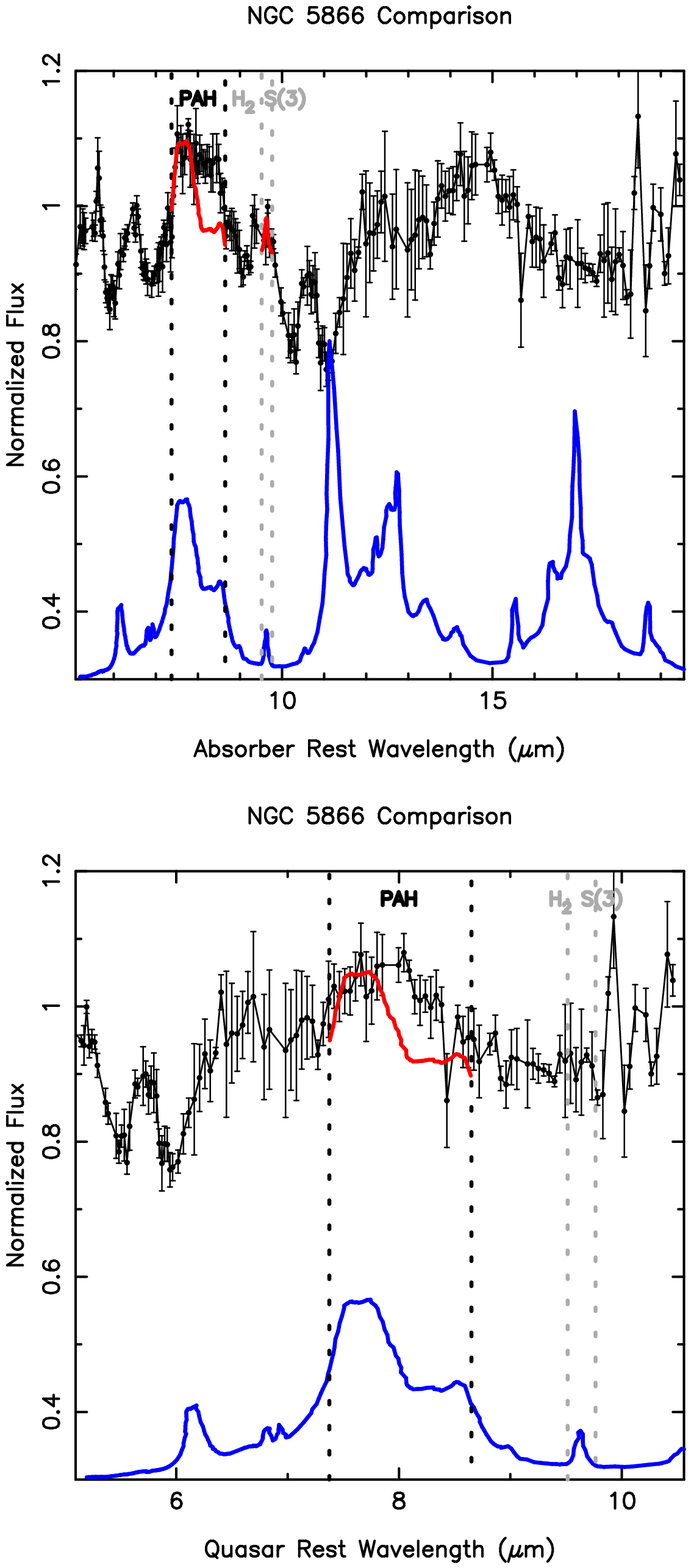}{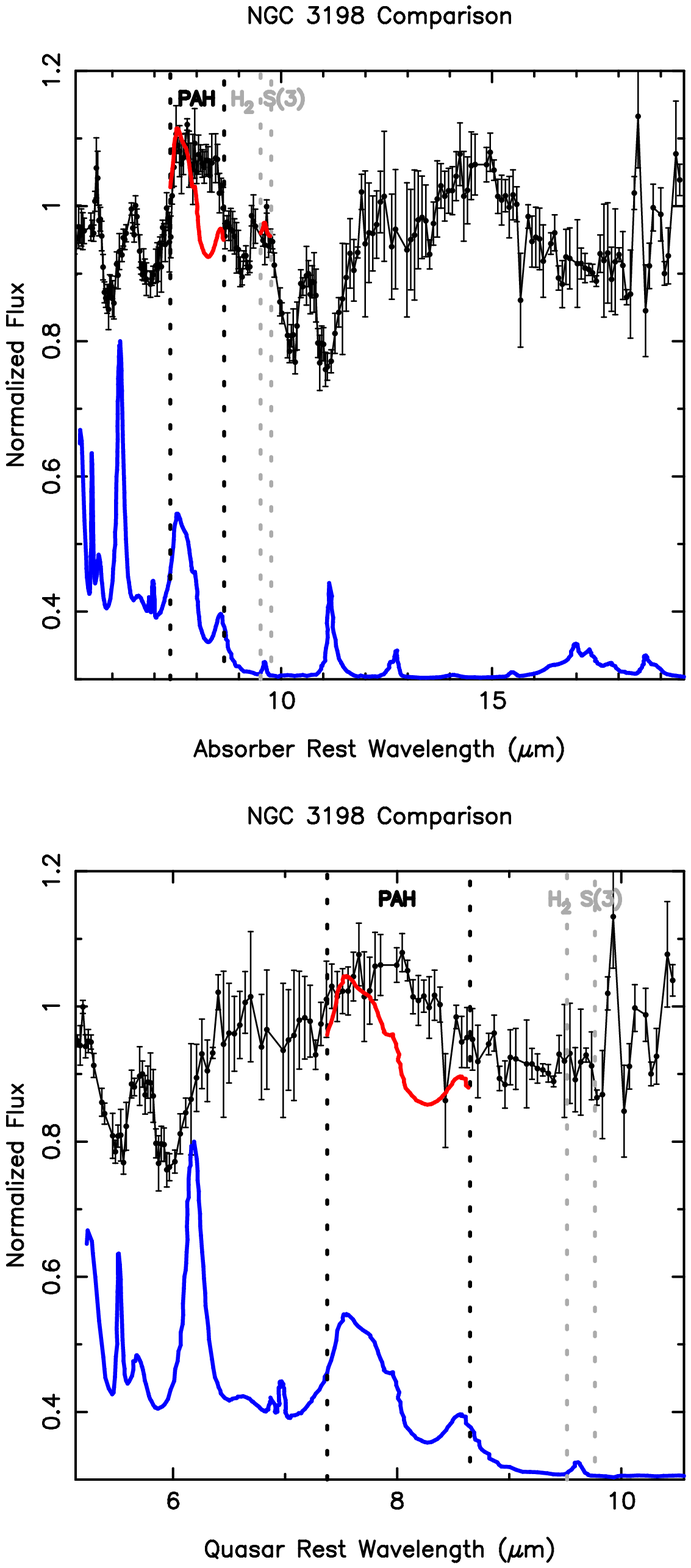}
\caption{Comparison of the PKS 1830-211 spectrum with the PAH-dominated, rest-frame, IRS spectra of star-forming galaxies NGC 5866 and NGC 3198, both noted to be silicate rich, taken from \citet{Smith}. 
These comparison spectra (shown in blue) have been normalized
by the dust+stellar continuum from \citet{Smith} and rescaled/shifted to extend from 0.3-0.8. The black vertical lines demarcate the region encompassing the 7.7 and 8.3~$\micron$ PAH features, while the light grey
vertical lines indicate the region spanned by the H$_2$S (3) feature, in the star-forming galaxies. In red, we show these prominent features in the star-forming galaxy spectrum rescaled and 
shifted upward to facilitate comparison with similar features in PKS 1830-211. (In the quasar rest-frame panels we omit the red rescaled/shifted H$_2$ S(3) feature.)
We note the alignment of the 7.7~$\micron$ complex PAH feature with the `emission' feature in the PKS 1830-211 in the z=0.886 absorber rest frame, as well as the match in alignment as well as in shape/breadth of the H$_2$ S(3) feature with the small inflection in our spectrum. We do not see evidence of the 11.3~$\micron$ feature in our spectrum, or of the lower wavelength PAH features near 6~$\micron$. However, as illustrated by a comparison between NGC 5866 and NGC 3198, there can be 
significant variations in the relative line strengths between PAH features in different galaxies.}
\label{PAHSil}
\end{figure}
\begin{figure}
\epsscale{1.0}
\plotone{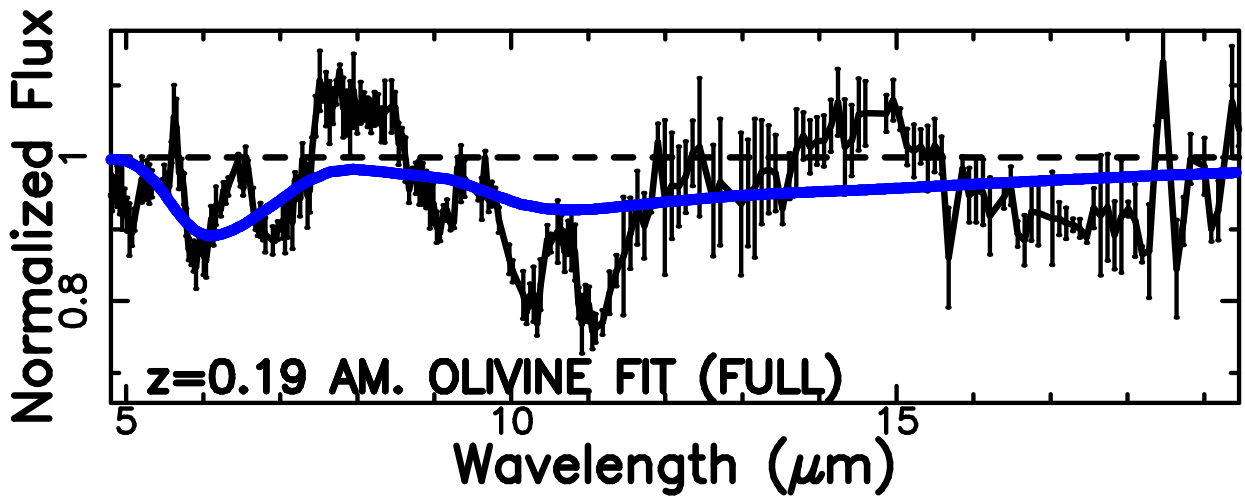}
\caption{Normalized flux profile for PKS 1830-211, in the z=0.886 absorber-rest-frame, overlaid with an amorphous olivine fit to the presumed z=0.19 foreground galaxy silicate absorption. 
While the purported 10~$\micron$ feature could instead be attributed to the 18~$\micron$ feature of the z=0.19 foreground galaxy, the corresponding 10~$\micron$ silicate feature for the foreground galaxy is absent.
Instead, there is an emission-type feature, relative to the surrounding absorption, present at the rest wavelength of the
z=0.19 galaxy 10~$\micron$ silicate feature. Therefore, we do not believe that the structure we see in the PKS 1830-211 spectrum originates from a superposition of 10~$\micron$ absorption from the z=0.886 quasar absorber and 
18~$\micron$ absorption from the z=0.19 foreground galaxy.}
\label{FGND}
\end{figure}
\begin{figure}
\epsscale{0.8}
\plotone{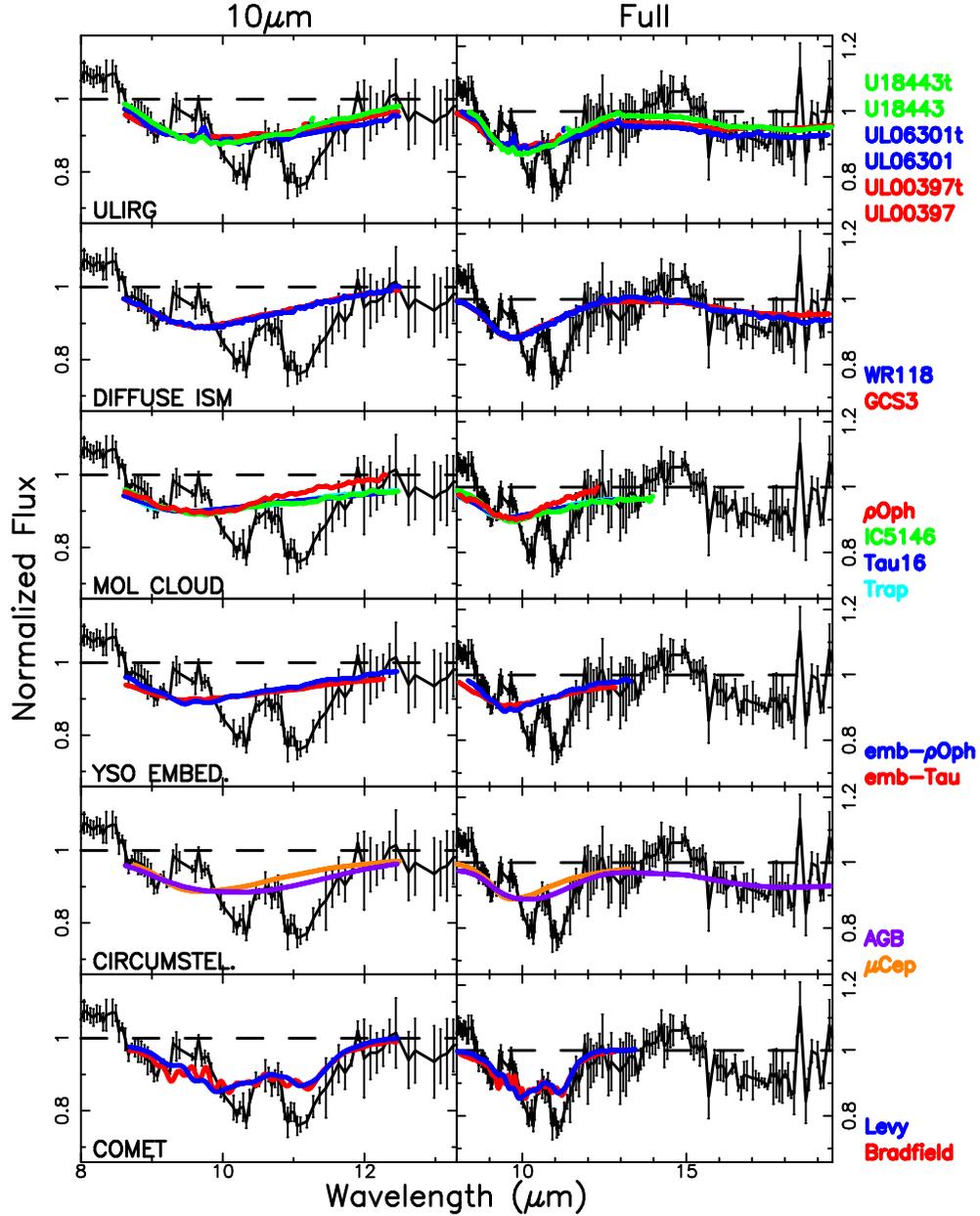}
\caption{Similar to Figure~\ref{fitsfigureOBS}, but showing the fits for an expanded repertoire of observationally-based template profiles, as detailed in Table~\ref{APtemplatesOBS}. The
best fits over the 10~$\micron$ feature are provided by the comets, ULIRGs, and AGB star, all of which tend to be relatively richer in crystalline silicate materials than the other sources.}
\label{AP-OBSFITTING}
\end{figure}
\begin{figure}
\epsscale{0.8}
\plotone{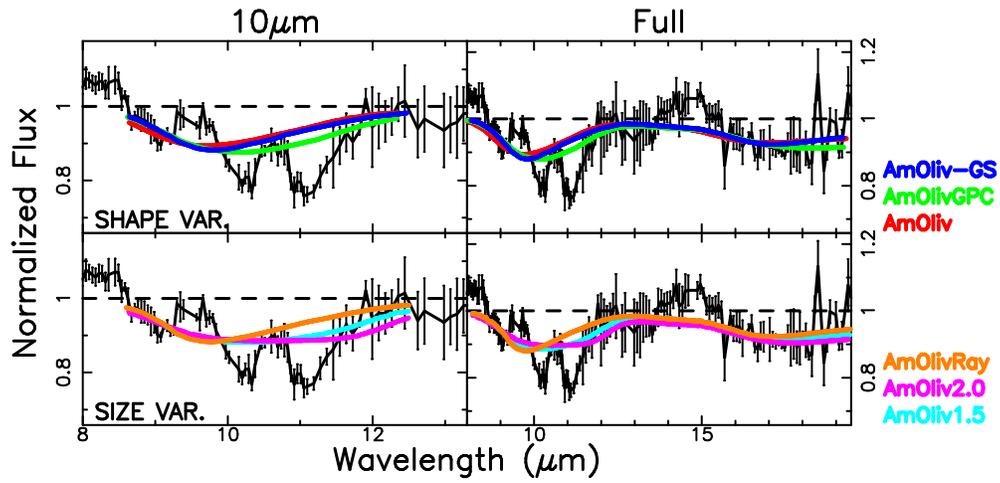}
\caption{Similar to Figure~\ref{fitsfigureLAB}, but showing fits for an expanded set of laboratory amorphous olivine templates, as detailed in Table~\ref{APtemplatesLAB}. The top panel shows variations with grain shape, while the bottom
illustrates variations with grain size. Small variations in these properties can have a marked impact on the profile shape, but they are not adequate to explain the substructure dominating the spectrum.}
\label{AP-LABAMFITTING}
\end{figure}
\begin{figure}
\epsscale{0.8}
\plotone{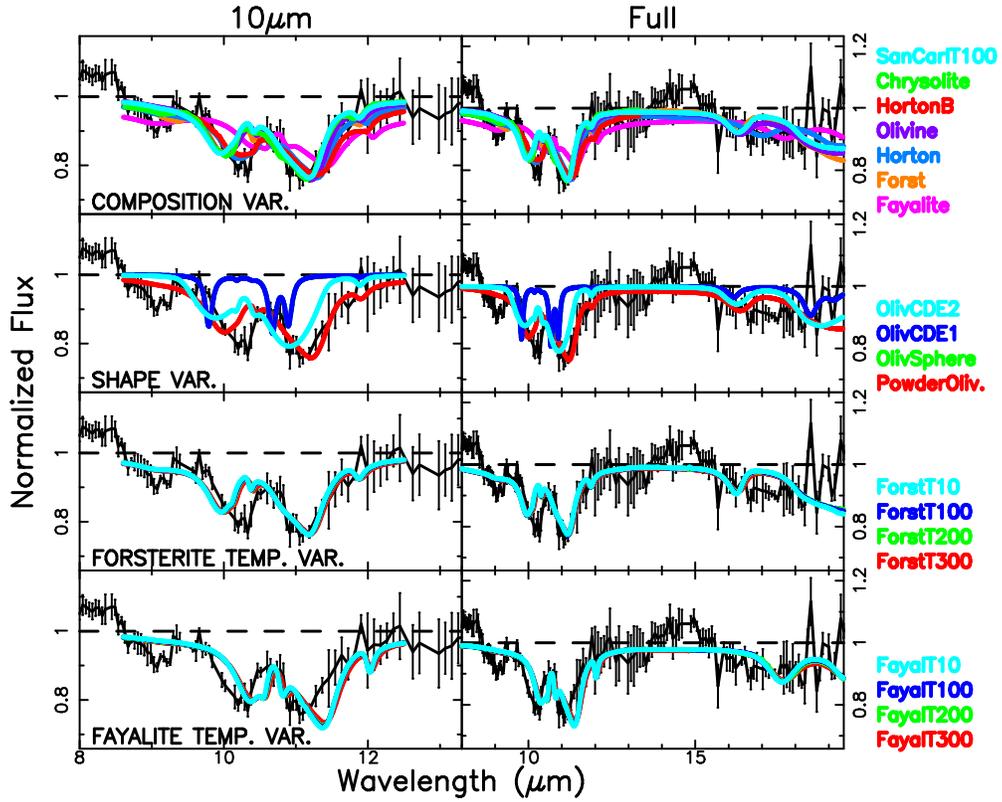}
\caption{Similar to Figure~\ref{fitsfigureLAB}, but showing fits for an expanded set of laboratory crystalline olivine templates, as detailed in Table~\ref{APtemplatesLAB}. In the top panel we illustrate variations 
in composition (i.e. Mg/Fe ratio), in the second panel variations in particle shape, and in the bottom two panels variations in temperature for forsterite and fayalite. As illustrated, small variations in composition and particle
shape can have a substantial impact on the profiles and derived fits. The variations with temperature differences are smaller in this spectral region, but become more pronounced at longer wavelengths.}
\label{AP-LABCRYFITTING}
\end{figure}
\begin{figure}
\epsscale{0.8}
\plotone{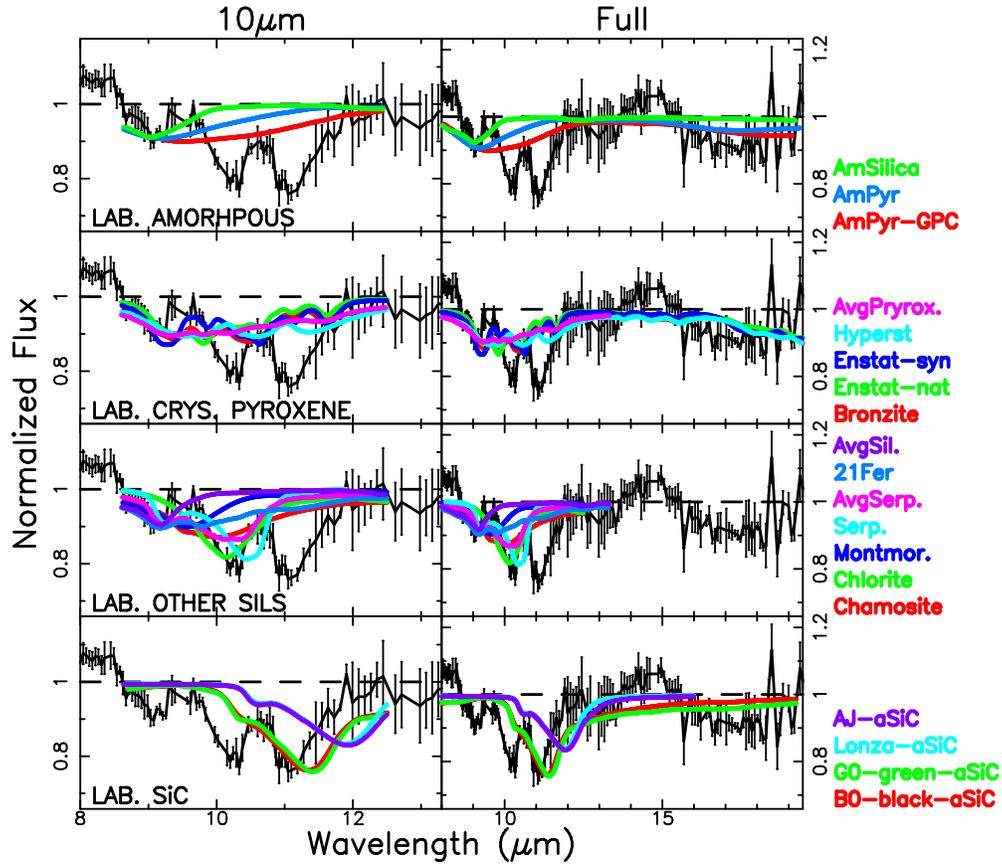}
\caption{Similar to Figure~\ref{fitsfigureLAB}, but showing fits for an expanded set of laboratory templates, as detailed in Table~\ref{APtemplatesLAB}. In the top panel we show amorphous silicates, followed
by crystalline pyroxenes, other laboratory silicates including phyllosilicates, and finally SiC. In comparison with Figure~\ref{AP-LABCRYFITTING}, none of these reproduce the substructure as well as crystalline olivine, although
some compounds can explain the leftmost substructure peak.}
\label{AP-LABFITTING}
\end{figure}

\end{document}